\documentclass{article}
\usepackage[english]{babel}
\usepackage[letterpaper,top=2cm,bottom=2cm,left=3cm,right=3cm,marginparwidth=1.75cm]{geometry}
\usepackage[authoryear]{natbib}

\usepackage{setspace}
\usepackage{graphicx}
\usepackage{hyperref}
\usepackage{booktabs}
\usepackage{pgfplotstable}
\usepackage{caption} 
\usepackage{comment} 
\usepackage{amsmath} 
\usepackage{ragged2e} 
\usepackage{longtable}
\usepackage{pdflscape}
\usepackage{tabularx} 
\usepackage{csvsimple-l3}
\usepackage{xurl}
\usepackage{multirow}   % For cells spanning multiple rows
\usepackage{makecell}   % For line breaks in cells (\makecell{})
\usepackage{ragged2e}               
\usepackage{calc} 
\usepackage{chngcntr}  % Needed for \counterwithin
\usepackage{threeparttable}
\usepackage{multirow}
\usepackage{pdfpages}

\hypersetup{
    colorlinks=true,
    linkcolor=blue,
    filecolor=magenta,      
    urlcolor=cyan,
    pdftitle={Overleaf Example},
    pdfpagemode=FullScreen,
    }

\providecommand{\keywords}[1]
{
  \small	
  \textbf{\textit{Keywords: }} #1
}

\pgfplotsset{compat=1.18} 
 
\title{Extracting O*NET Features from the NLx Corpus to Build Public Use Aggregate Labor Market Data \thanks{This work has been supported by Grant \# 2111-34962 from the Russell Sage Foundation, a grant from Washington Center for Equitable Growth, and data access from the National Labor Exchange Research Hub. Grant funders and The National Labor Exchange (NLx) Data Trust bear no responsibility for the analyses or interpretations of the data presented here. The opinions expressed herein, including any implications for policy, are those of the authors and not of the NLx Data Trust members. We are grateful to NLx and Paul Daniels, Marissa Hashizume, and Amber Gaither, and to Loyola University Chicago, particularly Kathleen Bobay, Ron Price, Jason Boyda, Joe Koral and the Walter F. Mullady, Sr. endowment, for computational infrastructure and support. We thank research assistants Guillaume Bolivard, Chlo\'e Clark, Krish Gandhi, Adam Goode, Quynh Hoang, and Snehil Sharad. We thank Kyle DeMaria, Luis Gonzalez, Lesley Hirsch, Phil Lewis, Ian Page, Micah Sanders and attendees of the National Labor Exchange Research Hub Connect conference for feedback. This publication uses the \href{https://www.onetcenter.org/}{O*NET 29.1 database of the U.S. Department of Labor's Education and Training Administration} and the \href{https://esco.ec.europa.eu/en}{ESCO v. 1.2.0 classification of the European Commission.} We publish code on our Github repository at \url{https://github.com/Job-Ad-Research-at-QSB-LUC/JAAT} and language models at \url{https://huggingface.co/loyoladatamining}.}}

\author{Stephen Meisenbacher\thanks{Technical University of Munich, School of Computation, Information and Technology}\and Svetlozar Nestorov \thanks{Loyola University Chicago, Quinlan School of Business} \and Peter Norlander\thanks{Loyola University Chicago, Quinlan School of Business. Corresponding author. pnorlander@luc.edu.}}

\begin{document}

\includepdf[pages=1]{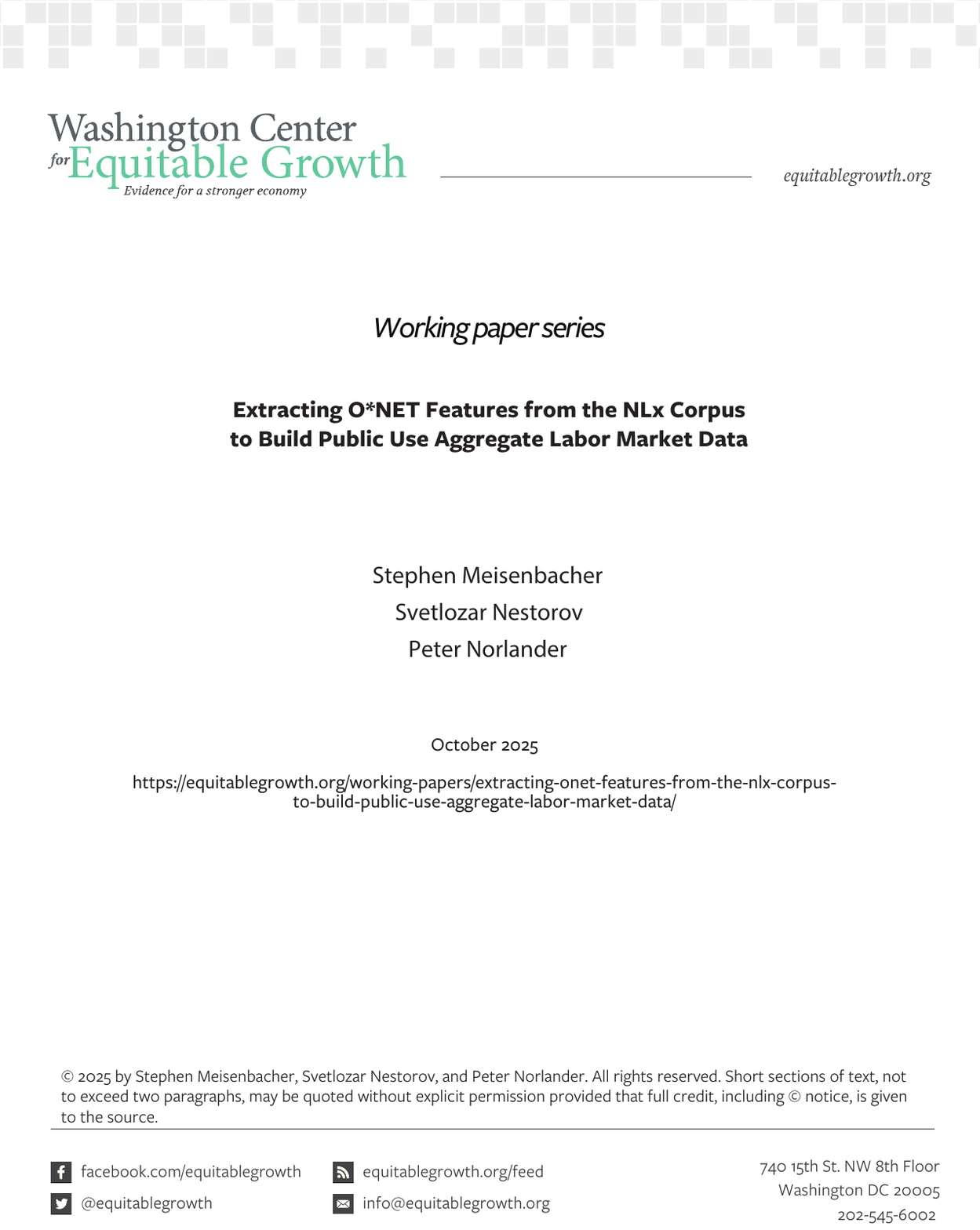}

\maketitle

\begin{abstract}
Data from online job postings are difficult to access and are not built in a standard or transparent manner. Data included in the standard taxonomy and occupational information database (O*NET) are updated infrequently and based on small survey samples. We adopt O*NET as a framework for building natural language processing tools that extract structured information from job postings. We publish the Job Ad Analysis Toolkit (JAAT), a collection of open-source tools built for this purpose, and demonstrate its reliability and accuracy in out-of-sample and LLM-as-a-Judge testing. We extract more than 10 billion data points from more than 155 million online job ads provided by the National Labor Exchange (NLx) Research Hub, including O*NET tasks, occupation codes, tools, and technologies, as well as wages, skills, industry, and more features. We describe the construction of a dataset of occupation, state, and industry level features aggregated by monthly active jobs from 2015 - 2025. We illustrate the potential for research and future uses in education and workforce development.


\end{abstract}

\newpage
\section{Introduction}

The availability of online job ads has contributed to significant advances in research and practice. However, ``data access restrictions '' and the ``lack of standardization across private and public data sources'' have until now limited the use of this data  \citep[p. 117]{committee_on_automation_and_the_us_workforce_an_update_artificial_2024}. The Occupational Information Network (O*NET) is the standard taxonomy of work that serves as a cornerstone for research and professional communities by providing an information architecture for the workplace \citep{onet}. However, measurement of tasks  ``needs substantial work,'' and O*NET's survey-based data collection method is updated slowly and not designed for longitudinal research \citep{committee_on_automation_and_the_us_workforce_an_update_artificial_2024}. We make two contributions aligned with the needs identified above and recommendations of the Department of Labor's Workforce Informatics Advisory Council \citep{hirsch_2024}.

First, we leverage the taxonomic structure of O*NET as a basis for feature extraction from job ad data, deconstructing a massive text corpus of 155 million job ads into billions of data points coded to elements of O*NET's content model. We develop transparent, high-accuracy, efficient, open-source natural language processing (NLP) tools to map language in job ads to standard O*NET features.  We provide these domain-specific, fine-tuned, and open-source machine learning (ML) and embeddings models that leverage encoder-only language models in a GitHub repository (the \href{https://github.com/Job-Ad-Research-at-QSB-LUC/JAAT}{Job Ad Analysis Toolkit (JAAT)}). Such models are generally more accurate than general-purpose large language models (LLMs), including in the job ad domain \citep{nguyen_rethinking_2024,zhang_skillspan_2022}, are more efficient and scalable than LLMs, and permit independent replication of results.

Second, we build and introduce a novel, large-scale description of aggregate workplace trends in the U.S. in the last decade. We create occupation, industry, state, and month level aggregate statistics from job ads provided by the National Labor Exchange (NLx) Research Hub. The NLx Research Hub's job ad corpus is ``the most accurate and comprehensive collection of real, online job openings in the United States'' and provides researchers and practitioners unparalleled access to real-time insight into the labor market. Our data is relevant to academic researchers, and workforce development and education planning professionals in community colleges and higher education. Compared to the aggregate dataset we build, we are unaware of any other dataset at present with as much structured information on as large a sample of job ads, and none that adopts O*NET's framework.  Aggregate data can be made available upon publication.

The layout of the paper is as follows. Section \ref{sec2} introduces NLx and O*NET data and background information on the uses and limitations of existing job ad data. Section \ref{sec3} summarizes methods and validation procedures. Section \ref{sec4} illustrates several potential uses of the data. Section \ref{sec5} concludes with limitations and directions for future work.

\section{O*NET and Job Ad Data}
\label{sec2}

We begin by describing the O*NET architecture for occupational information, and then summarize limitations of survey-based measurement and data frequency. We then describe job ad data, uses, providers, and limitations related to access, standardization, and transparency. This motivates the need for accurate, structured, timely labor market data from job ads -- and tools to build such data according to standard structures -- following methods consistent with scientific standards for replication and transparency.

\subsection{O*NET: The Occupational Information Network}

O*NET is a comprehensive database of occupational information tied to a content model that ``identifies the most important types of information about work and integrates them into a theoretically and empirically sound system.''\footnote{See \url{https://www.onetcenter.org/content.html}, The O*NET Content Model, accessed May 27, 2025.}  O*NET includes crosswalks and explicit relationships between 40 detailed tables of occupation, task, education, experience, tools, technologies, job titles, and more features of the workplace. Like other taxonomies in the sciences, O*NET is the product of efforts to develop, refine, and validate classification schemes, incorporating evolving individual and group judgments \citep{bowker_sorting_2000,abend_words_2023}. 

%O*NET's content model, displayed in Figure \ref{fig:onet_crm}, includes ``worker-oriented'' (top), ``job-oriented'' (bottom), occupation-specific (right-side), and cross occupation (left-side) features.

Table \ref{tab:onet_model_single_table} displays O*NET's content at a depth of two levels. ``Worker-oriented'' features are on the top three rows, ``job-oriented'' on the bottom three rows, occupation-specific features are on the rightmost columns, and cross-occupation features are on the leftmost columns. Below, we cycle through each major section of O*NET to describe our approach to acquiring data on each area. Where the level of detail within the O*NET database and the survey method of data collection have limited O*NET's comprehensiveness, we augment its tables with real job ad text from NLx to boost the available training data for ML, and supplement O*NET by cross-walking its skill elements to the more elaborated ESCO taxonomy of skills.

\begin{table}[htbp!]
\centering % Centers the entire table construct on the page
\small 
\setlength{\arrayrulewidth}{0.4pt} 
\newcolumntype{L}{>{\raggedright\arraybackslash}X}

\begin{tabularx}{\textwidth}{| L | L | L |}
\hline
\textbf{1 Worker Characteristics} &
\textbf{2 Worker Requirements} &
\textbf{3 Experience Requirements} \\
\hline

1.A Abilities &
2.A Basic Skills &
3.A Experience and Training \\

1.B Interests &
2.B Cross-Functional Skills &
3.B Basic Skills - Entry Requirements \\

1.C Work Styles &
2.C Knowledge &
3.C Cross-Functional Skills - Entry Requirements \\

& % Empty cell for 1.D equivalent position
2.D Education &
3.D Licensing \\
\hline\hline % Double horizontal line to separate the two panels
% --- BOTTOM PANEL HEADERS ---
\textbf{4 Occupational Requirements} &
\textbf{5 Occupation-Specific Information} &
\textbf{6 Workforce Characteristics} \\
\hline
% --- BOTTOM PANEL CONTENT ---
4.A Generalized Work Activities &
5.A Tasks &
6.A Labor Market Information \\

4.B Organizational Context &
5.C Title & % Original data has 5.C here
6.B Occupational Outlook \\

4.C Work Context &
5.D Description &
\\ % Empty cell for 6.C equivalent position

4.D Detailed Work Activities &
5.E Alternate Titles &
\\ % Empty cell for 6.D equivalent position

4.E Intermediate Work Activities &
5.F Technology Skills &
\\ % Empty cell for 6.E equivalent position

& % Empty cell for 4.F equivalent position
5.G Tools &
\\ % Empty cell for 6.F equivalent position
\hline
\end{tabularx}
\caption{O*NET's Content Model and structure covers nearly all elements related to work and provides scaffolding for extracting information from job ads.}
\label{tab:onet_model_single_table}
\end{table}

% \begin{figure}
%     \centering
%     \includegraphics[width=.9\textwidth]{pngs/onet_content_model.png}
%     \small 
% \justifying
% \\
% \emph{Note: }We extract features following the structure of the O*NET Content Reference Model. 
%     \caption{The O*NET Content Model}
%     \label{fig:onet_crm}
%     \medskip
 
% \end{figure} 

Within the six major elements of the content reference model lies a hierarchical structure with increasing specificity. There are over 600 elements in the content model at five levels of depth. Each element may contain a great level of additional detail. For example, tasks (5.A.) contains a list of more than 20,000 task statements (5.A.1.) given unique codes that are linked within the O*NET database to 2,072 Detailed (4.D.), 332 Intermediate (4.E.), and 41 General Work Activities (4.A.).  5.E contains a list of over 8,000 job titles and alternative titles that are mapped to 2018 Standard Occupation Codes (SOC). 

Based on surveys of workers, O*NET reports the Level, Importance, and Extent of specific elements within an occupation (see \href{ https://www.onetonline.org/help/online/scales}{O*NET Scales}). The Importance rating ``indicates the degree of importance a particular descriptor is to the occupation.'' The possible ratings range from `Not Important' (1) to `Extremely Important' (5).'' Importance data is available for Tasks, Knowledge, Skills, Abilities, Work Activities, and Work Styles.  Level ``indicates the degree, or point along a continuum, to which a particular descriptor is required or needed to perform the occupation.'' Level is on a 0-7 scale, and covers Knowledge, Skills, Abilities, and Work Activities. Relevance ``refers to the proportion of job incumbents who rated the provided task relevant to his/her job.''

\subsubsection{Uses of O*NET data}

O*NET's measures of occupational task intensity and levels are frequently used by labor economists in what has been called the ``task approach'' \citep{david2013task}. Work in this vein provides a richer view than traditional models of the interaction between worker skills, tasks on the job, and changes in work due to technological or trade shocks \citep{acemoglu2011skills}. Empirically, this often entails dis-aggregating occupations and jobs into the tasks or bundles of tasks (high-level work activities in the O*NET structure) that comprise the job, and studying how changes in the economy impact workers who perform activities that are `routine', `non-routine', `physical', `cognitive', and `interpersonal'  to research labor market trends \citep{deming_qje}.

Researchers often study the exposure of O*NET tasks or task bundles to a technological or other shock that is perceived to be changing the existing organization of work. One influential approach follows \cite{blinder_how_2009}'s study of the offshorability of jobs. Drawing from O*NET's measures of tasks, researchers calculate occupational exposure to a shock and estimate the potential impact on the labor force using a representative sample such as the Current Population Survey. This typically results in estimates of how many jobs are `offshorable' \citep{blinder_how_2009}, `teleworkable' \citep{dingel_how_2020}, `automatable' \citep{gathmann2024ai}, impacted by Large Language Models \citep{eloundou2024gpts}, etc.

Recent work by O*NET incorporates the use of ChatGPT and NLP methods to enhance the taxonomy \citep{lewis2024supplementing,klein2025identification,lewis2025adding}. Computer science researchers adopting O*NET's taxonomy also increasingly use machine learning and NLP methods to extract detailed task information from occupational text such as job descriptions (see, e.g., \citep{putka_evaluating_2023,rounds_using_2023}). \citet{handa2025economic}, for example, map requests from users of the large language model Claude to O*NET's list of task statements to identify the complementarity and automatability of specific tasks. Similarly, \cite{chatterji_how_2025} map user requests to ChatGPT to O*NET's work activities. 

\subsubsection{Limitations of O*NET data}

O*NET's survey based data collection provides superb indicative task information for each occupation, but it is not designed to be longitudinal \citep{david2013task}. Each occupation is updated infrequently, and sample size is small, with an average of 71 observations per occupation from a single point in time. Data collection for the last updated occupations occurred most recently in 2006, according to the metadata reported in O*NET Version 29.1. Over a decade ago, epidemiologists examining the suitability of O*NET data to determine occupational exposure to health and safety risk factors issued a cautionary note advising against its use \citep{cifuentes_use_2010}. Citing poor statistical power, infrequent data collection, and potential for confusion over concepts, the authors concluded that O*NET's task-based measurement of occupations, while promising in its design, lacked proven predictive value or convergent validity. 

It can be difficult to interpret values for each O*NET element. The calculation of occupational measures for Importance, Level, and Relevance may be ``opaque'' and  difficult to interpret \citep{david2013task}, and subject to researcher degrees of freedom \citep{cifuentes_use_2010}. While surveys directly inform the Work Values (1.B.2.), Work Styles (1.C.1.), and Work Activities (4.A.) of workers in specific occupations, the crosswalk between Work Activities and the calculation of values reported in other O*NET elements, including Abilities (1.A.) and Skills (2.A.), is theoretically driven. Assumptions of the O*NET model require all tasks and detailed work activities exist only within a single occupation. Theoretical assumptions driving calculations of some elements may not be empirically justified. 

A third limitation of O*NET for NLP use cases is insufficient detail on some elements (such as skills and organizational context) that are not elaborated at the same level of detail as others (such as task). Thousands of detailed and labeled text elements are often necessary to pursue accurate NLP analysis that follows a taxonomic knowledge structure. For the purpose of extracting structured data from job ads, O*NET's content model could serve as a foundation for many efforts, but in parts, lacks sufficient taxonomic elaboration or adequate text descriptions for text classification and extraction purposes.

\subsection{Job Ad Data}
Real-time large-scale online job ad data and other newer sources of information have significantly enhanced researchers' capabilities to understand labor markets in recent decades \citep{horton2015labor}. For practitioners, projects started over 30 years ago have continuously delivered online job ad data to frontline workforce development professionals to help job seekers in search, referrals, and matching \citep{eberts_frontline_2003}. For over a decade, labor market intelligence data from job ads have been used by employers in workforce planning, in education and curriculum planning, career planning, and economic development \citep{carnevale_understanding_2014}. Policy-makers, media, and the public also rely on aggregate job ad data to understand labor market trends. 

\begin{figure}[htbp]
    \centering
    \includegraphics[scale=0.76]{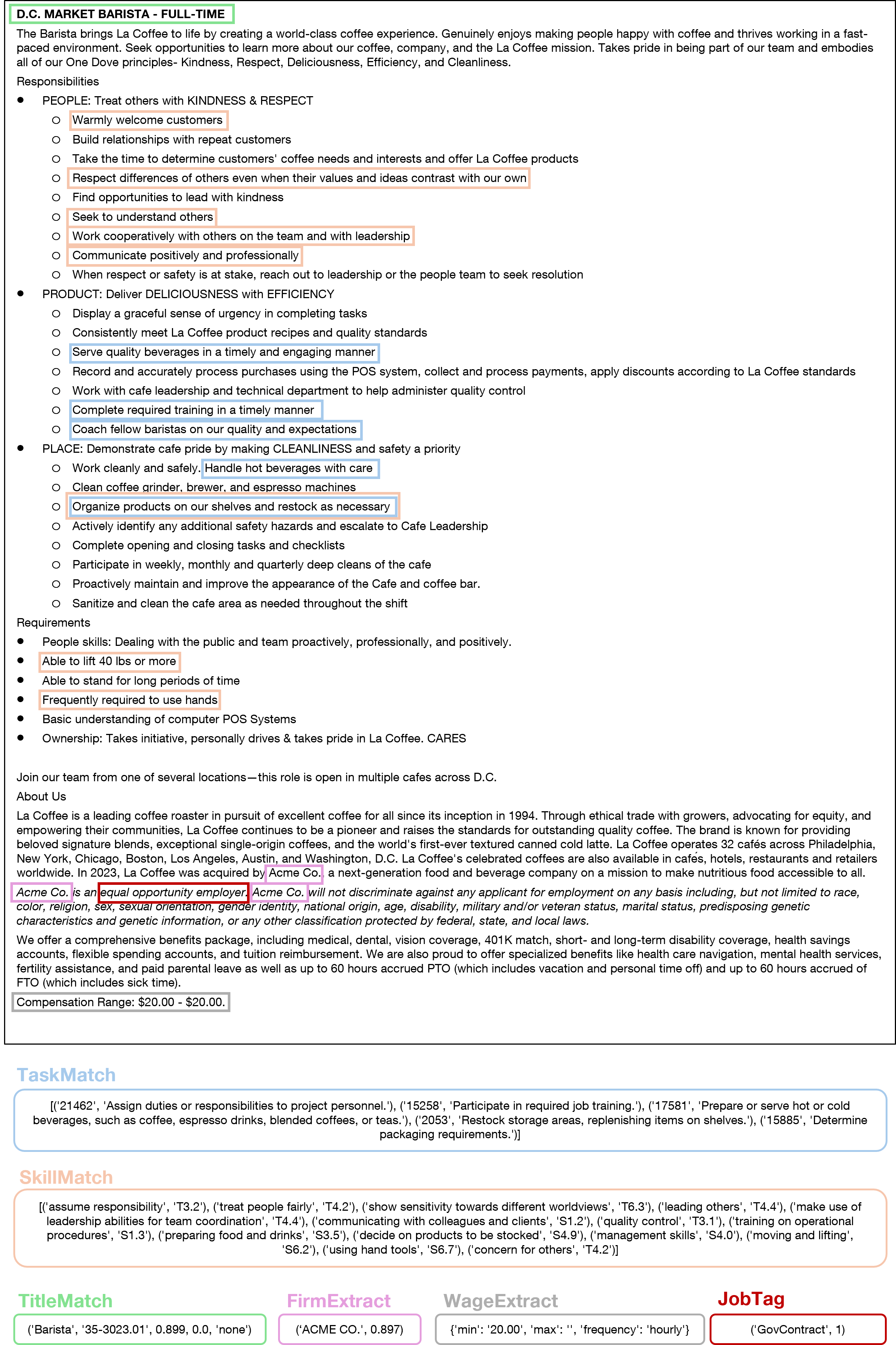}
    \\
    \small
    \justifying
    \emph{Note: }Each individual ML tool (TaskMatch, SkillMatch, TitleMatch, FirmExtract, JobTag) is built with custom, manually audited and validated training data. Actual JAAT outputs are displayed and mapped to their approximate locations in the original job ad. We obtain this ad by searching an online job search portal for ``coffee'' and anonymize the original employer name to \textit{La Coffee} and the parent company to \textit{Acme Co}.
    \caption{An illustrative job ad with features extracted by the Job Ad Analysis Toolkit (JAAT).}
    \label{fig:example}
\end{figure}

Job advertisements often contain granular information on the tasks and skills needed to do a job, required education, licenses, qualifications and preferences, and often include details of working conditions, wages, benefits and more. To illustrate the wealth of information a job advertisement contains that can be mapped to codes from O*NET,  Figure \ref{fig:example} displays a job ad and the actual codes extracted with the ML tools we develop and describe in Section \ref{sec:methods}. While Figure \ref{fig:example} highlights capabilities to extract occupation information, skills requirements, task detail, firm name and industry, and wage information, a great deal of additional information that we structure is not displayed. We separately describe how we use custom and standard dictionaries to capture additional elements of context below. 

\subsubsection{Uses of Job Ad Data}

Job ad data contributes to research on changing skills \citep{hershbein_recessions_2018,clemens2021dropouts}, labor market structure \citep{azar_labor_2022},  the polarization of job skills \citep{alabdulkareem_unpacking_2018}, the importance of language in jobs \citep{marinescu_opening_2020}, strategic management and recruitment strategy \citep{sauerwald_political_2024}, and many more areas. Despite this, aggregate job ad data and other labor market information from commercial sources used in academic papers is rarely made available. Exceptions include labor market concentration \citep{choi_data_2024} and outside options \citep{schubert2024employer} data. 

%Job ad data management practices have improved over time, but collections prior to 2015 have pro.

\subsubsection{Limitations of Job Ad Data}

All job ad data has limitations, summarized well in a technical report \citep{lancaster2019technology}. Researchers are often careful to acknowledge and adjust for these. As advertisements, they are employer's statements intended to attract workers, and may be less detailed than actual job descriptions, contain omissions, and inaccuracies. Online job ads are known to over-represent highly-educated workers and large firms, and to over- or under-represent certain occupations and industries. A single online job posting may represent no or multiple actual vacancies \citep{hashizume_timing}. 

\paragraph{Limitations of Proprietary Data.} Several companies license job ad data to academic researchers. Commercial providers typically sell access to structured data that has been built from job ad text without  disclosure of methods for creating structured information from text, or warranties or description regarding accuracy. In general, models used to build data for research using job ads are trade secrets and unavailable for independent use or testing. One notable exception is TechWolf and associated NLP researchers that have published multiple open-source synthetic and labeled training datasets and tools that adopt the ESCO framework for skills \citep{decorte_jobbert_2021,anand_is_2022, decorte_extreme_2023, decorte_career_2023, decorte_skillmatch_2024, decorte_efficient_2025,  decorte_multilingual_2025}.

%, prominently Lightcast (formerly Burning Glass Technologies or BGT), Revelio Labs, LinkUp, and WageScape

Rising use of proprietary data in academic research risks hindering scientific advances \citep{lazer_computational_2020}. Exaggerated industry claims about insight that is possible only through access to their ``big data'' may be attempts to monopolize the truth, de-emphasize worker and practitioner experience and knowledge, and devalue independent researcher analyses following traditional scientific methods \citep{maffie_mythology_2023}. As the national open-source taxonomy of work and occupations, O*NET provides invaluable insight and grounded data from worker interviews, but has not previously been combined with job ad data. Instead, data providers have developed bespoke libraries and definitions \citep{committee_on_automation_and_the_us_workforce_an_update_artificial_2024}. Generally, these taxonomies are not made readily available for inspection or public use and are difficult to cross-walk to standard sources like O*NET or ESCO. Because many taxonomies depend upon unsupervised learning and are not combined with theory or foundational taxonomies, design choices, such as the number of unsupervised clusters to form, can lead to arbitrary, incompatible, and confusing definitions of skills. For reasons of replication and equity, scientific research standards include making code and data public and `knowing your data source' \citep{aea_ethics}, and more generally, making research findable, accessible, interoperable, and reusable (FAIR) \citep{stall_make_2019}.

Only one independent technical analysis of NLx and a major commercially provided dataset is available: a University of Virginia team accessed both the Lightcast (formerly Burning Glass Technologies or BGT) and NLx data to test the suitability of each data source for use in workforce development \citep{lancaster2019technology}. The Lightcast data is the most frequently used in academic research and often described as representing the `near universe' of online job ads \citep{hansen2023remote}. Benchmarking Lightcast data against NLx in the UVA report finds that in a direct comparison of a sample of job ads in a region in a period, BGT has 24\% more observations than NLx. However, 29\% of BGT observations are duplicates while NLx has only 6\% duplicates.  Providers often state that they source their data from web scraping of employer webpages and job boards, often leading to duplication, and then undertake trade secret processes to de-duplicate and clean the data. 

According to the UVA analysis, the correlation coefficient between the number of observations in a region in the datasets is 0.996. Independent researchers’ findings, summarized in the University of Virginia report, are that accuracy for education, occupation and experience fields in BGT is under 80\%, there are missing values for 36\% of employer names, salary is provided for 7\% of observations, educational requirements are extracted for 53\% of observations, and experience for 52 percent. BGT data has more structured data fields than NLx: for example, BGT’s cleaned data includes an occupation family for 96.6\% of job ads, while NLx had 82.7\% at the time of the Virginia report.

\section{Data and Methods}
\label{sec3}

Since 2007, the National Labor Exchange (NLx) has been the leading platform for job ad distribution in the United States.  NLx is a not-for-profit partnership between the Direct Employers Association, which runs the national job ad syndication network, and the National Association of State Workforce Agencies (NASWA). NLx obtains data from over 300,000 employers that hire workers directly, and distributes job ads to a network of state workforce agencies and online job ad portals. Since 2021, with backing by the National Science Foundation and Bill and Melinda Gates Foundation, the NLx Research Hub has given researchers ``a trusted and transparent source of job vacancy data'' with a goal to ``make real-time job ad information a public utility for the first time, broadening opportunities for research, analytics and product development.''

Labor exchanges in the U.S. were established in 1933 under the Wagner-Peyser Act, and intermediate job-seekers and employers to facilitate efficient labor market matching while also creating opportunities to develop labor market insight from their operations \citep{balducchi_labor_2004}. Under the Vietnam Era Veterans’ Readjustment Assistance Act (VEVRAA), federal contractors must meet job posting requirements, including that postings be filed with state unemployment offices. NLx assists with recruitment-related compliance, as America's Job Bank (AJB) did before it.\footnote{Launched in 1995, AJB was an online job ads portal supported with funding from the U.S. Department of Labor with input and involvement from large employers and state workforce development agencies.  Free for employers and job-seekers, it was one of the most heavily trafficked websites on the early web. With more than 2.2 million monthly postings, 600,000 resumes, and 450,000 employers, it held what was the largest repository of online job ads at the time it was shuttered in 2007 \citep{frauenheim_what_2007}. The 1995-2006 archive of online job ads once managed by AJB was destroyed following defunding; attempts to recover the job ad text that was once part of AJB through Freedom of Information Act requests to the Minnesota, New York, and U.S. Department of Labor were unsuccessful.} 

Top recommendations in the November 2024 report of the Workforce Information Advisory Council, a group of 14 national leaders in workforce information, included strengthening the NLx, standardizing job postings data, creating pilot programs, and building tools and minimally viable data products for real-time use \citep{hirsch_2024}. Researchers can access NLx data via the NLx Research Hub. 

NLx's structured data fields are for the most part blank if the original creator of the job ad did not populate the field at the time of creation. The remainder of this section introduces the toolkit we develop to extract standardized data from job ad text. Section \ref{sec:methods} introduces the Job Ad Analysis Toolkit (JAAT).  Section \ref{sec:dictionaries} describes dictionaries of terms and knowledge maps we run through the job ads, including O*NET's tools and technologies dictionaries. Section \ref{sec:aggs} details the construction of additional variables necessary for creation of an aggregated dataset, including the `active month' used in the construction of time series data. 

Appendix \ref{sec:app_onet_data} describes the specific elements of O*NET structure we map to job ad features for extraction. Appendix \ref{sec:detailed_methods} provides additional detail on methodology and validation procedures. Appendix \ref{sec:robust} provides comparisons between aggregate data against benchmark Census and BLS sources. Appendix \ref{sec:all_dictionaries} lists custom dictionaries we develop. 

%Appendix \ref{sec:app_admin_report} provides a sample report for one occupation, illustrating the data that is available. 

\subsection{The Job Ad Analysis Toolkit (JAAT)}
\label{sec:methods}

The \href{https://github.com/Job-Ad-Research-at-QSB-LUC/JAAT}{Job Ad Analysis Toolkit (JAAT)} is an open-source collection of tools developed for extraction of standardized information from job ads. Table \ref{tab:model_overview} summarizes the models and other NLP tools built to create structured data from job ad text. JAAT features include SkillMatch (\ref{sec:titlematch}), TaskMatch (\ref{sec:skillmatch}), TitleMatch (\ref{sec:taskmatch}), FirmExtract (\ref{sec:firmextract}), WageExtract (\ref{sec:wageextract}), and JobTag (\ref{sec:jobtag}).  This section summarizes the methods and process followed, in general and in the construction of each tool, and provides out-of-sample validation test results that indicate the performance of key models.

We built models with a mindset in alignment with a recent report in the context of safety-critical systems recommending adoption of ``interpretable, traceable, highly accurate, and robust'' models; we also ``shift away from focusing strictly on algorithmic performance in isolation'' \citep{committee_on_using_machine_learning_in_safety-critical_applications_setting_a_research_agenda_machine_2025}. The suite of tools in JAAT are designed to transform job ad text into job ad data, and are capable of extracting high-quality data from hundreds of millions of job ads, including in low-resource and constrained computing environments. Training and classification of these models was done largely on a single NVIDIA Quadro RTX 8000 GPU. Typical processing times for a single model run (i.e., one JAAT tool) on the entire corpus are 10-14 days with this hardware. To speed inference, we acquired access to additional on-premise computational infrastructure.

\subsubsection{Research Methods and Process}
To scale a taxonomy with only a small number of labeled examples of text over a very large text corpus, we approached model construction with a trial-and-error mindset, engaging in experimentation and working in iterative cycles of building training data, fine-tuning models, testing model performance,  manually validating model results, and augmenting training data by ``humans in the loop'' \citep{https://doi.org/10.48550/arxiv.1811.10154}. Domain-specific ML and human-in-the loop processes improve performance, reduce biases, and provide labeled output with a high degree of correspondence with human understanding \citep{choudhury_machine_2020,Adadi2018,doi:10.1126/scirobotics.aay7120}. We track the performance of over 100 iterative stages of model construction in a laboratory log. We searched for and tested other open-source contributions, but saw a need to pursue \emph{de novo} processes and development to build a comprehensive toolkit.

We often begin by embedding an existing O*NET taxonomy or a newly labeled list of concepts as initial training data and ``augmenting'' or ``bootstrapping'' it by finding semantically similar text obtained from embedding text from a random sample of job ads. Data augmentation is exemplified in our introduction of SkillMatch (Section \ref{sec:skillmatch}). The ``augmented'' taxonomy is then adjusted with manual additions and deletions after hand-reviewing high-frequency results. This process dramatically increases available labeled training data beyond a small number of examples.  We perform strategic audits of each model and iteratively improve models -- in each iteration, we manually code a small random sample within stratifications of the cosine similarity to assess performance against ground truth. After one or more cycles of this process, we identify in a small manual audit a similarity score threshold where, above that threshold, the overall positive matches should achieve accuracy near 90 percent. For building aggregate data, we store only the results above this threshold.

Where no prior knowledge base of labeled text existed, we follow in the tradition of interpretive text analysis \citep{gephart1997hazardous}. Construction typically starts with keyword searches, and includes strategic manual audits of high-frequency keywords and text phrases, and random manual audits of human labeled output to ensure high content validity \citep{neuendorf2017content}. Once an initial list is developed based on interrogation of text, we begin the process described above of iteratively augmenting and constructing a large volume of labeled data.

In the absence of benchmark data, we perform post-hoc tests of model performance to assess convergent validity. We emphasize tests at a granular level that assess the ground truth of model output to labeled data from multiple independent sources. In addition, similar to the available information about the representativeness of proprietary job ad data \citep{hershbein_recessions_2018}, we also demonstrate convergent validity by comparing aggregate data from the NLx job ad corpus to Census and BLS sources in Appendix \ref{sec:robust}.

We encourage users to independently test and inspect JAAT model results. Upon release of the aggregate data, users should inspect results carefully and compare them to other statistics. Appendix \ref{sec:detailed_methods} provides additional detail on the methods used and known limitations with specific models. We provide the tools as is, as they are used in the construction of data. 

\begin{landscape}
\begin{table}[ht!]
\centering
\small
\begin{tabularx}{\linewidth}{
    l  % Module
    >{\raggedright\arraybackslash}X  % Tool
    l  % Base Model
    >{\raggedright\arraybackslash}X  % Type
    c  % # Parameters
    c  % Train Score
    c  % Validation Score
}
\toprule
\textbf{Module} & \textbf{Tool} & \textbf{Base Model} & \textbf{Type} & \makecell{\textbf{\#}\\\textbf{Parameters}} & \makecell{\textbf{Train}\\\textbf{Score}} & \makecell{\textbf{Validation}\\\textbf{Score}} \\
\midrule

% Note: multirow is now 4 because each of the 2 tools takes 2 rows
\multirow{4}{*}{\textbf{TaskMatch}}
 & Task / Not Task Classification & BERT-tiny & Fine-tuned (Binary) & 4.4M & 99.44 (F1) & 99.44 (F1) \\
 & \multicolumn{6}{l}{\small\url{https://huggingface.co/loyoladatamining/task-classifier-mini-improved2}} \\
 \cmidrule(l){2-7} % A light rule to separate tools within a module
 & O*NET Task ID Matching & GTE-small & Embedding & 30M & - & - \\
 & \multicolumn{6}{l}{\small\url{https://huggingface.co/thenlper/gte-small}} \\
\midrule

% Note: multirow is now 4 because each of the 2 tools takes 2 rows
\multirow{4}{*}{\textbf{SkillMatch}}
 & Skill / Not Skill Classification & BERT-small & Fine-tuned (Binary) & 29M & 98.15 (F1) & 98.32 (F1) \\
 & \multicolumn{6}{l}{\small\url{https://huggingface.co/loyoladatamining/skill-classifier-base}} \\
 \cmidrule(l){2-7}
 & ESCO Skill Matching & GTE-large & Embedding & 330M & - & - \\
 & \multicolumn{6}{l}{\small\url{https://huggingface.co/thenlper/gte-large}} \\
\midrule

% Note: multirow is now 6 because each of the 3 tools takes 2 rows
\multirow{6}{*}{\textbf{TitleMatch}}
 & Title to SOC Matching & GTE-small & Embedding & 30M & - & - \\
 & \multicolumn{6}{l}{\small\url{https://huggingface.co/thenlper/gte-small}} \\
 \cmidrule(l){2-7}
 & Hierarchy Scoring & DeBERTa-v3-base & Fine-tuned (regression) & 86M & 27.00 (MSE) & 34.08 (MSE) \\
 & \multicolumn{6}{l}{\small\url{https://huggingface.co/loyoladatamining/title_value}} \\
 \cmidrule(l){2-7}
 & Feature Classification & DeBERTa-v3-base & Fine-tuned (Multi-label) & 86M & 81.40 (Acc.) & 81.53 (Acc.) \\
 & \multicolumn{6}{l}{\small\url{https://huggingface.co/loyoladatamining/title_feature}} \\
\midrule

\textbf{FirmExtract}
 & Firm Name Extraction & DeBERTa-v3-base & Fine-tuned (Sequence) & 86M & 94.40 (F1) & 94.47 (F1) \\
 & \multicolumn{6}{l}{\small\url{https://huggingface.co/loyoladatamining/firmNER-v3}} \\
\midrule

% Note: multirow is now 6 because each of the 3 tools takes 2 rows
\multirow{6}{*}{\textbf{WageExtract}}
 & Wage Frequency Classification & BERT-tiny & Fine-tuned (Binary) & 4.4M & 96.82 (F1) & 96.85 (F1) \\
 & \multicolumn{6}{l}{\small\url{https://huggingface.co/loyoladatamining/is_pay}} \\
 \cmidrule(l){2-7}
 & Wage Extraction & DeBERTa-v3-base & Fine-tuned (Sequence) & 86M & 99.74 (F1) & 99.80 (F1) \\
 & \multicolumn{6}{l}{\small\url{https://huggingface.co/loyoladatamining/wage-ner-v2}} \\
 \cmidrule(l){2-7}
 & Wage Frequency Classification & DeBERTa-v3-base & Fine-tuned (Multi-class) & 86M & 99.20 (F1) & 99.64 (F1) \\
 & \multicolumn{6}{l}{\small\url{https://huggingface.co/loyoladatamining/pay-freq-v2}} \\
\midrule

\textbf{JobTag}
 & \makecell[l]{Job Feature\\Classification} & \makecell[l]{sklearn\\RandomForest} & Trained (Binary) & - & - & - \\
 & \multicolumn{6}{l}{\small\url{https://github.com/Job-Ad-Research-at-QSB-LUC/JAAT}} \\

\bottomrule
\end{tabularx}
\small
\justifying{\emph{Note: }An overview of the various language model-based tools used in the modules of the Job Ad Analysis Toolkit (JAAT). JAAT leverages a combination of pre-trained encoder-only embedding models, which are primarily used for semantic matching tasks, and fine-tuned language models, used for more specialized tasks. In the case of fine-tuning, we train a variety of models, including binary, multi-class, multi-label, and sequence classification models. The resulting models, their parameters sizes, and their training performance (on the selected validation metric) are included. Note that in the case of JobTag, we use simple RandomForest classification models.}
\caption{Job Ad Analysis Toolkit (JAAT) Models}
\label{tab:model_overview}
\end{table}

\end{landscape}

\subsubsection{SkillMatch}
\label{sec:skillmatch}

O*NET's skills data is built via a cross-walk from work activities, which we obtain independently from TaskMatch (described below). We sought an independent measure of skill requirements, and compared O*NET's skills taxonomy with skill taxonomies from the European Skills, Competences, and Occupations (ESCO) database, the OECD, and the World Economic Forum (WEF). We found the ESCO v. 1.2.0 database to be the most detailed labeled skills taxonomy, and manually developed crosswalks between 168 of its high-level skill codes and codes from O*NET, WEF, and OECD. We incorporated example text and labels from each of these taxonomies, and thereby increased the number of examples assigned to labels from the ESCO skills taxonomy.

\begin{figure}[htbp]
    \centering
    \includegraphics[scale=0.42]{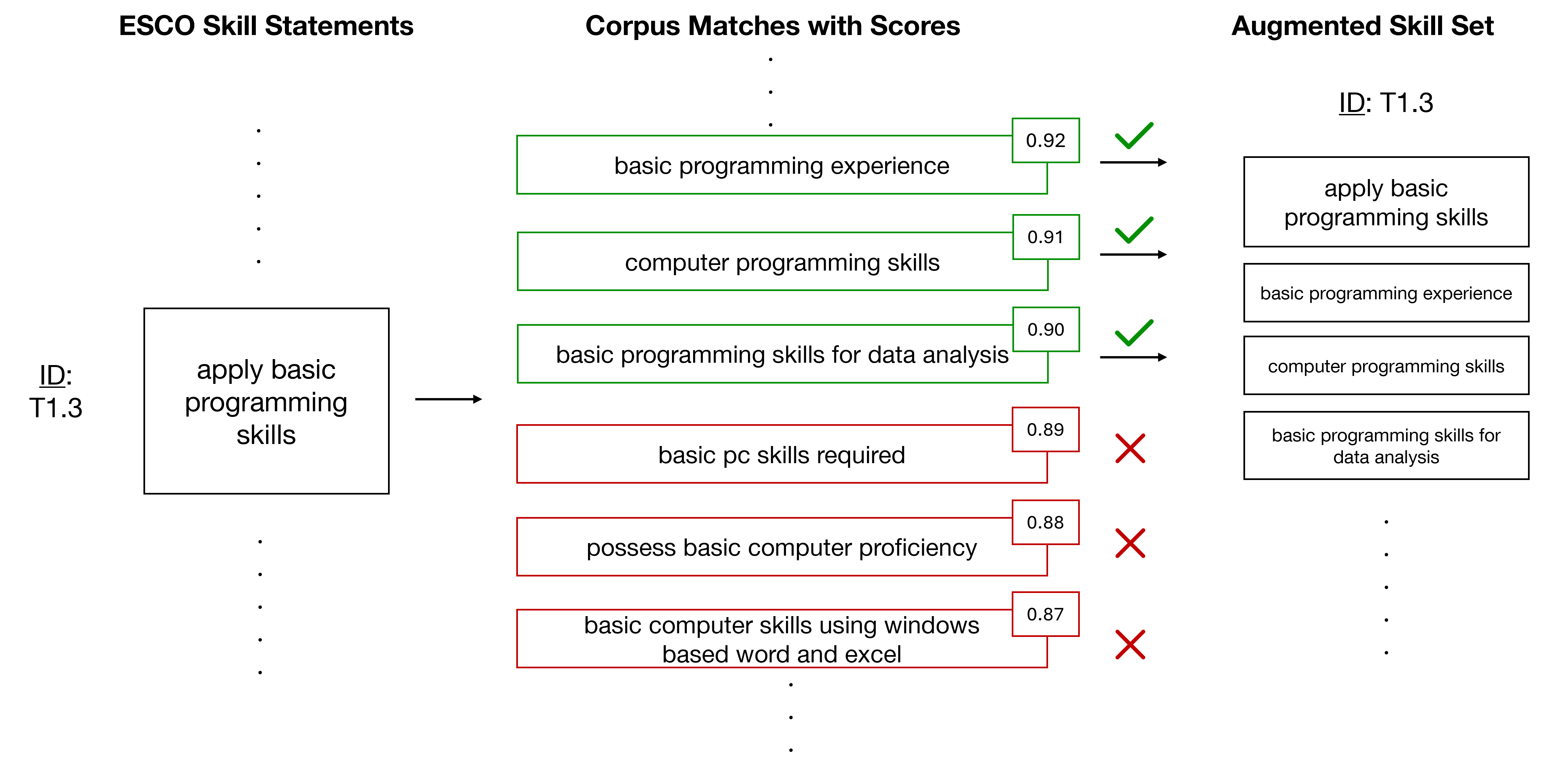}
    \small
    \justifying
    \emph{Note: } For each skill labeled in ESCO, we find the most semantically similar statements from a random sample of 100k job postings, above a certain similarity threshold (e.g., 0.9). These matches are then added with the original skill statement sets from ESCO, thus creating \textit{augmented} sets.
    \caption{An illustration of the data augmentation process}
    \label{fig:augment}
\end{figure}

SkillMatch is a two-stage model that first classifies ``skill sentences'', and then performs a semantic similarity search of positively identified skill sentences against a list of ESCO skills. Our training dataset began with the texts labeled by experts who developed the above mentioned taxonomies. These base texts were used to run an \textit{augmentation} procedure on a random sample of 100,000 job ads, where semantic matching was performed to find the most and least similar sentences. The most similar sentences, as measured by semantic (cosine) similarity of embeddings, were added to the original ESCO skill statements, thus creating an \textit{augmented} set. A depiction of this process can be found in Figure \ref{fig:augment}. Thus, we build a dataset with a roughly even split of $\sim$\,250k ``positive'' skill sentence examples and $\sim$\,250k ``negative'' not-skill example sentences.

\begin{figure}[htbp]
    \centering
    \includegraphics[scale=0.42]{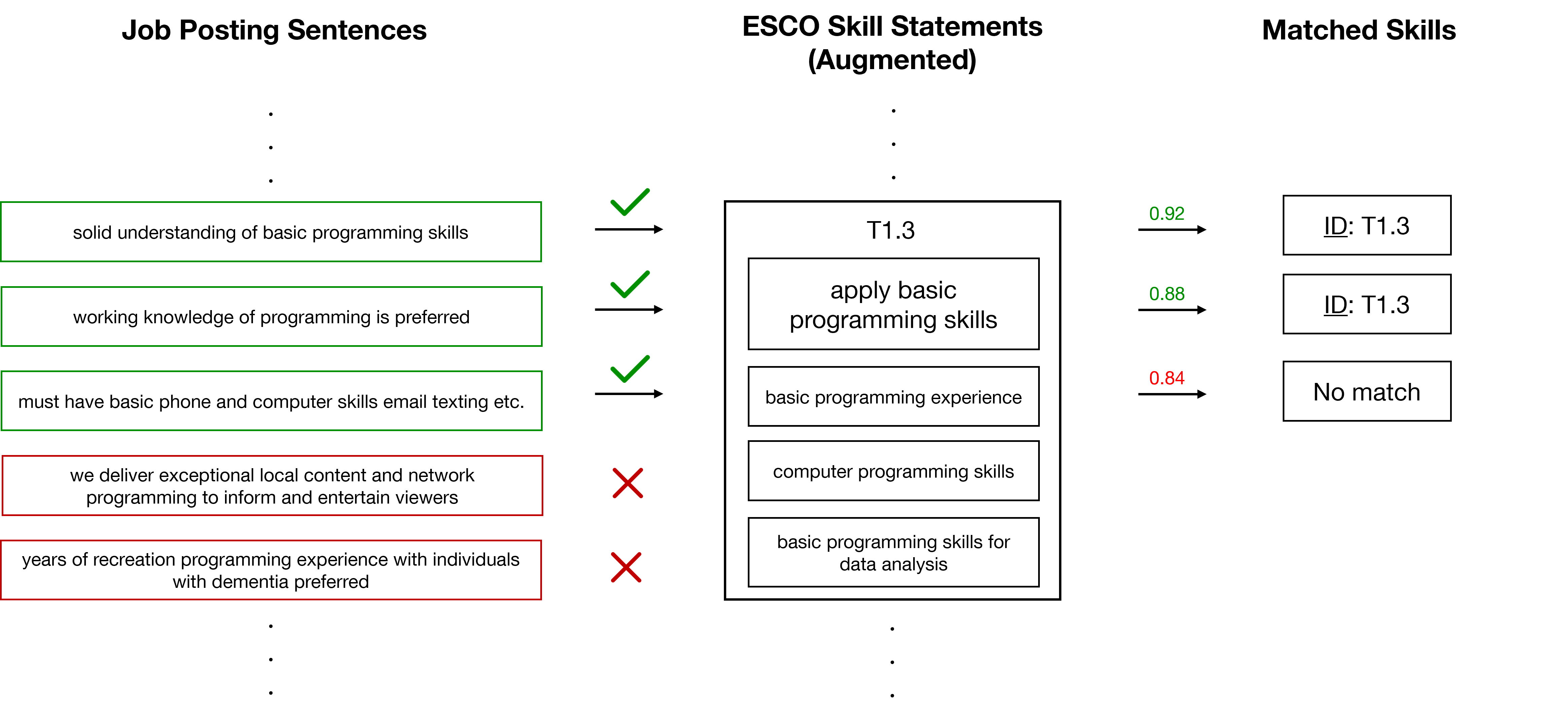}
    \small
    \justifying
    \emph{Note: } In the first stage, a binary classifier filters out sentences that do not represent a candidate skill sentence. Then, the remaining sentences are matched using embedding semantic similarity to the set of augmented skill statements per ESCO skill (see Figure \ref{fig:augment}). Only those matches exceeding a certain threshold (in our case, 0.87) are successfully matched to the skill set and its corresponding code.
    \caption{An overview of the SkillMatch process.}
    \label{fig:skillmatch}
\end{figure}

The first stage of SkillMatch uses this data to train a binary language model-based classification model intended to filter out non-skill sentences, reducing false positives and the computational overhead of running semantic matching over every sentence in the corpus. To fine-tune this model, we opted for  (\textsc{BERT-small}), due to initial testing that indicated the ``tiny'' version was not sufficient to capture the nuances of skill sentence classification. The resulting fine-tuned model achieved a 98.32 F1 score on the validation set. Accordingly, we used a larger, more capable embedding model (\textsc{GTE-large}) for the semantic matching portion of SkillMatch. Hand-coding small samples found that model accuracy dropped markedly below 0.87, was very high above 0.90, and that high-precision results could also be obtained between 0.87 and 0.89. Two independent raters coded 100 randomly selected observations within this range. Inter-rater reliability using Cohen's Kappa indicated moderate agreement ($\kappa$ = 0.58). This small strategic audit suggested that a threshold of 0.87 for cosine similarity would provide overall results that were ~90\% accurate. This threshold was employed as a default for SkillMatch. We ran SkillMatch on the corpus, discarding results below this threshold. An illustrative overview of the SkillMatch process is found in Figure \ref{fig:skillmatch}.

\paragraph{Summary of Model Performance.} Due to the two-stage pipeline of SkillMatch (also found in the ensuing TaskMatch in Section \ref{sec:taskmatch}), we sought to perform additional post-processing validation of the performance of both stages, namely in the binary classification of skill versus non-skill statements, and subsequently the semantic matching of skill statements to skill codes. We follow a two-part validation, leveraging the LLM-as-a-Judge paradigm \citep{10.5555/3666122.3668142} for an estimation of performance at scale, which is internally validated on a smaller sample of disputed results by two independent coders.

For the validation data, we use 213k job postings between the months of March and April 2022 from the Career One Stop platform \citep{careeronestop_2021}. All of these postings were run through our SkillMatch pipeline, where we saved the individual statement-level (sentence) decisions at both stages, i.e., the binary classification, and in the case of a skill statement, the matched skill code. The 213k job postings consisted of 5.34 million sentences, of which 2.78 million were marked by SkillMatch's classifier as being a skill statement. We formed the first validation set by randomly sampling 10k sentences marked as skill statements, and 10k marked as not. We then crafted a few-shot LLM prompt, with the task of deciding whether a given sentence was indeed a skill statement or not. This prompt is provided in Table \ref{tab:prompt1} of the Appendix. We use three LLMs for judging, two closed-source (\textsc{GPT-4o-mini} and \textsc{Gemini-2.0-flash}) and one open-source (\textsc{Llama-3.3-70B-Instruct}). The results of the LLM validation are presented in Table \ref{tab:clf_validation}.

\begin{table}[htbp]
\centering
\begin{tabular}{l|ccccc|cc|cc}
%\begin{tabular}{l|cccccccc}
% Validator & TPR & FPR & TNR & FNR & F1 & Agree & Strict Acc. & Lenient Acc. \\ \hline
% \textsc{GPT-4o-mini} & 0.581 & 0.419 & 0.883 & 0.117 & 0.685 & \multirow{2}{*}{0.888} & \multirow{2}{*}{0.699} & \multirow{2}{*}{0.811} \\
% \textsc{Llama-3.3-70B} & 0.733 & 0.267 & 0.821 & 0.179 & 0.767 &  &  &  \\ \hline \hline
% Human 1 & x & x & x & x & x & \multirow{2}{*}{x} & \multirow{2}{*}{x} & \multirow{2}{*}{x} \\
% Human 2 & x & x & x & x & x &  &  & 
\hline 
\multirow{2}{*}{Validator} & \multicolumn{5}{c}{SkillMatch vs. LLM} & \multicolumn{2}{|c|}{LLM Reliability} & \multicolumn{2}{|c}{Accuracy} \\ 
 & TPR & FPR & TNR & FNR & F1 & Agree & $\kappa$ & Strict & Lenient \\  \hline 
\textsc{Gemini-2.0-Flash} & 0.717 & 0.283 & 0.815 & 0.185 & 0.754 & \multirow{3}{*}{0.859} & \multirow{3}{*}{0.807} & \multirow{3}{*}{0.682} & \multirow{3}{*}{0.811}  \\
\textsc{GPT-4o-mini} & 0.581 & 0.419 & 0.883 & 0.117 & 0.685 &  &  &  \\
\textsc{Llama-3.3-70B} & 0.733 & 0.267 & 0.821 & 0.179 & 0.767 &  &  &  \\ \hline \hline
\end{tabular}
\vspace{5pt} \\
\small
\justifying{
\emph{Note:} We provide True Positive, False Positive, True Negative, and False Negative rates, as well as the resulting F1 scores. In addition, we indicate the overall agreement, the inter-rater reliability ($\kappa$), and resulting accuracy scores for SkillMatch in a strict setting (SkillMatch corresponds to \textit{all} coders) or a lenient setting (corresponds to at least one).}
\caption{Validation results for LLM-as-a-Judge on SkillMatch binary classification.}
\label{tab:clf_validation}
\end{table}
%\caption{Validation results for LLM-as-a-Judge and human coders on SkillMatch binary classification. We provide True Positive, False Positive, True Negative, and False Negative rates, as well as the resulting F1 scores. In addition, we indicate how often the pairs of coders (LLMs or humans) agree, and the resulting accuracy scores in a strict setting (SkillMatch corresponds to \textit{both} coders) or a lenient setting (corresponds to at least one).}

False negatives in the first stage of SkillMatch are particularly concerning. We conduct a small-scale investigation with two independent human coders into 170 disagreements between SkillMatch and LLM results to assist in adjudication. Table \ref{skill_human_lab} provides results. These indicate promising future directions using LLM-as-a-judge to label training data.

\begin{table}[htbp]
\centering
\begin{tabular}{l|c|cc|cc}
\hline
LLM Results & Validator & Not Skill &  Skill & \multicolumn{2}{c}{Human Reliability} \\ 
 & & & & Agree & $\kappa$ \\\hline
\textsc{Strict LLM Agreement - Not Skill} & SkillMatch & 0 & 50 & \multirow{3}{*}{0.740} & \multirow{3}{*}{0.313} \\ 
\textsc{Strict LLM Agreement - Not Skill} & Human 1 & 44 & 6 & &   \\
\textsc{Strict LLM Agreement - Not Skill} & Human 2 & 33 & 17 & & \\ \hline
\textsc{Lenient LLM Agreement - Not Skill} & SkillMatch & 72 & 20 & \multirow{3}{*}{0.696} & \multirow{3}{*}{0.291} \\
\textsc{Lenient LLM Agreement - Not Skill} & Human 1 & 36 & 56 & &  \\
\textsc{Lenient LLM Agreement - Not Skill} & Human 2 & 16 & 76 & &  \\ \hline
\textsc{Lenient LLM Agreement - Skill} & SkillMatch & 28 & 0 & \multirow{3}{*}{0.929} & \multirow{3}{*}{0.472} \\ 
\textsc{Lenient LLM Agreement - Skill} & Human 1 & 3 & 25 & &  \\
\textsc{Lenient LLM Agreement - Skill} & Human 2 & 1 & 27 & & \\ \hline
\end{tabular}
\vspace{5pt} \\
\small
\justifying
\emph{Note:} Overall agreement between humans and LLMs in this small sample of disputed results is 40.6\%  ($\kappa$ = 0.23). Independent human coders agree overall with one another in 74.7\% of cases ($\kappa$ = 0.489). For 50 cases of strict LLM agreement that a sentence is not a skill sentence (and SkillMatch disagrees), human coders agree with one another in 37 of those cases, and of those, agree with the LLMs in 87\% of those  cases. For 120 sentences where at least one LLM suggests a skill is within the sentence, overall human agreement that it is a skill sentence is 86\% ($\kappa$ = 0.67).
\caption{Human Ratings in Disputed Cases}
\label{skill_human_lab}
\end{table}

From the 2.78 million sentences that were flagged as being skill statements, we also validate the second-stage of SkillMatch's semantic matching process, where each sentence is matched to the most similar skill code (via the code's title), and only the match results above a chosen threshold of similarity (in our case, 0.87) are kept. To illustrate how this process performs outside of its run on the full corpus, we choose a random sample of 1000 match results at all similarity scores in the range of [0.8, 1.0], rounded to two digits. In the case where 1000 results do not exist, we take the complete (maximum) number of results for that score. This resulted in a final validation set of 16597 statements, each with a corresponding matched skill.

These statements were evaluated via LLM-as-a-Judge using a second prompt, found in Table \ref{tab:prompt2}, which tasked the LLMs to provide a binary decision on whether the matched skill was an appropriate match or not given the skill statement. Two independent human coders audit a smaller set of results, with random sampling of labeled sentences within stratifications by the similarity score. The results of this validation round are presented in Table \ref{tab:match_validation}.

\begin{table}[htbp]
\centering
\resizebox{\linewidth}{!}{
\begin{tabular}{l|ccccccccc}
% & \multicolumn{9}{c}{\textbf{Cosine Similarity Scores}}  \\ \hline
                            & \textbf{0.8--0.84}    & \textbf{0.85}  & \textbf{0.86}  & \textbf{0.87}  & \textbf{0.88}  & \textbf{0.89}  & \textbf{0.90}  & \textbf{0.91-- 0.95}   & \textbf{0.96--1} \\ \hline
(1) Freq. Distribution          & 0.11          & 0.18  & 0.21  & 0.19  & 0.13  & 0.09  & 0.05  & 0.04          & 0.00  \\ \hline
\textsc{(2) Gemini-2.0-Flash}   & 0.59           & 0.79   &  0.86  & 0.93   & 0.93   & 0.96   & 0.97   & 0.99          & 1.0   \\
\textsc{(3) GPT-4o-mini}        & 0.19          & 0.26  & 0.35  & 0.44  & 0.57  & 0.67  & 0.73  & 0.90          & 0.99  \\
\textsc{(4) Llama-3.3-70B}      & 0.49          & 0.65  & 0.73  & 0.82  & 0.87  & 0.90  & 0.91  & 0.96          & 1.0   \\ \hline
(5) N LLM (k)               & 3.93          & 1     & 1     & 1     & 1     & 1     & 1     & 5             & 1.67  \\ \hline
\textsc{(6) Majority Agree}    & 0.54          & 0.39  & 0.29  & 0.81  & 0.86  & 0.90  & 0.91  & 0.96          & 1.0  \\ 
\textsc{(7) Strict Agree}    & 0.65         & 0.41  & 0.25  & 0.89  & 0.90  & 0.95  & 0.96  & 0.99          & 1.0  \\ \hline
(8) Human 1                     & 0.35          & 0.55  & 0.77  & 0.87  & 0.92  & 0.85  & 0.85  & 0.99          & 1.0   \\
(9) Human 2                     & 0.14          & 0.60  & 0.51  & 0.56  & 0.56  & 0.72  & 0.85  & 0.84          & 1.0   \\ \hline
(10) N (Hand Labeled)            & 91            & 20    & 39    & 39    & 39    & 58    & 59    & 67            & 5     \\ \hline \hline
\end{tabular}
}
\vspace{5pt} \\
\small
\justifying
\emph{Note:} Results are given per similarity score (columns). Row 1 indicates the frequency distribution of 2.8 million skill sentences, rows 2-4 provide values for each LLM represent the percentage of correctly matched skills as judged by the LLMs compared to SkillMatch results. Row 5 provides the number of sentences (in thousands) evaluated by LLMs. The percent of SkillMatch results in agreement with the majority of LLMs is provided in row 6, and row 7 displays strict agreement (for 11,582 observations where all LLMs agree). Overall, the majority LLM results have 88\% agreement with SkillMatch when using a 0.87 threshold. Strict LLM results agree with 94\% of SkillMatch results using the 0.87 threshold. Rows 8 and 9 represent 2 independent human evaluators, blinded to both LLM and similarity score results. Overall, rater 1 and 2 agree on 78\% of evaluated cases.
\caption{Validation results for LLM-as-a-Judge and human coders on SkillMatch semantic matching.}
\label{tab:match_validation}
\end{table}

\paragraph{Overall Estimate of Ground Truth} 

\begin{figure}[htbp]
    \centering
    \includegraphics[scale=0.22]{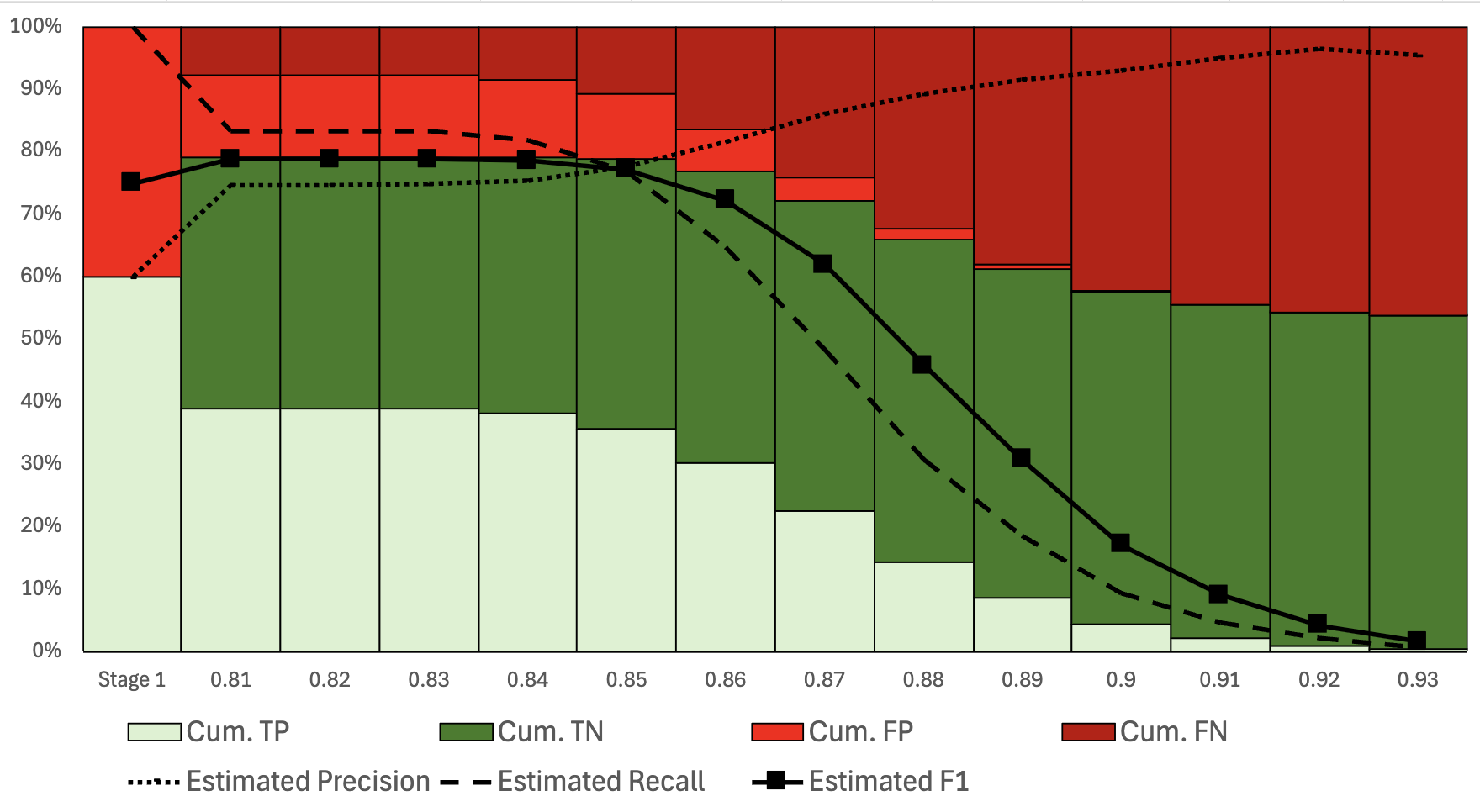} \\
    \vspace{5pt}
    \small
    \justifying
    \emph{Note:} This visual uses the distribution by match score and estimates to simulate the tradeoffs between recall and precision at different thresholds. Storing all results above a threshold that is too low (below 0.84) returns many false positives (light red, top left), lowering precision. Storing only the data above a high threshold (above 0.88) drops many false negatives (dark red, top right), lowering recall. For the construction of data, we chose a threshold (0.87) above which results are high-precision; this simulation suggests that the overall precision for data returned by SkillMatch is 0.86, recall is 0.58, and the the F1 score is 0.7.
    \caption{Estimated Precision, Recall, F1 Score and TP, FP, TN, FN Distribution of SkillMatch.}
    \label{fig:skillmatch_simulation}
    \end{figure}

Figure \ref{fig:skillmatch_simulation} illustrates the simulated effect of Stage 1 of SkillMatch and choosing a threshold between 0.81 and 0.93 on the proportion of True Positive, False Positive, True Negative, and False Negative sentences. This figure uses the observed distribution of 2.8 million skill sentences by match score as coded by SkillMatch, and average LLM estimates for accuracy in stage 1 and at each threshold. 

The post-processing validation demonstrates the overall performance of procedures we followed in augmenting a small number of labeled items in a taxonomy. We estimate that overall, the accuracy of positive text labels from SkillMatch is 86 percent, and that from the 5.34 million sentences in 234k job postings, SkillMatch returns approximately 1.2 million true positive skill statements above match score 0.87 coded to a ESCO skill label, and 195,000 false positives. This exercise also demonstrates that threshold selection and the two stage model work as intended: absent stage 1, stage 2 with no threshold would return 3.2 million true positives and 2.1 million false positives. Given the desire for manageable volumes of high-precision data, the initial choice of a threshold could have been lower, but appears to have been well-reasoned. 

\subsubsection{TaskMatch}
\label{sec:taskmatch}

TaskMatch provides detailed structured information from job ads about the work performed on the job. Each of over 20,000  O*NET task statements has a unique identifier, which is linked via a hierarchical taxonomy to detailed, intermediate, and general work activities, which can be cross-walked via O*NET to taxonomies of skills and abilities. TaskMatch bridges highly precise hand-created task statements  based on interviews with workers (i.e., those found in O*NET) and the ability to generalize these statements to language in job ads that describe job duties.  The semantic matching process we introduce above is applied to O*NET task statements and ``candidate'' task statements from job ads, and allows for expert-curated knowledge from O*NET to be scaled efficiently over large corpora. 

As with SkillMatch, TaskMatch is a two-stage model that first identifies task sentences in job ads. After augmentation, the final training dataset for the first stage consists of nearly 150,000 texts (44k task, 106k not task). An efficient, compact version of a \textsc{BERT} model (\textsc{BERT-tiny}) was fine-tuned on the training dataset for one epoch to produce a binary task classification model. This model was chosen due to its compactness (17 MB) and ability to be run efficiently (even on CPU).  The fine-tuned BERT model achieved an F1 score of 99.44 on the held-out validation set during training.  Only statements identified to be task statements by the binary text classification model were considered in the semantic matching process described below. 

To build the second stage of TaskMatch, we embedded O*NET’s task statements and searched for similar task sentences identified by the binary classifier from a random sample of 100,000 NLx job ads. In pre-run validation, we performed a manual audit on a small random sample within bins of the similarity score. We identified that above an embedding match score of 0.90 (i.e., cosine similarity), we obtained excellent precision scores (7 false positives / 165 reviewed), and the quality deteriorated below that level (65 false positives / 90). Discarding the results below 0.90 meant dropping 60\% of the results (decreasing recall), but provides confidence that retained results are highly accurate. 

\paragraph{Summary of Model Performance.}
In a similar manner as was done with SkillMatch, we post-validate TaskMatch on a randomly selected sample of job ads from the 5.34 million sentence of our Career One Stop corpus. The first part of the validation once again leveraged LLM-as-a-Judge to evaluate the performance of the binary classification step, which predicts whether a given sentence contains a task statement or not. We run a similar LLM process as with TaskMatch, using three LLMs to independently evaluate a sample of 10k sentences marked as task statements and 10k sentences marked as not. The results of this validation stage can be found in Table \ref{tab:clf_validation_task}.

%These are cross-validated with a smaller sample hand-coded by the same two human experts. 

\begin{table}[htbp]
\centering
\begin{tabular}{l|ccccc|cc|cc}
\hline 
\multirow{2}{*}{Validator} & \multicolumn{5}{c}{TaskMatch vs. LLM} & \multicolumn{2}{|c|}{LLM Reliability} & \multicolumn{2}{|c}{Accuracy} \\ 
 & TPR & FPR & TNR & FNR & F1 & Agree & $\kappa$ & Strict & Lenient \\  \hline 
\textsc{Gemini-2.0-Flash} & 0.811 & 0.189 & 0.784 & 0.317 & 0.800 & \multirow{3}{*}{0.842} & \multirow{3}{*}{{0.706}} & \multirow{3}{*}{0.718} & \multirow{3}{*}{0.876}  \\
\textsc{GPT-4o-mini} & 0.714 & 0.286 & 0.887 & 0.113 & 0.782 &  &  &  \\
\textsc{Llama-3.3-70B} & 0.814 & 0.186 & 0.812 & 0.189 & 0.813 &  &  &  \\ \hline \hline
\end{tabular} \\
 \vspace{5pt}
\small
\justifying
\emph{Note:} We provide True Positive, False Positive, True Negative, and False Negative rates, as well as the resulting F1 scores. In addition, we indicate the overall agreement, the inter-rater reliability ($\kappa$), and resulting accuracy scores for TaskMatch in a strict setting (TaskMatch corresponds to \textit{all} coders) or a lenient setting (corresponds to at least one).
\caption{Validation results for LLM-as-a-Judge on TasklMatch binary classification.}
\label{tab:clf_validation_task}
\end{table}

We also validate the matching stage of TaskMatch, taking a random sample of 1000 matched tasks per two-digit match score in the range [0.81, 1.00] (no observations below 0.81), and calculating the resulting metrics per score. From the 2.05 million sentences marked as task statements by the binary classifier, this resulted in a validation set of 18,051 statements matched to a task. These results are presented in Table \ref{tab:match_validation_task}. Based on these results and the distribution by similarity score, we estimate the overall precision for retained TaskMatch data is 0.85, recall is 0.56, and F1 is 0.68.

\begin{table}[htbp]
\centering
\resizebox{\linewidth}{!}{
\begin{tabular}{l|ccccccccc}
 & \textbf{0.81--0.84} & \textbf{0.85} & \textbf{0.86} & \textbf{0.87} & \textbf{0.88} & \textbf{0.89} & \textbf{0.90} & \textbf{0.91-- 0.95} & \textbf{0.96--1} \\ \hline
(1) Freq. Distribution & 0.00 & 0.01 & 0.04 & 0.10 & 0.17 & 0.21 & 0.18 & 0.28 & 0.01 \\ \hline
(2) \textsc{Gemini-2.0-Flash} & 0.12 & 0.21 & 0.36 & 0.40 & 0.52 & 0.66 & 0.73 & 0.90 & 1.00 \\
(3) \textsc{GPT-4o-mini} & 0.02 & 0.11 & 0.15 & 0.22 & 0.35 & 0.38 & 0.52 & 0.78 & 0.99 \\
(4) \textsc{Llama-3.3-70B} & 0.06 & 0.25 & 0.33 & 0.42 & 0.53 & 0.64 & 0.72 & 0.87 & 1.00 \\ \hline
(5) N LLM (k) & 2.05 & 1 & 1 & 1 & 1 & 1 & 1 & 5 & 5 \\ \hline
(6) \textsc{Majority Agree} & 0.95 & 0.81 & 0.73 & 0.65 & 0.53 & 0.40 & 0.68 & 0.87 & 1.0 \\
(7) \textsc{Strict Agree} & 0.98 & 0.90 & 0.83 & 0.73 & 0.56 & 0.42 & 0.72 & 0.90 & 1.0 \\ \hline
\end{tabular}
} \\
 \vspace{5pt}
\justifying
\small
\emph{Note:} Results are given per similarity score (columns). Row 1 indicates the frequency distribution of 2.05 million task sentences, rows 2-4 provide values for each LLM represent the percentage of correctly matched tasks as judged by the LLMs compared to TaskMatch results. Row 5 provides the number of sentences (in thousands) evaluated by LLMs. The percent of TaskMatch results in agreement with the majority of LLMs is provided in row 6, and row 7 displays strict agreement (for 14,987 observations where all LLMs agree). Majority LLM results have 84\% agreement with TaskMatch when using a 0.90 threshold. Strict LLM results agree with 89\% of TaskMatch results using the 0.90 threshold.
\caption{Validation results for LLM-as-a-Judge on TaskMatch semantic matching}
\label{tab:match_validation_task}
\end{table}

\paragraph{Overall Estimate of Ground Truth} Based on a similar to analysis to that done for SkillMatch, we estimate that overall accuracy of positive text labels from TaskMatch is 85\% in our data: from the 5.34 million sentences, TaskMatch would return approximately 816,000 true positive task labels, and 145,000 false positives. As with SkillMatch, we could have adopted a lower threshold, but our original approach again proves to generate large volumes of high-quality data.

\subsubsection{TitleMatch}
\label{sec:titlematch}
TitleMatch disambiguates job title features, returning standard SOC-O*NET codes, estimated hierarchal level, and other features. In this section, we describe occupation matching and performance of occupation matching and hierarchy models. Additional detail is in Appendix \ref{sec:titlematch_detail}. 

O*NET's sample of reported and alternate job titles and associated occupation codes form the basis of our model that matches job titles to occupation. However, job titles are not perfect indicators of occupations. Within O*NET's reported titles, for example, there are 9 potential different occupations for the job title ``data analyst.'' Despite this, we follow economists \citep{atalay_new_2018,atalay_evolution_2020}, epidemiologists (SOCCER) \citep{russ_computer-based_2014,russ_evaluation_2023} computer scientists \citep{gasco_overview_2025}, and independent researchers (SOCkit) \citep{howison_replication_2022,howison_recommending_2023,howison_extracting_2024} in building a computational model that returns occupation codes from job titles.  

We preserve all job title-SOC code combinations from O*NET in the training data, even when a title appears under multiple codes. The first step of TitleMatch involves a semantic matching procedure using a \textsc{GTE-small} embedding model. O*NET sample titles are used as a foundation, to which an instance in question is matched, following a simple nearest neighbor selection. For TitleMatch, we do not choose a minimum similarity threshold, thus always returning the best-matched title (and its corresponding occupation code from O*NET).

\paragraph{Summary of Model Performance.} Although benchmark job title-SOC labeled data does not exist, we use administrative data to test occupational coding by TitleMatch and Sockit. The Department of Labor releases disaggregated Labor Condition Application Disclosure Data that employers are required to complete to lawfully place foreign-born guest workers at a worksite.\footnote{See \url{https://www.dol.gov/agencies/eta/foreign-labor/performance}. Accessed September 8, 2025.} These data include employer’s self-reports of job titles mapped to occupation codes \citep{gibbons_monopsony_2019}.  

For high-skilled, seasonal worker, agricultural, and permanent resident programs, we combine 7.5 million employer filings from 2008 – 2024. We reduce these into unique non-null combinations of job titles and occupation codes, restricting the dataset to title-code pairs with more than 5 observations that include a six-digit occupation code that exists in the SOC system (n = 77,562, weighted = 2.86 million employer filings). The dataset contains occupation codes for 661 of the 867 SOC codes. No tool could match all these job title-SOC combinations from job titles alone, as codes vary within job title in the administrative data: the average number of different six-digit occupation codes per unique title in the dataset is 20.8. Job titles appear in multiple occupations for many valid reasons, including that job titles do not perfectly indicate occupations, and that human raters often disagree. Strategic behavior may also affect the selection of occupation in the LCA data, as guest worker minimum wages are set to the prevailing wage within an occupation and region \citep{devaro_wage_2021}. 

\begin{table}[ht]
\centering
\small
% S column aligns numbers on the decimal point. table-format=1.2 means 1 digit before, 2 after.
\begin{tabularx}{\textwidth}{
  >{\raggedright\arraybackslash}X  % First column is left-aligned and expands
  c
  c
  c
}
\toprule
\textbf{Test / Tool} &
{\textbf{2-Digit SOC}} & % Braces are needed to make \textbf work in an S column
{\textbf{4-Digit SOC}} & 
{\textbf{6-Digit SOC}} \\
\midrule
Sockit	& 0.53 &	0.39 &	0.22 \\
Sockit (Wtd.) &	0.62 &	0.51 &	0.29 \\
Sockit Matches Any Occ Within Title	 & 0.65 &	0.54 &	0.39 \\
Sockit Matches Any Occ Within Title (Wtd.) &	0.74 &	0.67 &	0.56 \\
\addlinespace % Adds a little extra space to separate groups
TitleMatch	& 0.62 &	0.49 &	0.32 \\
TitleMatch (Wtd.) & 0.72 &	0.62 & 0.47 \\
TitleMatch Matches Any Occ Within Title	& 0.75	& 0.64	& 0.49 \\
TitleMatch Matches Any Occ Within Title (Wtd.)	& 0.86 &	0.78	& 0.67 \\
\bottomrule
\end{tabularx}
\\
\vspace{5pt}
\justifying{\emph{Note: } Weighted (Wtd.) results reflect the frequency of appearance of a unique combination of job title and occupation code in the data. Weighted by the number of employer filings of a given job title and occupation code combination, TitleMatch matches 72\% of the LCA data at the two-digit level, 62\% at the four-digit level and 47\% at the six-digit level. In terms of matching any of the occupations listed by an employer within a given job title, TitleMatch occupation codes match 86\%, 78\%, and 67\% of cases weighted by frequency at the 2- 4- and 6-digit levels respectively, indicating the frequency with which TitleMatch results match analysis done by human expert coders filing LCAs.}
\caption{TitleMatch and SockIt: Percent Job Title-SOC Match with LCA Data}
\label{tab:titlematch_performance}
\end{table}

Table \ref{tab:titlematch_performance} reports the result of comparing TitleMatch and Sockit using the LCA data. TitleMatch consistently outperforms Sockit at returning occupation codes that match those assigned to job titles in the LCA data. Appendix \ref{sec:titlematch_detail} reports similar results of a test against a collection of newspaper job ads from 1950-2000 \citep{atalay_evolution_2020}.

%This analysis could be used to improve TitleMatch. 

\paragraph{Hierarchy and Other Features from Titles.} TitleMatch also returns a hierarchy value and features of the job advertised in the title. Hierarchy values and features are extracted using distinct fine-tuned DeBERTa-v3-base models. Hierarchy values returned are a number within a range [-10,60], as described in Appendix Table \ref{tab:combined_hierarchy}, where -10 represents trainees and interns,  0 represents a non-managerial role, 10 represents a first-level manager, and increasing levels of managerial responsibility increment by tens up to the Chief Executive Officer (60).

\begin{figure}[htbp!]
\centering
\setlength{\tabcolsep}{5pt}% A
%\begin{tabular}{p{0.48\textwidth} p{0.48\textwidth}}
\includegraphics*[width=0.75\textwidth]{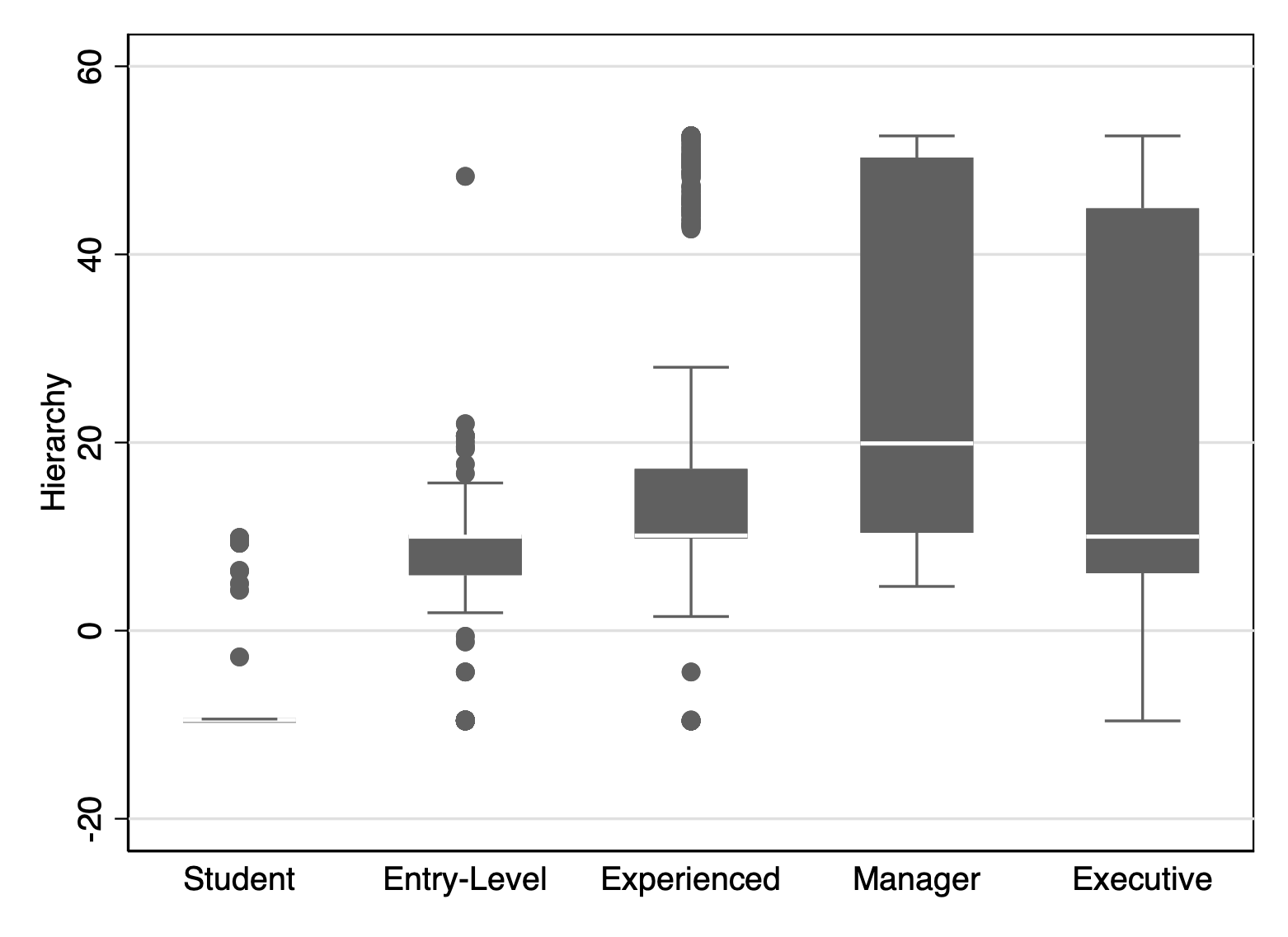}
%\includegraphics*[width=0.45\textwidth]{pngs/kdens_title_hierarchy_simp.png} \\
%\scriptsize{(a) Boxplot: TitleMatch Hierarchy Prediction Matched to NYC Career Levels} &
%\scriptsize{(b) Density Plot: TitleMatch Hierarchy Prediction Similar to NYC Career Levels} \\
\\
\vspace{5pt}
\justifying{\emph{Note: } NYC job ad career levels are on the horizontal axis. Inspection shows that `commissioner' appears frequently in NYC executive rank postings, but was not in the hierarchy coding model training data. We note this for future improvements.}
\caption{TitleMatch Hierarchy Prediction Matched to NYC Job Ad Career Levels}
\label{fig:hierarchytitles}
%\end{tabular}
\end{figure}

We assess the accuracy of the hierarchy match by running TitleMatch on 3,219 New York City job ads downloaded on March 24, 2025 \citep{nyc_job_ads}. NYC job ad metadata includes five career levels (Student, Entry-Level, Experienced, Manager, and Executive). Figure \ref{fig:hierarchytitles} illustrates results. With the exception of executive titles, the boxplot illustrates that the distribution of the model's predicted hierarchy level corresponds to student, entry-level, experienced, and managerial positions. Overall, the correlations between TitleMatch's hierarchy level and the NYC job postings minimum salary range (0.41), top salary range (0.49), and career level (0.48), are consistent with a moderate positive association between this measure and important characteristics of the job. In many cases where wage information is unavailable, this measure may be informative in combination with occupation and other information.

%For Panel B, we map TitleMatch predictions to the Career Levels used by NYC Job Ads - we divide by 10, round to the nearest whole number, and truncate the hierarchy level returned by TitleMatch to match the five levels of the NYC system.  

%Other features extracted from titles are described in Appendix Table \ref{tab:dict_title}. 

\subsubsection{FirmExtract}
\label{sec:firmextract}

FirmExtract retrieves the firm name from the text description of job ads, with additional capabilities to clean and standardize firm names, and perform a similarity match to other sources of firm name information. NLx metadata is missing 38.7\% of firm names for the 2015-2025, similar to the 36\% missing found in research using the Lightcast data \citep{hershbein_recessions_2018,lancaster2019technology}. We train a custom NER (Named Entity Recognition) model (“firmNER”) to extract names from job ad text. FirmNER is created by fine-tuning a \textsc{DeBERTa-v3-base} model on quality labeled data – a large sample of job ad data with firm name present in the metadata. 

In the next steps of FirmExtract, the extracted sequence representing the firm name is standardized and fuzzy matched to an existing collection of known firm names in the United States. We  standardize all extracted firm names using common firm record-linkage cleaners \citep{wasi_record_2015}. This cleaning protocol standardizes firm names that can be subject to multiple spellings: ``Seven-Eleven'', ``7-11 Inc.'', etc. We then fuzzy match firm names from job ads to firm names in a yearly file  of U.S. establishments licensed from Data Axle for 2015-2023. Data Axle's information includes a unique establishment ID, and indicates relationships between establishments, subsidiaries, and parent companies. Data Axle fields also include industry (SIC and NAICS) for all firms, and sales volume, and number of employees for many observations.

\begin{figure}[!htbp]
    \centering
    \includegraphics[width=.9\textwidth]{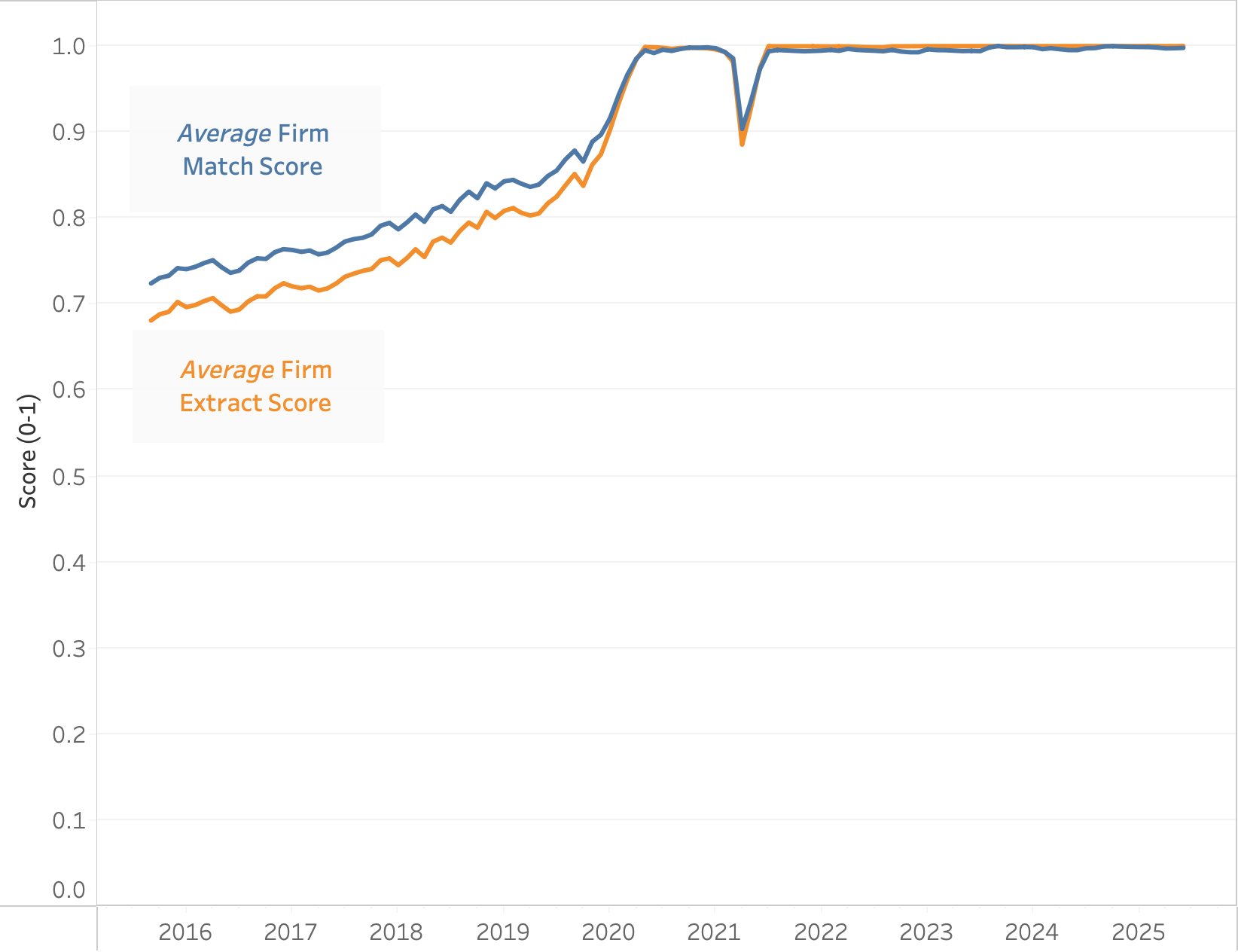}
    \small 
\\
\justifying
\emph{Note: }The percent of job advertisements matched to a specific firm (top) and the model score / confidence in the match (bottom).
    \caption{Firm Availability in the Data}
    \label{fig:has_firm}
 \end{figure}

\begin{figure}[!htbp]
    \centering

    \includegraphics[width=.9\textwidth]{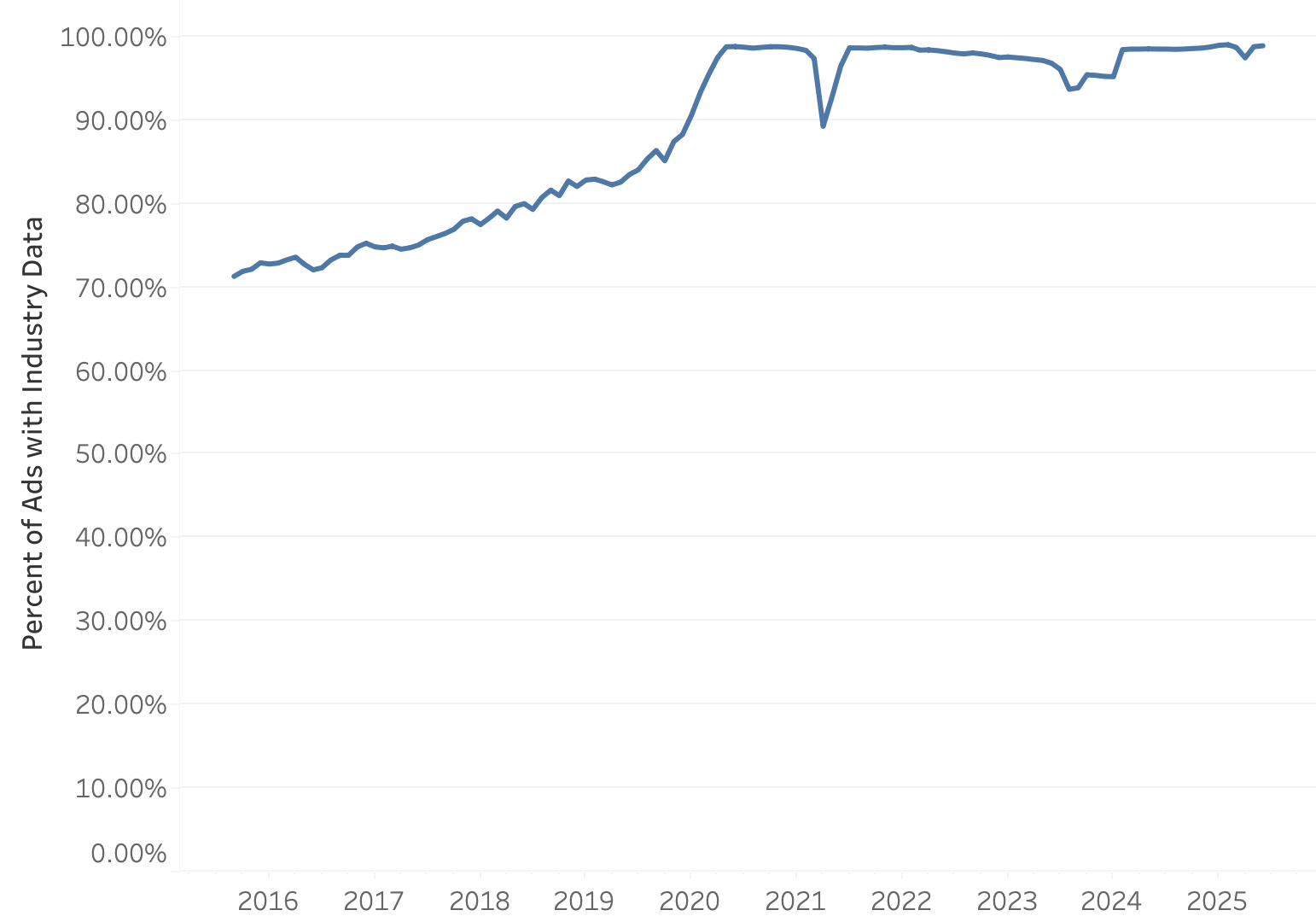}
    \small 

    \caption{Industry Availability in the Data}
    \label{fig:has_industry}
 \end{figure}

Figure \ref{fig:has_firm} displays the average confidence score of the extraction and the match for the duration covered, and illustrates that improvements in NLx data collection over time lead to major improvements in performance. Figure \ref{fig:has_industry} provides percentage of job ads each month that are matched to a unique firm ID, and thus industry NAICS code. Firm names are available for approximately 75\% of job ads in the NLx corpus prior to major improvements in data collection by NLx in 2018, after which we are able to obtain a firm name and a link to industry for nearly 100\% of job ads. 

%percentage of job ads each month that could be matched to a unique firm ID, and thus industry. The bottom panel 

\subsubsection{WageExtract}
\label{sec:wageextract}

WageExtract retrieves pay frequency, minimum, and maximum wages from the unstructured text of job ads. We developed WageExtract by identifying sentences in a random sample of 100,000 job ads that contain a list of regular expressions plausibly related to wages. We developed regular expressions to extract wages from these sentences, and manually audited and corrected each scenario present in the training data. We then constructed a training dataset that distinguishes between sentences containing wage information, and those that do not. Using this, we fine-tune a lightweight \textsc{BERT-tiny} binary classification model, which quickly and efficiently identifies sentences with potential wage information. This model achieves a 96.8\% F1 score on the validation set. 

We then fine-tuned a \textsc{DeBERTa-v3-base} model for sequence classification in order to extract the spans of text containing the wage statements from the identified wage sentences. In particular, we use custom tags to delineate whether an identified span refers to a lower range wage value (MIN) or upper range wage value (MAX). The resulting model achieves an F1 of 99.8\% on the validation set, which measures the accuracy of predicting the correct \textit{spans} containing wage information. Given the model outputs, we design a simple parsing algorithm to separate the extracted spans into distinct MIN and MAX return values. In addition to the nominal wage values, we also train a multi-class \textsc{DeBERTA-v3-base} classification model to extract the pay frequency expressed by the wage values (\textit{hourly}, \textit{weekly}, \textit{monthly}, or \textit{annually}). These labels for the training dataset were obtained likewise via crafted regular expressions, and manually checked for validity. The training dataset represented a subset of 22k examples, and the resulting model predicts pay frequency with a F1 of 99.6.

\begin{figure}[h!]
  \centering
  \includegraphics[scale=0.3]{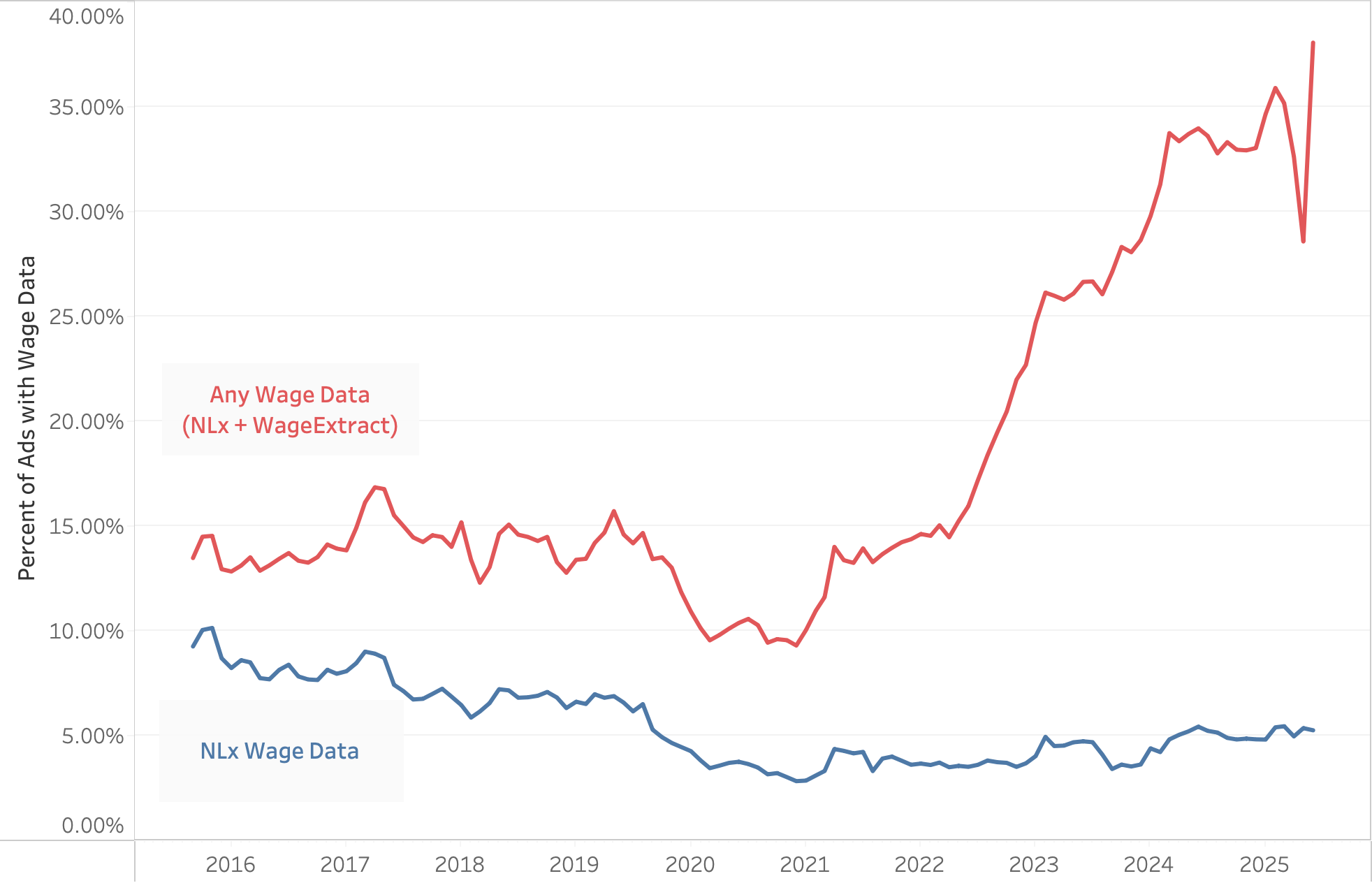}
      \caption{WageExtract: The percent of postings with wage information.}
  \label{fig:wage_extract}
\end{figure}

We combine the results of WageExtract with NLx's structured wage information and post-process results to remove outliers and standardize the wage as an annual salary, using either the point wage provided or the midpoint of the wage range provided. For the duration studied, NLx structured data includes a minimum or maximum wage for 4.62\% and 4.15\% of job postings. With WageExtract, we obtain wage information for far more observations. Figure \ref{fig:wage_extract} illustrates that the availability of wage information in our data hovers between 10\% and 13\% before 2022, and dramatically increases beginning in 2022. In our dataset, the percent of job ads with wage data in the text reaches 39.6\% in May 2025. 

For comparison purposes, an analysis of structured data provided by Lightcast, \cite{batra2023online} report that 14\% of job ads had any wage information between 2012 and 2017, and 8\% had point data. Using data from Lightcast, \cite{hazell2022national} state that 5\% of job ads include point wage data from 2010-2019. 

\subsubsection{JobTag (CRAML)}
\label{sec:jobtag}

The JobTag module of the Job Ad Analysis Toolkit (JAAT) classifies job ad text into user-defined categories using niche classifiers built with the Context Rule Assisted Machine Learning (CRAML) tool \citep{meisenbacher_classifiers_2022, meisenbacher_transforming_2023}. In particular, the nine classifiers are Random Forest classifiers trained on data built by expert validated rules that are run on ``context windows'' relevant to each niche class. For example, the class ‘union’ loads a classifier that first identifies whether a job ad contains a specific keyword indicating a section of a job ad is plausibly related to labor unions. For example, if the ``union'' keyword appears in a job ad, then the classifier will be run on the keyword in its relevant context -- the six words to the left and right of the keyword -- to determine if the job ad language truly indicates the presence of a labor union (as opposed to a credit union, etc.). 

\begin{figure}[h!]
  \centering
   \includegraphics[width=\textwidth]{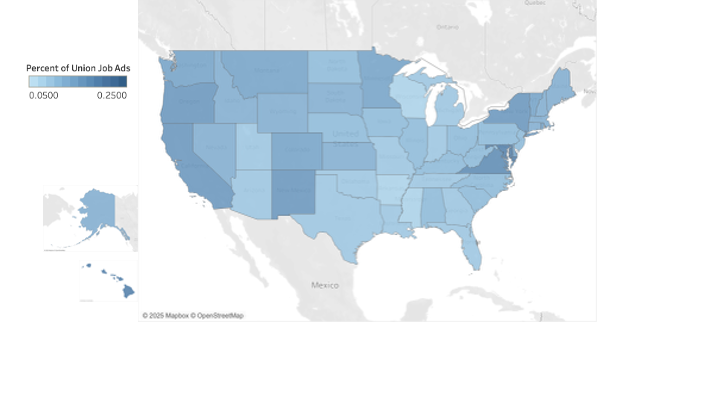}
    \caption{Mentions of Labor Unions as a Percent of Job Ads by State in 2024.}
  \label{fig:ind_contractor}
\end{figure}

As one example of a job tag feature, Figure \ref{fig:ind_contractor} illustrates state-level variation in the appearance of labor union mentions in job ads, as a percent of all monthly active job ads appearing in each state in 2024. Users should compare these data to other benchmark sources \citep[See pg. 13][and also see Appendix \ref{sec:robust}]{bls2025unions}.

This high-speed, flexible, and expandable method is used for pre-defined classifiers included in the JobTag module. JobTag illustrates the merits of CRAML’s domain-specific classifiers that are fine-tuned on expert-curated context rules.  The JobTag module is extensible in that it can support any number of newly added classifiers, accomplished via the definition of a new class and its keywords, and then via the validation of extracted context windows based on these keywords. In this way, should other researchers or practitioners develop and publish niche classifiers, this module allows for coverage of novel, emergent, and specialized interest in data extraction from job ads. 

% 9.9\% according to the BLS, and 11.8 \% in the job ads. 
% https://www.bls.gov/news.release/pdf/union2.pdf

\subsection{Dictionaries}
\label{sec:dictionaries}
We exact match terms using pre-existing and novel dictionaries that correspond to elements of O*NET's taxonomy of work. Custom dictionaries we develop are presented in Appendix \ref{sec:detailed_methods} Table \ref{tab:dict_title} for titles, and Appendix \ref{sec:all_dictionaries} Tables \ref{tab:dict_benefits} - \ref{tab:dict_dbc}, and include dictionaries for benefits, education, shifts, and drug, background and criminal background checks. 

%Standard dictionaries from the management literature are described in this section and presented in more detail in Appendix \ref{sec:std_dictionaries}.

To execute dictionary-based strategies, we use patent-pending analytic engines that scale large ‘knowledge maps’ with unique concept identifiers and association rules over unstructured text with exact matching  \citep{price_systems_2024}. `Knowledge maps' find and match one or more keywords to a standard label or code at high speed. Capable of addressing negation and complex association rules such as the presence of multiple unique concept identifiers within a specified span, lists of terms, such as the O*NET dictionary of 21,841 tools and technologies, are run against the corpus and return UNSPSC codes associated with the presence or absence of the dictionary term(s) within each job ad.

\begin{figure}[h!]
  \centering
  \includegraphics[width=\textwidth]{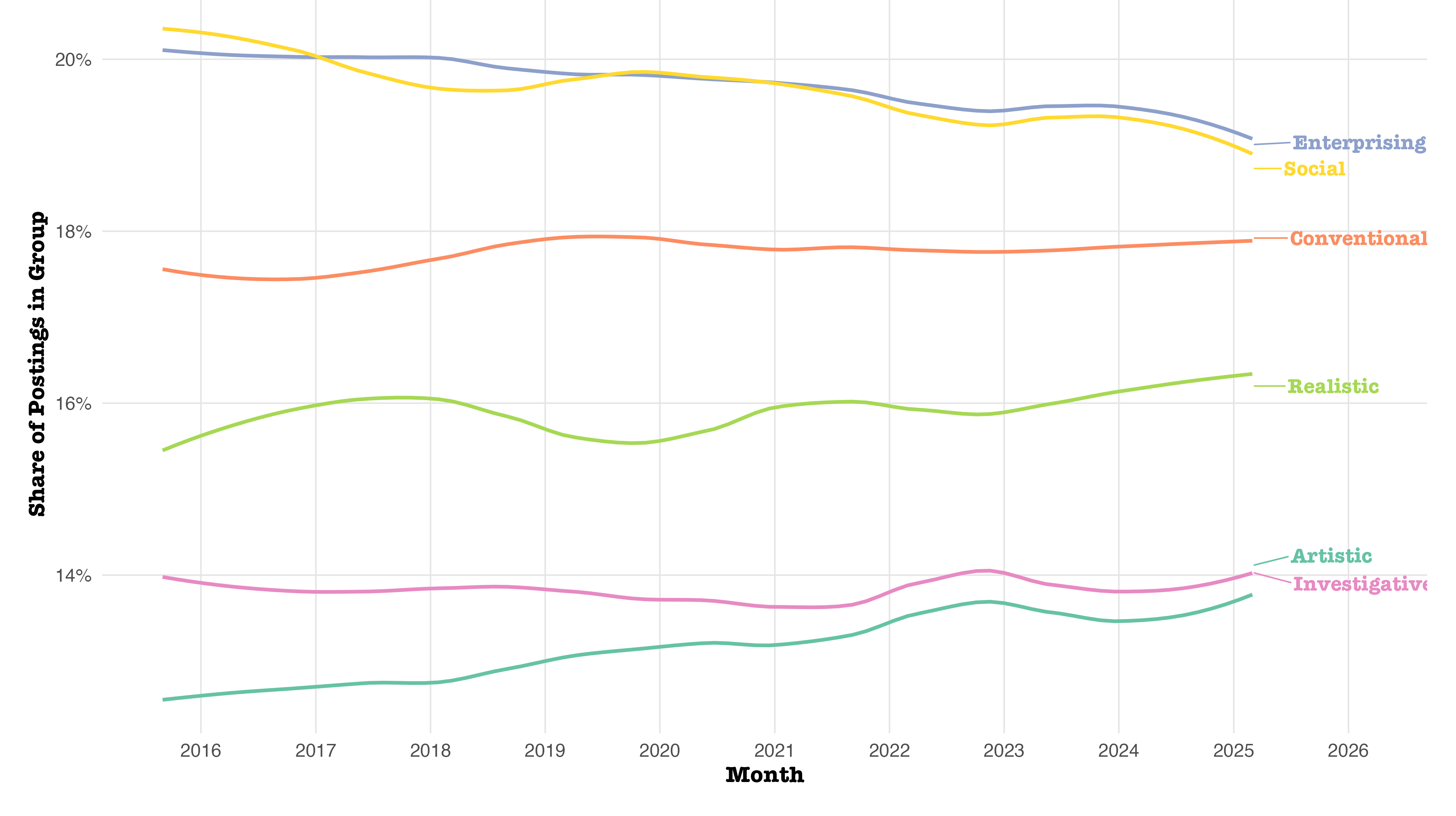}
  \\
  \small
  \vspace{5pt}
  \emph{Note: } This figure is smoothed and uses monthly data aggregated by date compiled. Social and Enterprising terms remain dominant, while Artistic terms increase as a percent of all RIASEC terms in job ads.
  \caption{RIASEC Keywords as a Percent of All RIASEC keywords.}
  \label{fig:riasec}
\end{figure}

We visualize results for one O*NET dictionary in an abbreviated fashion here to illustrate how counting words may be of use to other researchers. General occupational interests used in vocational interests and career planning based on Holland's (\citeyear{holland_making_1997}) RIASEC (Realistic, Investigative, Artistic, Social, Enterprising, and Conventional) framework are captured using a dictionary of RIASEC keywords developed by O*NET \citep{rounds_updating_2022}. Figure \ref{fig:riasec} illustrates that as a share of the total RIASEC keywords extracted, there has been an approximately 1\% decline in the share of enterprising and social keywords, and a 1\% increase in artistic keywords between 2015 and 2025.

%, and other novel dictionaries include a dictionary of diversity, equity, and inclusion terms based on a subset of the list of terms recently reported to the subject of scrutiny by the federal government \citep{yourish_these_2025}

\begin{figure}[h!]
  \centering
  \includegraphics[width=\textwidth]{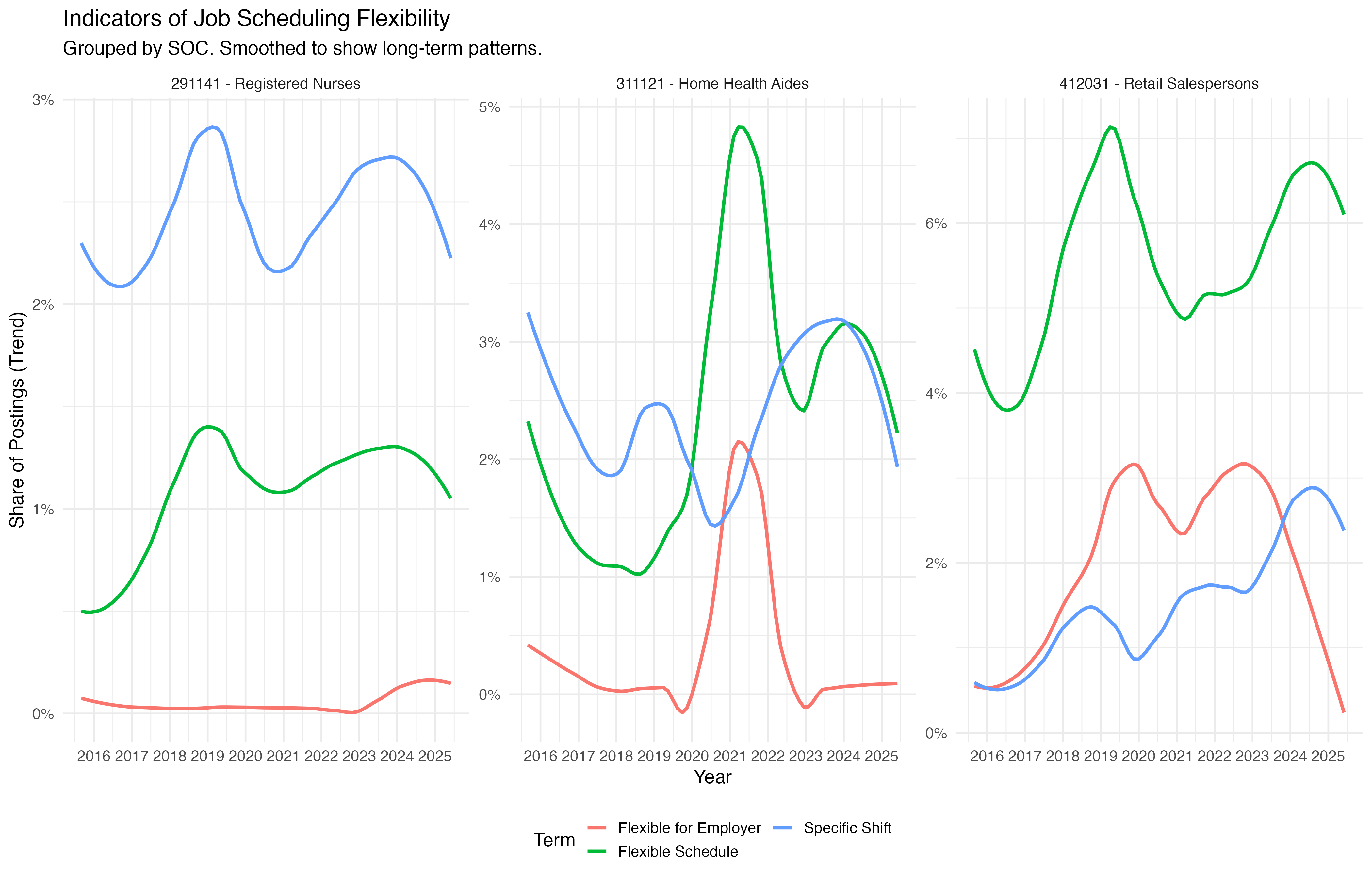}
  \small
  \vspace{5pt}
  \justifying
  \emph{Note: } This figure is smoothed and uses data aggregated by date compiled. 
    \caption{Indicators of Job Flexibility.}
  \label{fig:nurse_health_retail}
\end{figure}

Several novel dictionaries indicate various aspects of scheduling predictability and flexibility of the job. Figure \ref{fig:nurse_health_retail} illustrates several indicators of job flexibility and predictable schedules for three large occupations as coded by TitleMatch: Home Health Aides, Nurses, and Retail Salespersons. Specific shift includes phrases associated with a specific, predictable shift; flexible schedule indicates several types of unpredictable and flexible schedules, including those that indicate a willingness to accommodate workers' preferences; flexible for employer indicates a desire for workers who can work hours that the employer prefers. The results suggest a rise in flexible schedules and predictable shifts in the last decade across these three large occupations. While registered nurses and home health aides generally have a low percent of postings with expectations that the worker be flexible for employer needs compared to retail sales, there was a significant increase in expectations for flexibility around employer needs for home health aides during the 2020-2022 time period. Additional use of dictionaries for extraction is described in Appendix \ref{sec:onet_occ_reqs} in discussion of management practices, and Appendix \ref{sec:all_dictionaries} and custom dictionaries we develop.

%presented in the Appendix \ref{sec:std_dictionaries} 

\subsection{Aggregation}
\label{sec:aggs}

We aggregate data at month, occupation, industry, and geographic levels in order to build data that is usable for research and practitioner purposes. Occupation and industry aggregations at the 2-digit, 4-digit, and 6-digit level are performed with the output of TitleMatch and FirmMatch, respectively, as described above. We create sums, means, and percentile variables to reflect the underlying data within a ``month'' that we create as described below.

NLx has improved systems for collecting and storing job ad data over time. A major data warehouse upgrade in 2021 added comprehensive job history tables that track more precise windows of dates when job ads were posted.  For periods prior to 2015, additional job postings are available, but less reliable.

%\subsubsection{Building a Time Series: Month}
\subsubsection{Data Processing and Transformation}
\label{sec:date}

We extract data from files provided monthly by the NLx. All jobs included in a given monthly file were closed (taken offline) during that month.  The actual closing date is given in a field named \emph{date\_compiled}.  For example, the January 2025 monthly job ad file includes all job ads that were closed in January 2025. The values of \emph{date\_compiled} for all job ads in the January 2025 file range from January 1, 2025 to January 31, 2025.

%For months after September 2015,  NLx provides monthly `job' files with one row per unique job ad, grouped by the month of a job ad's last \emph{date\_compiled}. The NLx field \emph{date\_compiled} provides ``the last date [a job ad] appeared in the job feed.'' 

Analyzing NLx data, \cite{hashizume_timing} finds half of job postings from  Fortune 500 firms are available for 37 days or less. As the monthly file contains only the job postings in the month in which they are last posted, its contents include many postings that were also posted online in earlier monthly periods. In Appendix \ref{sec:robust_dates}, we provide more detail on our analysis of dates. As each monthly file can also be subject to large fluctuations (especially prior to 2021), we seek to smooth the data appropriately, accurately reflect that many postings that close in a month were on display in earlier months, and reduce the potential for noise in a given monthly jobs file to drive results.

\subsubsection{Monthly Active Jobs}
\label{sec:maj}

\begin{figure}[!ht]
    \centering
    \includegraphics[width=\textwidth]{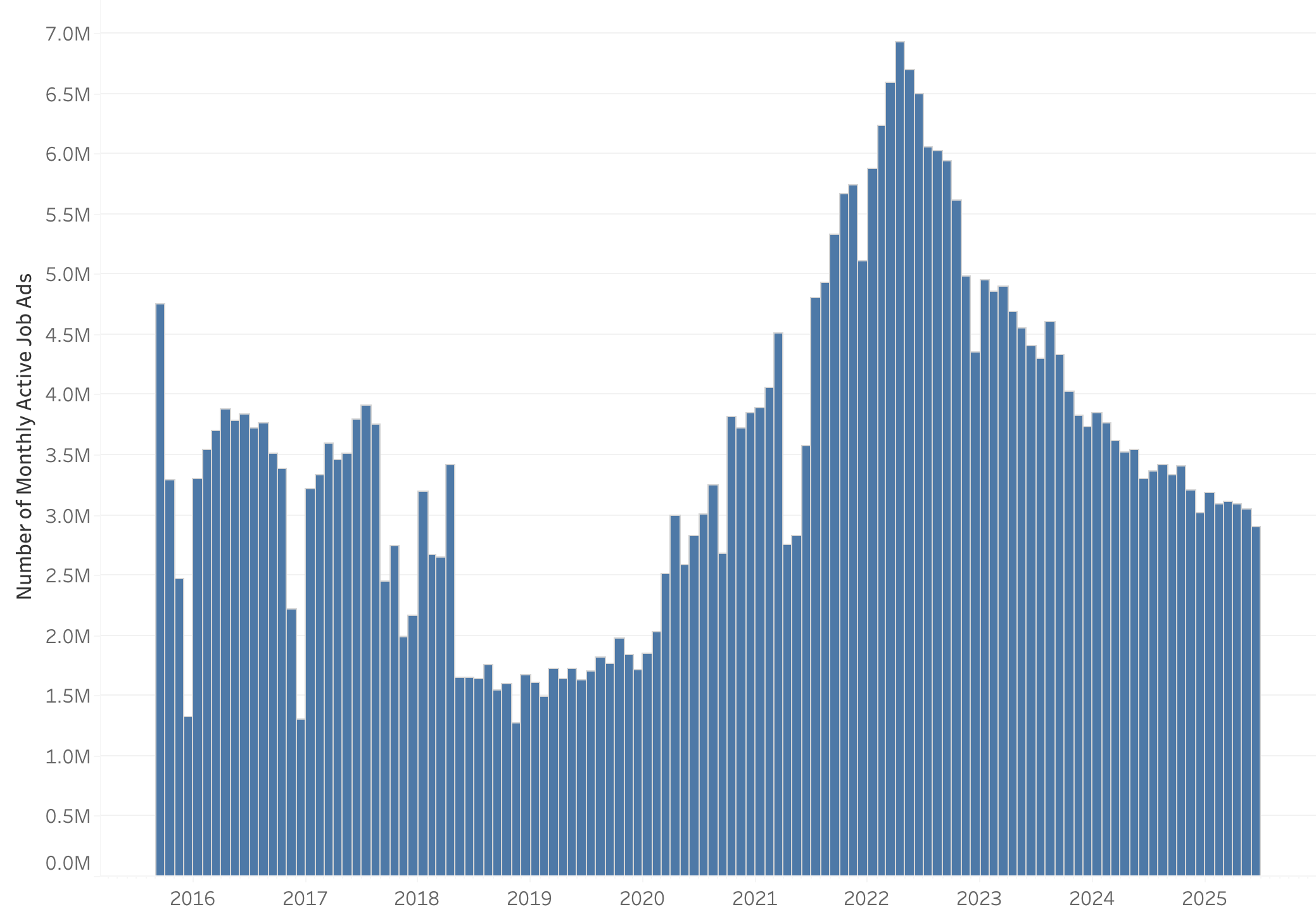}
    \caption{Number of Monthly Active Job Ads}
    \label{fig:distr_maj}
\end{figure}

We build our aggregate data using the concept of monthly active jobs (\emph{MAJ}).  A job is considered active during all months within the span of its \emph{date\_acquired} and \emph{date\_compiled}. Prior to 2021, there are several months with abnormally large numbers of jobs acquired, and other months with no monthly jobs acquired. As described in Appendix \ref{sec:robust_dates}, we develop a solution and create the (\emph{MAJ}) to address the problem. Figure \ref{fig:distr_maj} presents the distribution of monthly active jobs we use for construction of aggregate data. Except where otherwise noted, figures are aggregated by (\emph{MAJ}).

% \subsubsection{Geocoding: State and Zip Code to Commuting Zone}
% \label{sec:geo}

% For geographic analysis, we standardize the state field from NLx metadata, and use the zip code from the NLx metadata to construct  1990 commuting zone level data. We link zip codes from job ads to county FIPS codes provided by \cite{din2020crosswalking}.  If a zip code includes more than one county FIPS, we assign the zip code to the county with the greater proportion of businesses.  We link from FIPS codes to 1990 commuting zones / local labor market areas \citep{tolbert1996us}.

\subsubsection{Convergent Validity of Aggregated Data}
\label{sec:agg_converge}

\begin{figure}[htbp]
    \centering
    \includegraphics[scale=0.4]{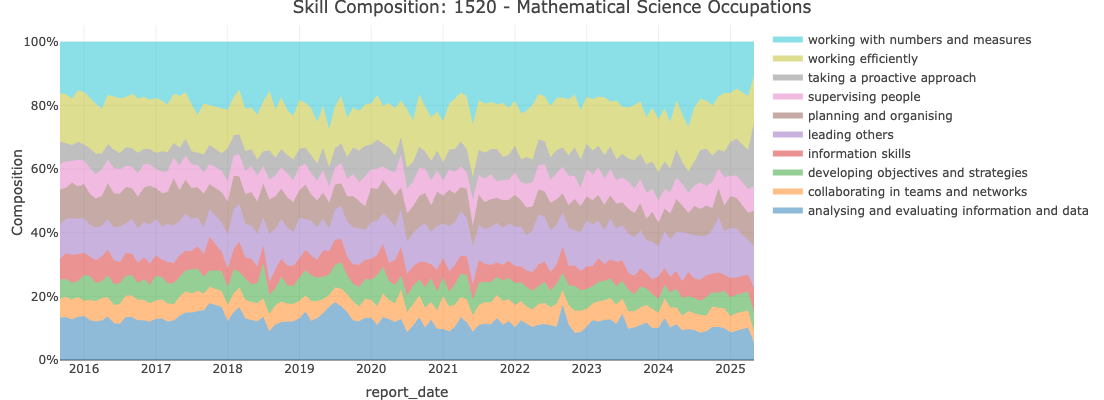} \\
    \vspace{15pt}
    \includegraphics[scale=0.4]{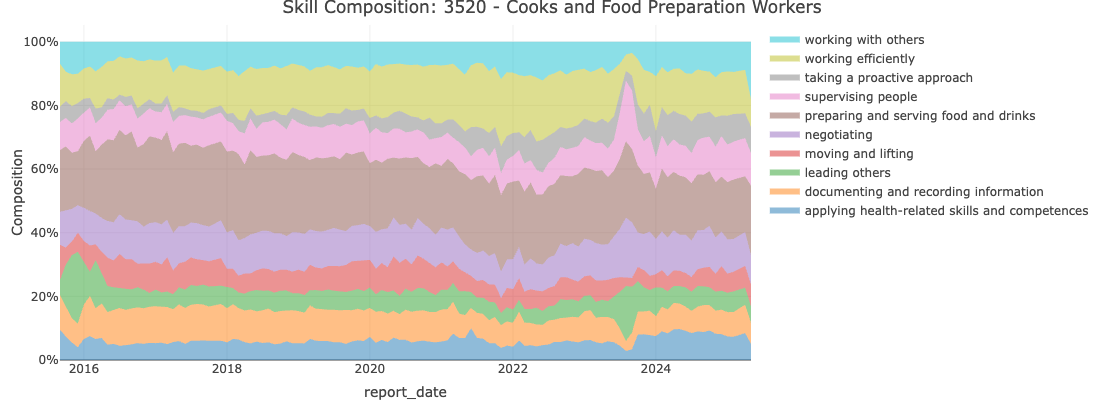} \\
    \vspace{15pt}
    \includegraphics[scale=0.4]{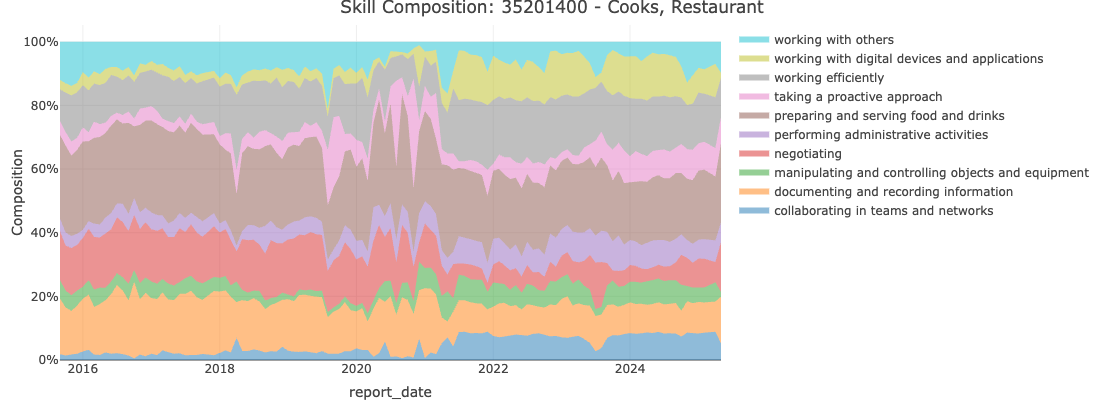}
    \small
    \justifying{\emph{Note:} These figures are aggregated by date compiled. The stability of skills in the Mathematician occupation can be contrasted with post-pandemic changes in skills desired in the Cooks occupation over a 10 year period. Changes are magnified at the detailed occupation level. Examining Restaurant cooks, after the COVID-19 pandemic, there is an observed increase in the demand for collaboration and digital skills.}
    \caption{Top 10 Tasks by Month for Mathematicians, Cooks, and Fast Food Cooks}
    \label{fig:cooks_skills}
\end{figure}

The convergent validity of each JAAT tool can be evaluated in combination with with aggregate data from other tools. Scrutinizing skill output by occupation, for example, combines data from two independently constructed models, SkillMatch and TitleMatch, trained with different models on different data from different parts of a job ad. Figure \ref{fig:cooks_skills} illustrates the top 10 SkillMatch results for two occupations at the minor group level -- Mathematicians and Cooks and Food Preparation Workers -- and Fast Food Cooks at the detailed occupation. Top skills for Mathematicians demonstrate month-to-month fluctuations but remain relatively stable over time. The top skill is ``working with numbers and measures'', with ``analyzing and evaluating information and data'', and ``information skills'' also in the top ten.

For Cooks and Food Preparation Workers, ``preparing and serving food and drinks'' is the top skill over the duration. There is a post-pandemic increased emphasis on working efficiently, but no apparent shock in the immediate pandemic aftermath. At the detailed occupation level, for restaurant cooks, the pandemic appears to be a seismic event, with durable aftershocks: while great volatility occurs between 2020 - 2022, ``collaborating in teams and networks'' and ``working with digital devices'' emerge as significant and enduring top 10 post-pandemic skills desired by employers.

\paragraph{TaskMatch: Inspecting Detailed and Aggregate Results}
To enable further independent assessment of the validity of tools, we developed a data exploration tool to inspect the top tasks from each month by major and detailed occupation codes, industry, and state. This allows for easy comparison to O*NET data.  As an illustration, Table \ref{tab:firefighter} provides the top 15 tasks extracted for the Firefighter occupation (33-2011.00) and, by comparison, the most important tasks in the O*NET data. There are subtle differences between O*NET data for this and other occupations that indicate that, for certain purposes, job ad informed O*NET task data might be preferred to O*NET's survey-based data.

%We restrict the data in the application to show only cells with more than 200 job postings.

\begin{table}[htbp]
  \centering
  \begin{threeparttable}
    
    \small % Use a smaller font for the entire table content
    
    % --- First Table (from Job Ads) ---
    \justifying 
    \textbf{A. Top Firefighter Tasks Extracted by TaskMatch from NLx Job Ad Data}
    \par\medskip
    \begin{tabular*}{\textwidth}{@{\extracolsep{\fill}} >{\raggedleft\arraybackslash}p{2cm} p{13cm}}
    \toprule
    \textbf{Count}  & \textbf{Task} \\
    \midrule
    15611 &  Participate in firefighting efforts. \\
    8226  &  Drive and operate fire fighting vehicles and equipment. \\
    6619  &  Conduct wildland firefighting training. \\
    4017  &  Clean and maintain fire stations and fire fighting equipment and apparatus. \\
    2441  &  Work with or remove hazardous material. \\
    2164  &  Rescue and evacuate injured persons. \\
    2118  & Conduct fire, safety, and sanitation inspections. \\
    1961  &  Communicate fire details to superiors, subordinates, or interagency dispatch centers, using two-way radios. \\
    1867  &  Develop training materials and conduct training sessions on fire protection. \\
    1529  &  Interview and hire applicants. \\
    1473  &  Assign duties to other staff and give instructions regarding work methods and routines. \\
    1460  &  Operate safety equipment and use safe work habits. \\
    1305  &  Direct, and participate in, forest fire suppression. \\
    995   &  Maintain knowledge of fire laws and fire prevention techniques and tactics. \\
    905   &  Supervise activities of other forestry workers. \\
    \bottomrule
    \end{tabular*}
    
    \par\bigskip % Add more vertical space between the tables
    
    % --- Second Table (from O*NET) ---
    \justifying 
    \textbf{B. Top Firefighter Tasks from O*NET Data}
    \par\medskip
    \begin{tabular*}{\textwidth}{@{\extracolsep{\fill}} >{\raggedleft\arraybackslash}p{2cm} p{13cm}}    \toprule
    \textbf{Importance} & \textbf{Task} \\
    \midrule
    93 & Rescue survivors from burning buildings, accident sites, and water hazards. \\
    91 &  Dress with equipment such as fire-resistant clothing and breathing apparatus. \\
    90 &  Assess fires and situations and report conditions to superiors to receive instructions, using two-way radios. \\
    90 &  Move toward the source of a fire, using knowledge of types of fires, construction design, building materials, and physical layout of properties. \\
    90 &  Respond to fire alarms and other calls for assistance, such as automobile and industrial accidents. \\
    89 &  Create openings in buildings for ventilation or entrance, using axes, chisels, crowbars, electric saws, or core cutters. \\
    88 &  Drive and operate fire fighting vehicles and equipment. \\
    88 &  Inspect fire sites after flames have been extinguished to ensure that there is no further danger. \\
    87 &  Position and climb ladders to gain access to upper levels of buildings, or to rescue individuals from burning structures. \\
    87 &  Select and attach hose nozzles, depending on fire type, and direct streams of water or chemicals onto fires. \\
    86 &  Operate pumps connected to high-pressure hoses. \\
    84 &  Maintain contact with fire dispatchers at all times to notify them of the need for additional firefighters and supplies, or to detail any difficulties encountered. \\
    84 &  Collaborate with other firefighters as a member of a firefighting crew. \\
    83 &  Patrol burned areas after fires to locate and eliminate hot spots that may restart fires. \\
    83 &  Collaborate with police to respond to accidents, disasters, and arson investigation calls. \\
    \bottomrule
    \end{tabular*}
   
    \par\medskip
    
    % --- Table Notes ---
    \begin{tablenotes}[flushleft]
      \small 
      \justifying \item \textit{Note:} For the September 2015 - June 2025 period, 294,651 tasks are extracted from 26,987 job postings in the firefighter occupation. This table reports the top 15 based on data from an aggregated summary table on the top tasks for each occupation and month (by date compiled).
    \end{tablenotes}

  \end{threeparttable}
  \caption{The Top Fifteen Tasks for Firefighters in NLx Job Ad Data and O*NET Data.}
  \label{tab:firefighter}
\end{table}

\begin{figure}[htbp]
    \centering
    \includegraphics[scale=0.4]{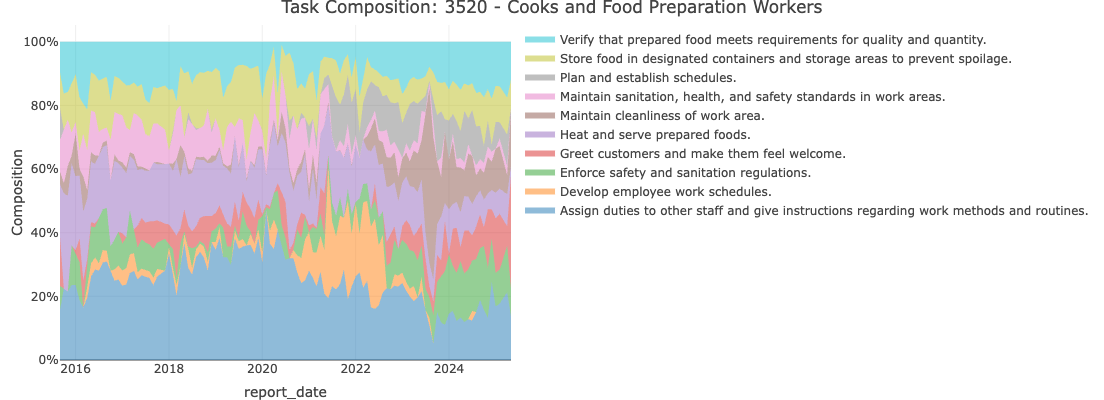}
    \\
    \small
    \justifying{\emph{Note:} This figure is based on data aggregated by date compiled. The increased share of tasks involving managing employee work schedules (orange) during the 2020-2022 period aligns with post-pandemic labor shortages and the ``great resignation.''}
    \caption{Top 10 Tasks by Month for Cooks}
    \label{fig:cooks_tasks}
\end{figure}

Data exploration of subgroups can provide additional demonstrations of the convergent validity of TaskMatch and other tools. As discussed above, a limitation of O*NET is the representation of occupational tasks based on data collection from a single point in time. Many labor market observers assume that there is change in tasks within occupations over time. Data exploration enables investigation of trends and change over time, and assists in assessing convergent validity in combination with knowledge of specific occupational and industry on trends involved. For example, Figure \ref{fig:cooks_tasks} illustrates change in the top 10 tasks over time for the minor occupation group ``3520 - Cooks and Food Preparation Workers,'' an occupation affected greatly by pandemic health concerns and post-pandemic labor shortages often referred to as the ``Great Resignation.'' In the 2020-2022 time period, task statements ``Maintain sanitation, health, and safety standards'' and ``Developing employee work schedules'' grew as a share of the top 10 tasks sought by employers, corresponding to trends described in trade publications \citep{littman_2021}. 

\paragraph{WageExtract}

\begin{figure}[h!]
  \centering
    \includegraphics[width=\textwidth]{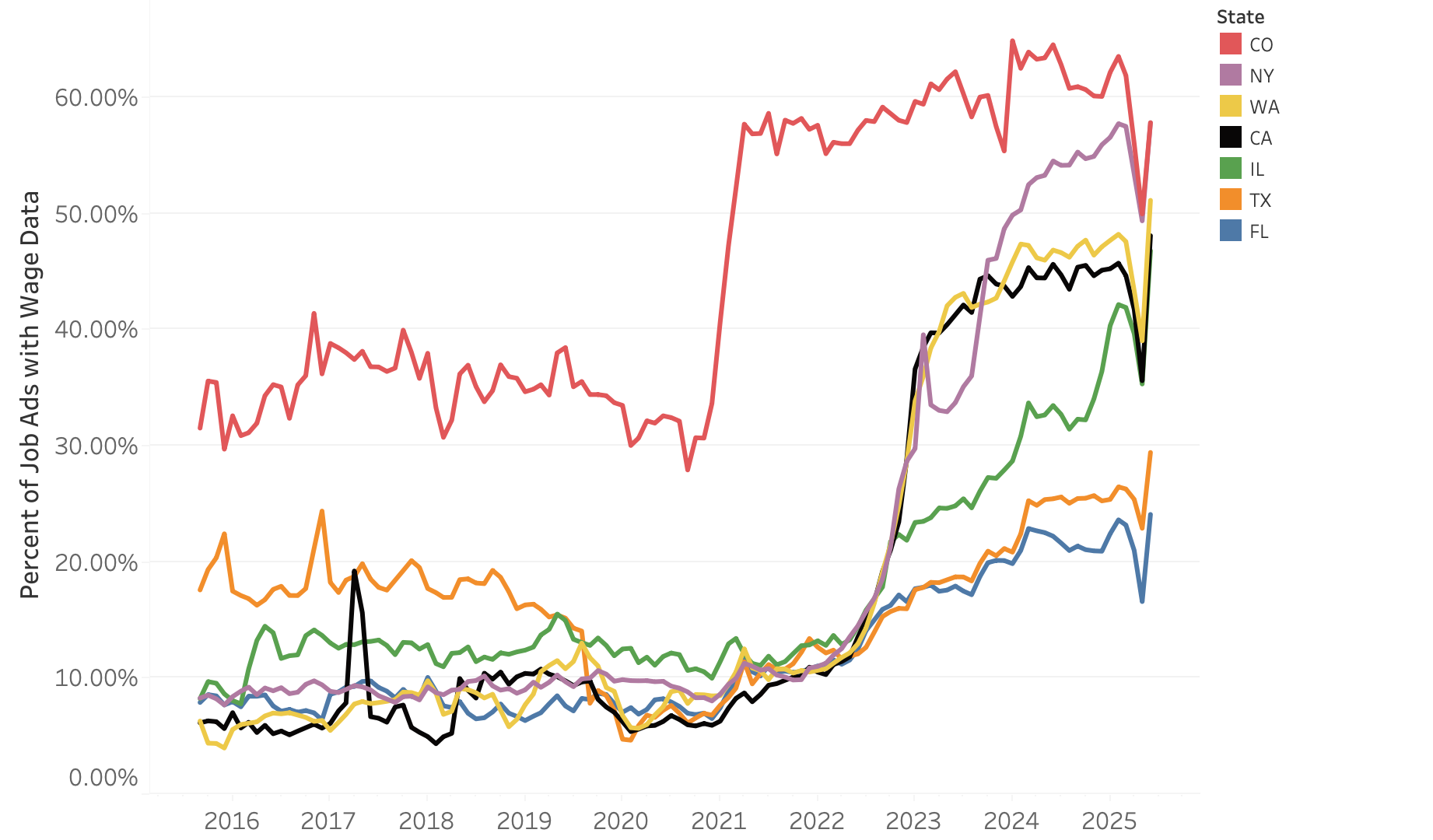}
  \small
  \justifying
  \emph{Note: } Colorado enacted the first state-wide pay transparency law in 2019, which took effect January 2021. Other states have followed: Washington and California laws took effect January 2023; New York in September 2023; Illinois' law took effect January 2025. With each law, the percent of job ads including a wage that we extract increases significantly following passage. Florida and Texas did not pass laws.
\caption{WageExtract: Wage Information Availability in Select States}
  \label{fig:pay_transparency}
  \end{figure}

Using BGT data, \cite{arnold2022impact} report that availability of pay information increases by 30 percentage points following the implementation of Colorado's pay transparency law and reaches 70\% by the year following the law taking effect. We find similar results in our dataset, for Colorado and other states. Figure \ref{fig:pay_transparency} demonstrates that availability of wage information varies significantly by state, and increases significantly following the passage of pay transparency laws.  

\section{Results}
\label{sec4}

Our aggregate data is capable of providing both telescopic and wide-angle evidence of labor market demand changes, with an unprecedented number of features of work within and across occupations over time, and by industry, and geography. We highlight several results for researchers and practitioners, including the ability to capture labor market change and shocks in granular detail, as well as unprecedented large-scale description of national trends in skills, tasks, and management practices. Each result presented here requires more in-depth research. Users of this information should proceed with an awareness of the known limitations of this new data, which are described more fully in the concluding section.

\subsection{Applications for Research}

To illustrate significant national trends and shocks in labor market demand by employers, we demonstrate: 1) meaningful shocks that alter demand for specific task bundles in the last ten years; 2) changes in the measured linguistic complexity of job postings over the 10 year period; and 3) national trends involving the gradual rise of project-based work, which is detected at the task level, especially after the pandemic, and the rise of interpersonal work activities as a share of postings.

%Aggregate data released and tools in the JAAT toolkit are available for research purposes. Trained on job advertisement text and custom built, the JAAT toolkit can be especially helpful for researchers examining and comparing the extracted and structured information in job ads and other sources of information. While the data released contain a great deal of information that may be useful for specific studies, JAAT tools may be of assistance when  working with job ad data for specific study designs. While not built here for other purposes, JAAT tools can also retrieve useful information from other labor related corpora.  Here, we demonstrate how the aggregate data can be of use in the study of how technological and other shocks affect the organization of work and jobs as an aid to exploration and discovery as well as an input to future work focused on causal inference. 

\subsubsection{A Wide-Angle Lens on National Trends}

In a turbulent ten year period of economic change including a global pandemic, hot labor markets, supply-chain disruptions, and technological shocks, trace evidence of trends that gradually transform the labor market, and shocks that quickly change employer demand, should be evident in the 2015-2025 time frame. Many research agendas depend upon access to large-scale and real-time data to understand these changes and shocks.

Tight pre-pandemic and post-pandemic labor markets (referred to as the ``great resignation'') exemplify periods of high labor demand and extremely low unemployment. The release of ChatGPT on November 30, 2022, heralding rapid advances in generative AI, and the COVID-19 pandemic, beginning in March 2020, are examples of shocks -- sudden unexpected events with far-reaching implications. Such shocks can have short and long term impacts. Over a longer period of time, generative AI is likely to transform occupations as workers adopt technology to replace their effort on certain tasks \citep{chatterji_how_2025,handa2025economic}. COVID-19 closed many workplaces to all but ``essential workers'' in the short-term, with effects on the composition of available jobs, and likely had lasting impacts on the organization of work at different levels of education, in specific sectors, and occupations. 

Event studies and plausibly exogenous technological and policy shocks are rare in research with job postings. \cite{horton_death_2025} demonstrate that it is possible to trace the extinction of technologies through job postings, and \cite{sauerwald_political_2024} trace the impact of board members' prior political experience on recruitment policies toward foreign-born workers after President Trump's first assumption of power in 2017. However, no prior public research-use dataset has been able to follow change at a highly-granular level, including change within occupation at the task-level, in combination with thousands of features such as technologies, tools, using the standard taxonomies for skill and other features.

\begin{figure}[ht!]
\centering
\includegraphics[width=\textwidth]{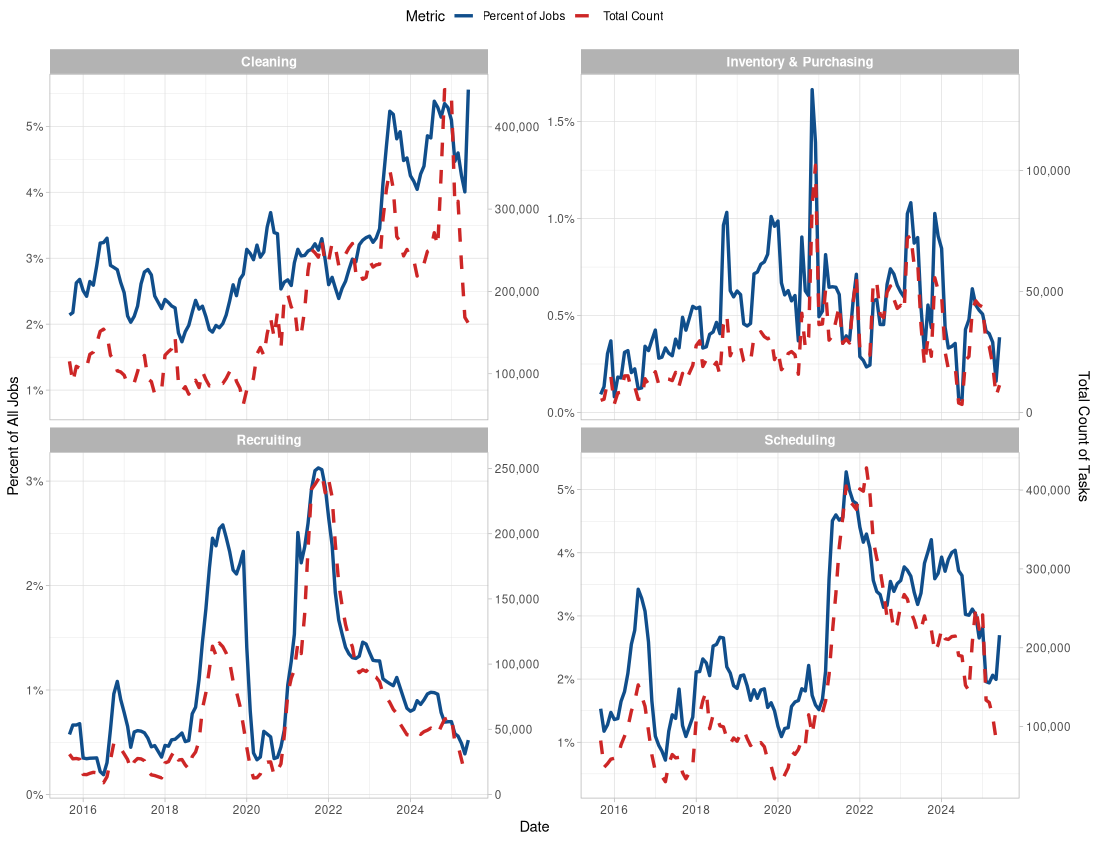}
\\
\vspace{5pt}
\justifying
\emph{Note: } This figure illustrates task bundle trends, comparing absolute numbers and proportional contribution over time. It is based on Task ID data aggregated by month compiled. Cleaning tasks are at the top left (Task IDs 23536, 20779, 9515, 20790, 23102, 23557). Inventory and purchasing tasks are in the top right (Task IDs 15704, 1333, 72). Scheduling tasks are in the bottom right (Task IDs 23695, 18658, 1106). Recruiting and hiring tasks are in the bottom left (Task IDs 22954, 23170, 17611, 9711, 18858, 21313).  
\caption{National trends in task bundles related to cleaning, inventory and purchasing, recruiting and scheduling.}
\label{fig:nat_trends}
\end{figure}

Figure \ref{fig:nat_trends} presents a time series of tasks related to cleaning, inventory and purchasing, recruiting, and scheduling from the 2015-2025. This figure presents both count and percent (as a percent of job postings in a month) to illustrate that the composition of job postings in the labor market and number of jobs involving a task being offered are different but both relevant. Moving clockwise from the top left quadrant of Figure \ref{fig:nat_trends}, we first illustrate change in cleaning tasks. A national increase in the count of cleaning-related tasks within job postings is evident after 2020 and the pandemic. However, the proportion of jobs including cleaning tasks does not increase until 2023. This may be due to changes in the composition of labor demand during the period of pandemic restrictions: while there was increased demand for cleaning among jobs that were open, the overall labor market demand may have been biased against in-person jobs that require cleaning. 

Pandemic-induced shifts in consumer demand and international supply chains disrupted inventories following the onset of COVID-19. As a percent and a count, tasks related to inventory and purchasing increased dramatically in months immediately before and after January 2021. The bottom right panel illustrates a dramatic increase in scheduling related tasks beginning in 2021, a period referred to as ``the great resignation'' and one in which employers made efforts to retain workers. Flexible scheduling and stable work hours are a top worker amenity, especially for women and other groups, with significant implications for labor market participation \citep{bell_job_2020}. The bottom left quadrant displays tasks involving recruitment and hiring. Recruitment peaks pre-pandemic and post-pandemic are interrupted by a plunge in recruitment related tasks during the peak pandemic period.

% cleaning_ids <- c(23536, 20779, 9515, 20790, 23102, 23557)
% schedule_ids <- c(23695, 18658,1106)
% recruit_ids <- c(22954, 23170, 17611, 9711, 18858, 21313)
% inventory_ids <- c(15704, 1333, 72)

\begin{figure}[htbp!]
\centering
\includegraphics*[width=\textwidth]{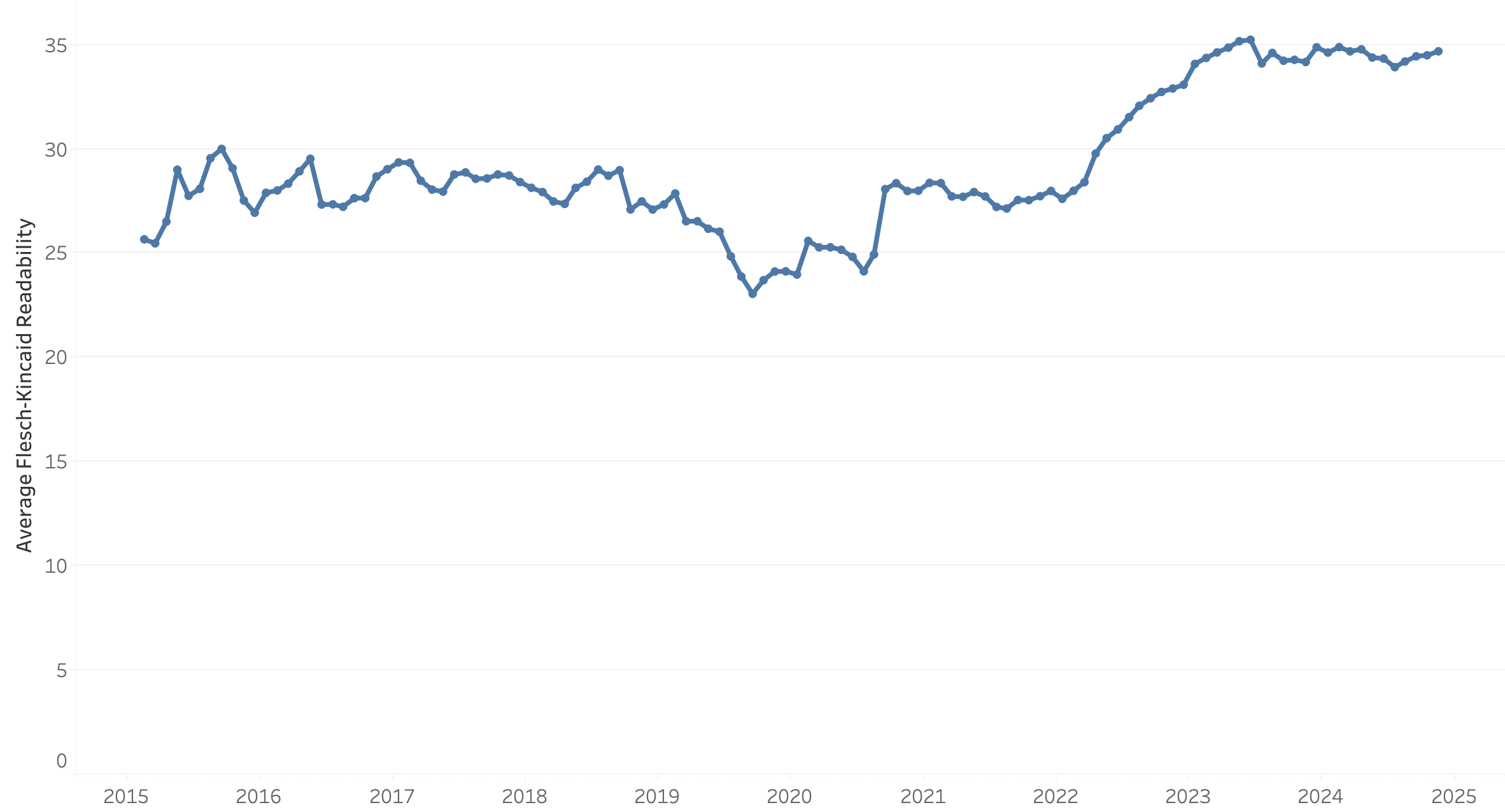}
\caption{National Changes in the Readability of Job Ads}
\small
    \justifying{\emph{Note:} Higher values of the Flesch reading-ease score correspond to more readable text. We notice a dip during the pandemic (possibly due to changes in the composition of postings) and an increase in 2022 (possibly due to the adoption of LLMs in recruitment).}
\label{fig:fk_read_natl}
\end{figure}

We construct the Flesch-Kincaid reading-ease score for all job ads and provide the average by occupation and other groups. Figure \ref{fig:fk_read_natl} demonstrates a decline in the readability of job ads in the months following the COVID-19 pandemic, likely due to changes in the composition of jobs posted in that time period and a shift in labor market demand toward jobs requiring higher levels of education. In the months after November 2022, there is a sharp increase in readability of job postings nationally, which may suggest the adoption of LLM tools that can aid recruiters in writing the postings, and job seekers in accessing jobs. 

\begin{figure}[htbp]
    \centering
    \includegraphics[scale=0.55]{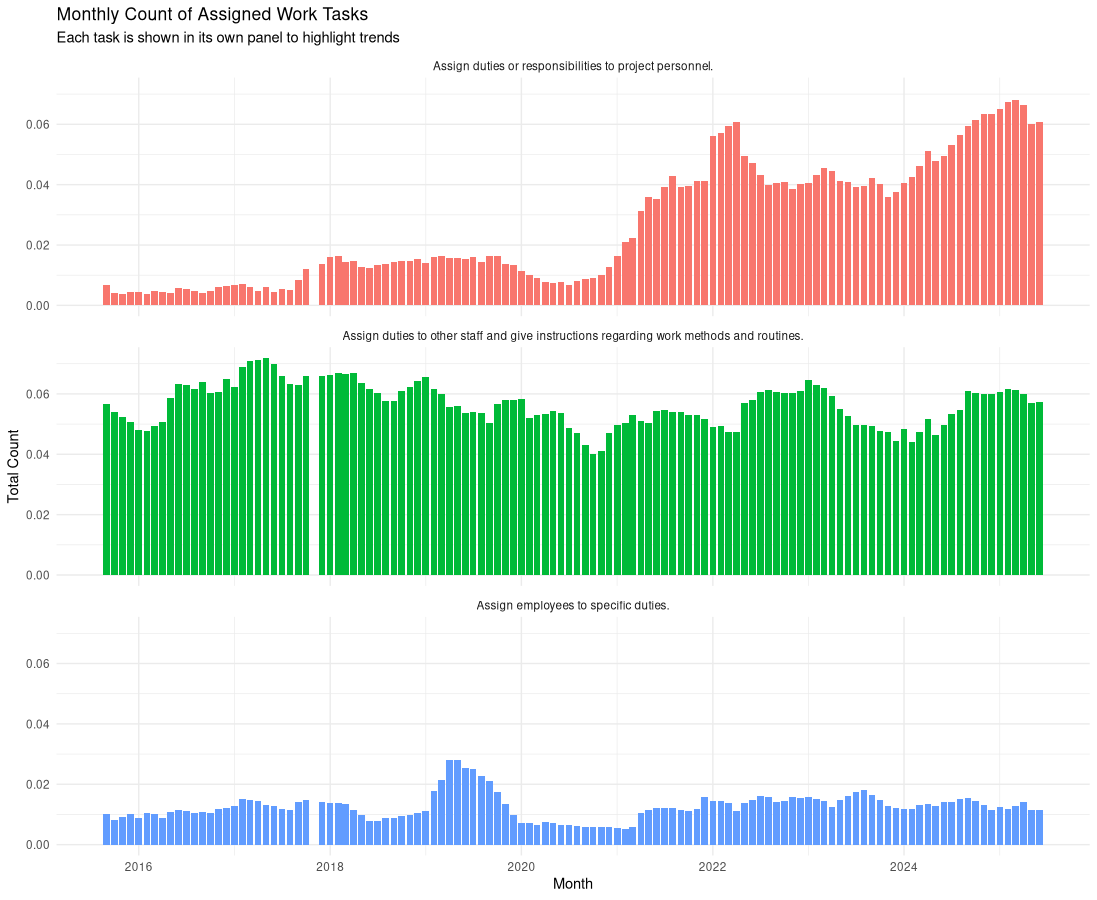} 
    \small
    \justifying
    \emph{Note:} The top panel illustrates the growth of assignment of duties to project personnel. The middle panel illustrates that assignments of work to staff is relatively stable. The bottom panel illustrates relative stability of assignment of employees to duties. This figure is based on task data aggregated by month of date compiled.
    \caption{TaskMatch: Monthly Count of Specific Tasks}
    \label{fig:tasks_projects}
\end{figure}

Examining the 10 most common task statements with a decade of data, we note a significant growth in the share held by Task ID 21462 ``Assign duties or responsibilities to project personnel.'' This growth has occurred in conjunction with  an overall stability in Task IDs related to assigning work to employees or staff, e.g., Task IDs 659, ``Assign employees to specific duties.'', and Task ID 9583 ``Assign duties to other staff and give instructions regarding work methods and routines.'' Figure \ref{fig:tasks_projects} Panel A (top) illustrates that, for all occupations, the project related Task ID is 0.6\% of Task IDs in 2015 versus 6.0\% in 2025, while assignments of work to staff or employees is relatively flat over the time period, moving from 5.1\% to 5.6\%. The subtle difference that the model picks up between unique Task IDs related to projects versus staff or employees merits deeper investigation, and would be washed away at even the next level of aggregation, the detailed work activity, where these Task IDs fall under DWA 4.A.4.b.4.I13.D06 with the label ``Assign duties or work schedules to employees.''

\begin{figure}[htbp!]
\centering
\includegraphics*[width=\textwidth]{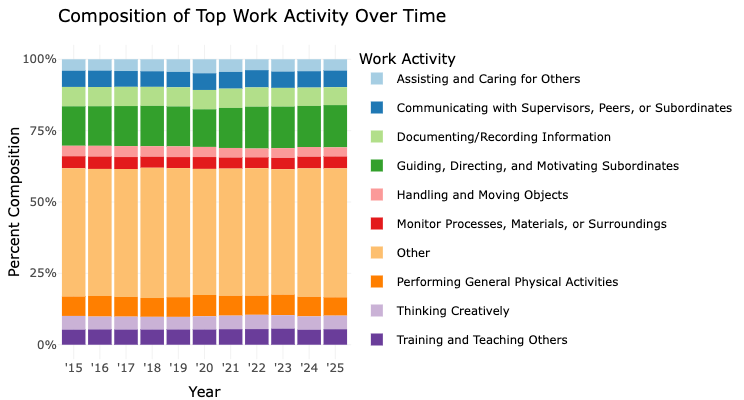}
\small
\justifying
\emph{Note: }This figure is based on aggregation by year of date compiled. 
\caption{TaskMatch at the Activity Level: Increased Emphasis on People Management}
\label{fig:activities_people}
\end{figure}

The last major trend we note is the rise of interpersonal work activities, including managing people, communication, and providing consultation. This change is visible at the work activities level. As seen in Figure \ref{fig:activities_people}, for all occupations, the percent of jobs involving supervision of people increases from 13.8\% to 14.8\% from 2015-2025. Again, the change can be seen more dramatically within detailed occupations. In the software developer occupation, managing people grows from 13.7\% to 17.5\% of work activities over the duration. 

\subsubsection{A Telescope for Finer-Grained Analysis}
\label{task_trends}

\begin{figure}[htbp!]
\centering
\includegraphics*[width=\textwidth]{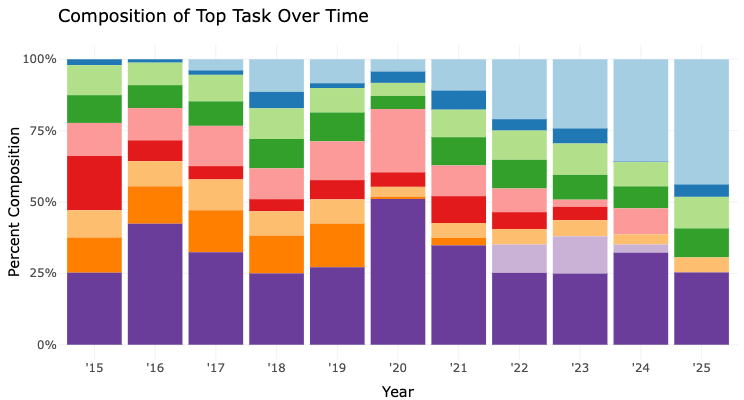} 
\vspace{5pt}
\small
\justifying
\emph{Note: }This figure is based on aggregation by year of date compiled. The growing share of in project-based work tasks is depicted at top in light blue. 
\caption{Top Ten Tasks Over Time in the Software Developer occupation (15-1252.00).}
%At bottom are the Top 10 Activities for Software Developer. Guiding, Directing, and Motivating Activities are depicted in pink and Communicating Activities in dark blue.}}
\label{fig:software_developer_activity}
\end{figure}

Our dataset is capable of tracking change within detailed occupations, as suggested above in the rise of health-related tasks and scheduling for cooks and food preparation workers post-pandemic in Section \ref{sec:taskmatch} and Figure \ref{fig:cooks_tasks}. The opportunity to track fine-grained changes provides additional means to assess labor market change. Taking one example from the prior section, the shift toward project-based work is particularly dramatic in specific occupations. Figure \ref{fig:software_developer_activity} illustrates that, for Software Developers, assigning duties to project personnel did not appear as a top 10 task in 2015, and grew to 43.8\% of top tasks for the occupation by 2025. Further investigations by occupation and industry, especially when paired with subject matter knowledge and expertise, could suggest whether this is a shift in the organization of work toward flatter, specific purpose project-based teams, and/or fissuring of employment and outsourcing \citep{weil_fissured_2014}. Measurement of contingent work, independent contracting, and management of vendor staff rather than employees of the direct firm, is a challenging issue \citep{dey_rise_2025}. 

\begin{figure}[htbp!]
\centering
\includegraphics*[width=\textwidth]{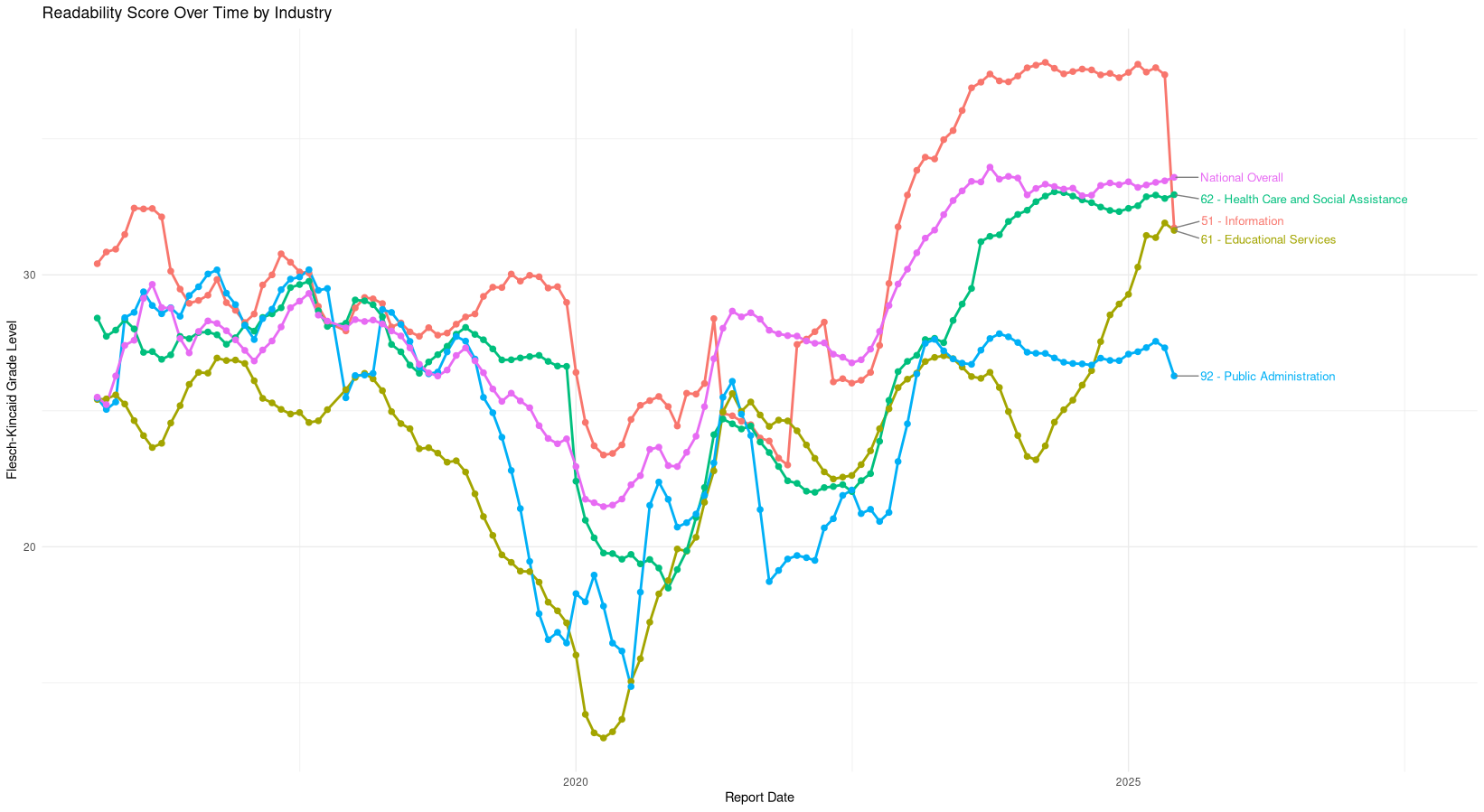} 
\small
\justifying
\emph{Note: }This figure is based on aggregation by month of date compiled. 
\caption{Growth of ML in the Computer and Information Research Scientist Occupation}
\caption{Illustrative Sectoral Changes in the Readability of Job Ads}
\label{fig:fk_read_info}
\end{figure}

As another example from the above section, adoption of generative AI technology in postings may be less or more prevalent in specific industries and/or geographies.  Figure \ref{fig:fk_read_info} features the change in readability of job ads nationally in contrast to different sectors including information, public administration, education, and health care. Information sector job postings become more readable faster than the national trend following the launch of generative AI tools, while adoption in the education and health care sectors are slower but converging toward the national level, while public administration postings have a smaller change. 

\begin{figure}[htbp!]
\centering
\includegraphics[width=\textwidth]{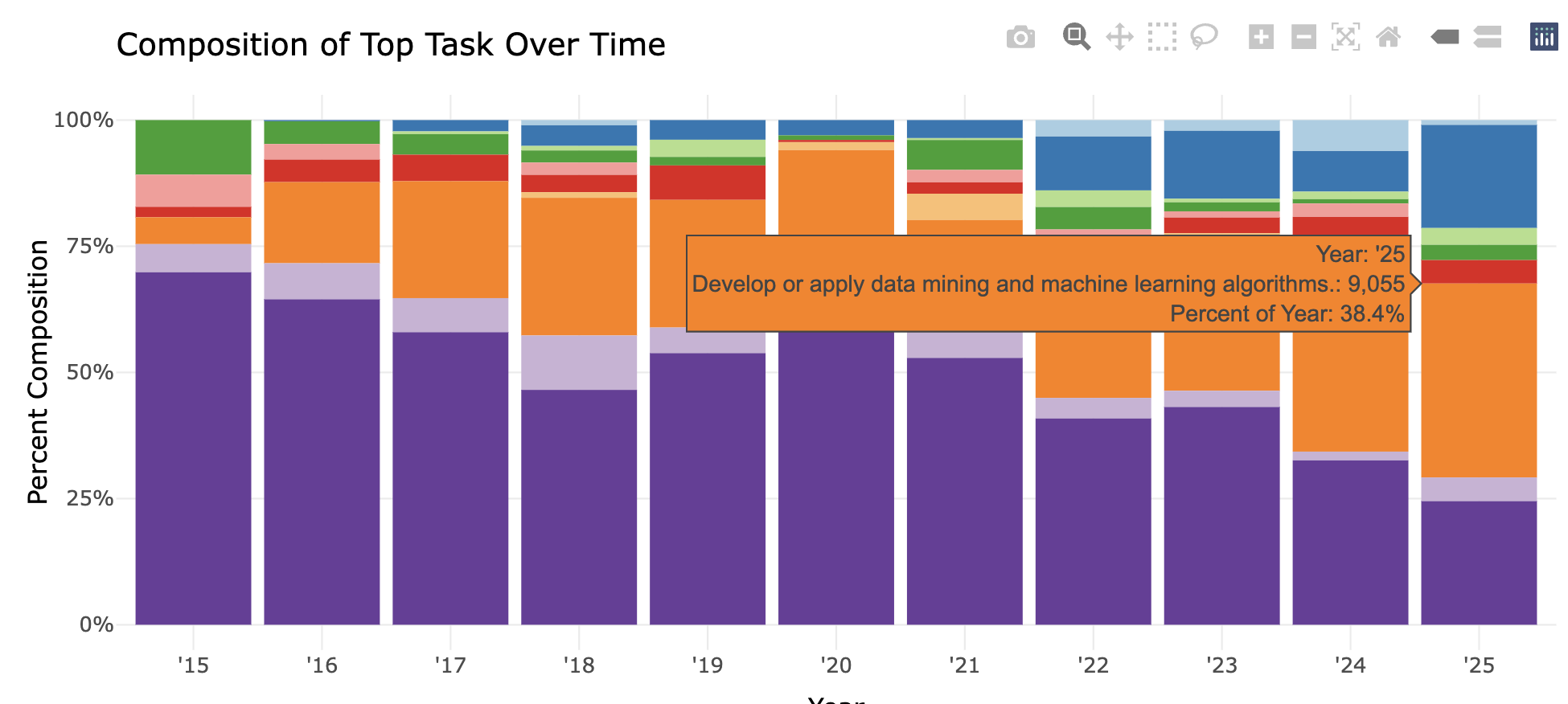}
\small
\justifying
\emph{Note: }This figure is based on aggregation by year of date compiled. 
\caption{Growth of ML in the Computer and Information Research Scientist Occupation}
\label{fig:cs_change}
\end{figure}

Along with the shock to the entire labor market with the widespread adoption of large language models (LLMs), our dataset allows us to see historical changes in key occupations that may presage broader transformations. For example, machine learning algorithms and data mining are understood to be essential to advances in computer science, AI, and LLMs. Using TaskMatch output, Figure \ref{fig:cs_change} indicates that within the Computer and Information Research Scientists occupation (15-1221.00), there has been considerable growth in job ads indicating tasks that involve ML algorithms and data mining.  

\begin{figure}[h!]
  \centering
  \includegraphics[width=\textwidth]{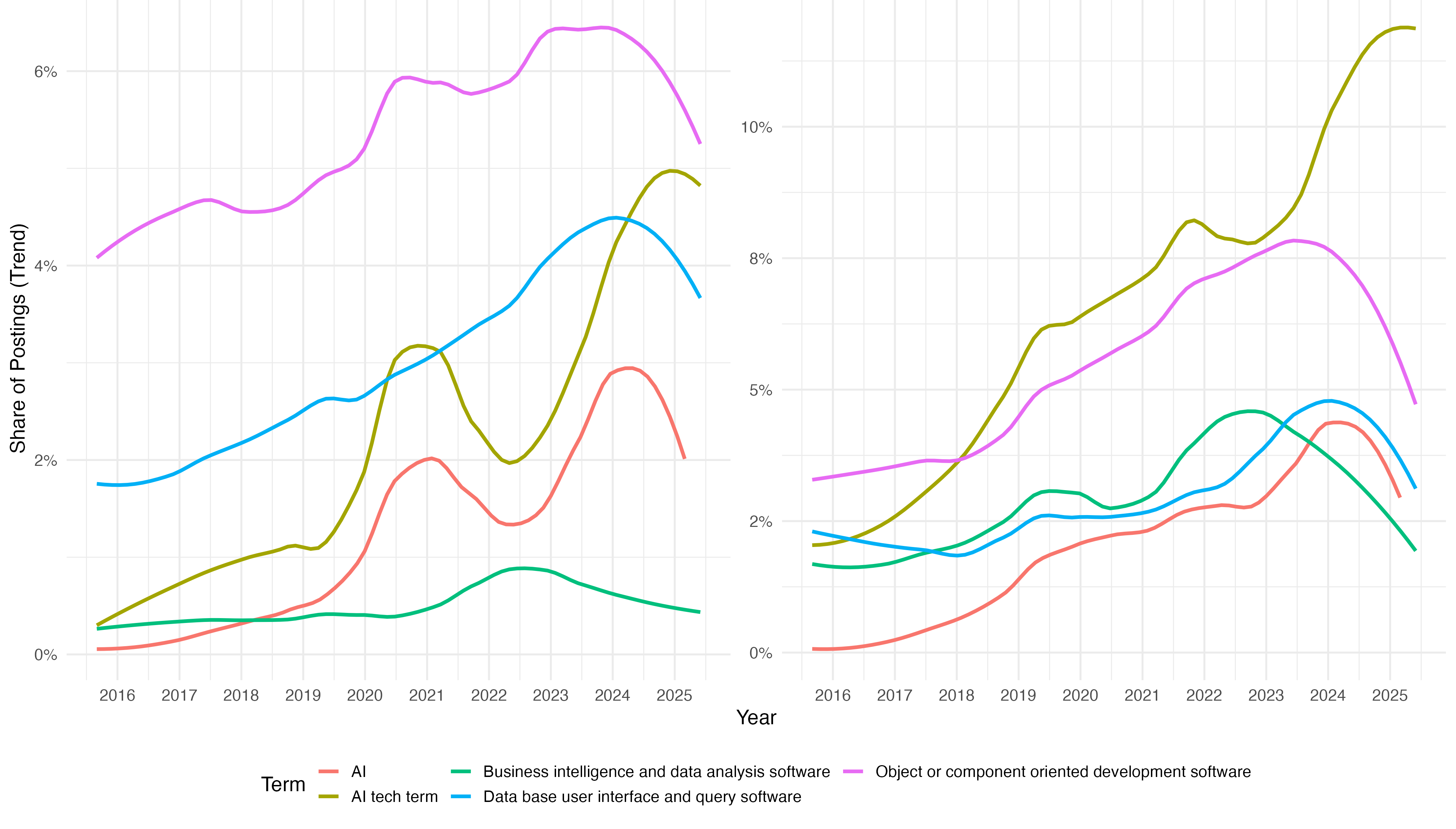}
    \small
\justifying
\emph{Note: }This figure is smoothed and based on aggregation by month of date compiled. \\
\vspace {8pt}
\\
  \begin{tabular}{p{0.5\textwidth} p{0.5\textwidth}}
  (a) SOC 1512 - Computer Occupations &
(b) SOC 1520 - Mathematical Occupations.
  \end{tabular}
     \caption{AI and Related Technologies in Computer and Mathematical Occupations.}
  \label{fig:ai_techs}
\end{figure}

This gradual, decade-long rise in tasks related to machine learning algorithms and data mining in the computer and information research scientists occupation is an example of a discovery otherwise not readily available for researchers to access from current data sources. We also leverage the available list of tools and technologies from O*NET, and custom dictionaries built to track mentions of ``Artificial Intelligence'' and its variants (AI, e.g.), as well as a more detailed list of technical terms related to AI (e.g., ``machine learning'', ``neural networks''), to assess changes in technology adoption.  Figure \ref{fig:ai_techs} illustrates that as a percent of all job postings in the computer (Panel A) and mathematical occupations (Panel B), products related to databases appear most frequently for much of the duration, with technical AI terms rising rapidly in the mathematical sciences and exceeding the percent of data analysis software more recently. 

Visualizing the aggregate data can serve as a vehicle for exploration suggesting areas that require additional in-depth research. Aggregate data will be made available upon publication.

\section{Conclusion, Future Directions, and Limitations}
\label{sec5}

This paper contributes new tools and data that point to a high-potential directions for advancing research and practice. By using O*NET's taxonomy as a basis for extraction of data from online job ads provided by NLx, we address limitations related to access and standardization of job ad text data. We highlight future directions related to expanding access to online job vacancy data and improving software and machine tools, and improving our understanding of labor markets and management practices. We then discuss known limitations of the models and data built,  the unavailability of benchmark or training data, and potential for disputes where conceptual clarity is missing.

First, the NLx Research Hub can advance the creation of a more vibrant and open ecosystem for labor market information \citep{hirsch_2024}. While this project demonstrates potentialities of NLx job ad data and O*NET, it ``takes a village'' to build a robust data ecosystem that share and manage data well, as information scientists have demonstrated \citep{borgman2021takes}. Independent teams of researchers, practitioners, and the private sector have the potential to improve upon and use common data resources, accelerating the production of better data products and information \citep{lane2020democratizing,Lane2024Democratizing}. Models of multiple levels of government and public-private-philanthropy-research collaborations demonstrate a path forward to create data infrastructure, protect sensitive information, and prove value \citep{cunningham2021value}.

The Job Ad Analysis Toolkit is available now for other researchers to extract structured information from the unstructured text of job ads. Other research groups could similarly pursue contributions that improve upon measurement.``Horse race'' competitions between independently developed purpose-built models against benchmark data in this domain would dramatically advance the field. While trained on job ad data, the JAAT may also be of use in extracting information from other labor market corpora. 

%We plan to continue to update the data monthly until at least December 2025 under current funding and plans.  

%Measurement and assessments of job quality, DEI practices, firm corporate social responsibility, and environmental, social, governance policies -- often featured in job postings and one of the features of organizational context we capture -- often suffer from opaque analysis, with measures obtained via obscure methods, with low convergent validity and far less coverage than is possible from job ads \citep{whelan2023we, eccles2017integrate, van2021research}. 

%Validation tests reported in the manuscript and examination of aggregate data produced valuable new insight.

Steps taken after the extraction and aggregation of data, and reported in this manuscript, provide new information and direction for future improvements to JAAT models. Indeed, many of the limitations discussed below were identified during additional testing of JAAT tools and examination of aggregate data output. Because of an emphasis on interpretable and traceable methods in developing explainable AI \citep{Adadi2018}, most limitations of the work described in this paper can be addressed in the future.

The data and tools we create could be used to further refine upon standard taxonomies for understanding work. There is a potential path for using the data to make improvements to the standard taxonomy of occupations and work. In other scientific fields,  centralized repositories, dictionaries, taxonomies, and ontologies have received more funding and attention over many more years than comparable resources such as O*NET. Refining the taxonomies used in workforce development using job postings data could unlock enormous value \citep{zweig_job_2026}. 

Our aggregate data will be released upon publication, and may be useful to developers, practitioners, and researchers. The aggregate data we build can support many more detailed investigations, some suggested in the manuscript body. Especially useful would be studies of management practices. Existing measurement of management practices is often reliant on a single respondent survey at an establishment. Research still faces this issue 25 years after \cite{gerhart2000measurement} identified the substantial measurement errors due to this method. While imperfect, our results from TaskMatch and extraction of tools and technologies, especially in combination with textual sources of information in the public domain (or as managed by a trustee with a mission to support research), have great potential to advance knowledge and practice.

%Future work could yield benchmark datasets for model building and testing in the job ad classification domain. 

%High-frequency missed matches between JAAT output and LLM judges or LCA data can be manually audited, and added in the future to training data. For TitleMatch, low-frequency job-title occupation code combinations could be removed to aim for a model that returns the modal occupation code for a given title.

\subsection{Limitations}

%We then discuss limitations related to precision and recall of the models used,  the unavailability of benchmark or training data in this space, and potential disputes where conceptual clarity is lacking.

Our estimates of precision and recall indicate that both false positives and negatives occur in our data. Our post-validation work demonstrates we could have used a lower threshold and still retrieved high-quality data from SkillMatch and TaskMatch. However, working in a computational environment with limited data storage and computation, we focused efforts. Given the volume of data construction, beyond the validation results we report, we cannot make representations about the accuracy of individual features extracted or provide a confidence level for individual subgroups in the aggregated dataset. We note results can be especially noisy in smaller subgroups.

We also acknowledge the underlying limitations of our JAAT modules and the data augmentation processes used to build up these modules, particularly in their reliance on pre-trained embedding models and cosine similarity-based matching. We did not perform a comprehensive comparative analysis of available models, but instead chose performative models based on past experience and the current state-of-the-art (at the time of initial development). Nevertheless, the performance of JAAT is bound by the limits of these embedding models; as these models continue to improve, better and different results may be obtainable from the same training data. 

Users of the data should assess its suitability for specific purposes. Researcher judgment in interpreting aggregate output is required. As with all online job ads data, there are many challenges. We perform no cleaning of suspect identifications in post-processing and do not hide model output, and caution users to apply judgment and audit model outputs on their own data. We notice, for example, TaskMatch returns tasks related to recruitment interviews for many occupations where recruitment is likely not a task that the worker will perform on the job, but one that a candidate must undertake during the job screening process. 

As another example, one limitation discussed above is that a job title can fit into multiple occupations. Other elements such as tools, technologies, skills, required education, career level, and industry are necessary for a more robust occupation coding model that considers more information than job title alone. We preserve the best available prediction of occupation in this work, but with the data extracted, better prediction of occupation is possible. 

%by incorporating results of the LCA analysis and additional information we extract from job ads. 

A vibrant NLP ecosystem depends upon standard benchmark labeled data to permit reliable comparison of model performance and independent training of models.  European efforts with ESCO labeled data include a recent skills and titles challenge \citep{gasco_overview_2025}. However, due to limitations in researcher access to job postings, there is no benchmark data for assessing the accuracy of NLP tools for O*NET coding. This limits the ability of researchers to develop models independently using the same training data, or to compare our results against an agreed upon benchmark of labeled data. While aspiring to FAIR standards \citep{stall_make_2019}, and providing the ML models and description of how they are constructed, we cannot release all information. We cannot share training data due to agreements that protect row-level disaggregated information in the NLx corpus. This means that underlying training data for models we build, which include short excerpts of real job ad text and their associated labels, cannot be shared.

Because we use established dictionaries and taxonomies as the principal approach to extraction and do not interrogate all of the text features extracted -- whether the appropriate label is used for a given feature is not addressed. 
The meaning of labor market concepts and the appropriate labels can be subjects of great debates. Even the most exhaustive taxonomies are known to be non-exhaustive in their coverage of features as well \citep{bowker_sorting_2000}, and we pursue novel creation of only a small number of JobTags or dictionaries. Results of these are necessarily ad hoc:  another researcher might achieve different results. Control over the answer to questions such as ``is a hotdog a sandwich?'' might appear unimportant, but many important legal and business cases revolve around similar questions \citep[pgs. 3-27]{abend_words_2023}, and researchers and practitioners often have diverse use cases. 

%For example, the meaning of terms such as Corporate Social Responsibility are vague and can change rapidly \citep{van2021research}, and conceptual schemes based on keyword counting are not very rich \citep{duriau_content_2007}.
%With any ML model, there may be issues arising from perceived inaccuracies or disputes about what goes in the training data.

%As a result, 

\subsection{Conclusion}
We build custom ML models and tools to create an aggregated dataset to understand change in the workplace using job ad descriptions and data from the National Labor Exchange Research Hub and O*NET's structured taxonomy as a basis feature extraction. Potential applications for researchers and practitioners are described. Intended to overcome limitations in O*NET and job ad data, the data and tools built here are suggested to have high potential for future use in research and practice. Further research is needed. 

\clearpage
\bibliographystyle{aea}
\bibliography{onet}

@article{pencle_whats_2016,
	title = {What's in the {Words}? {Development} and {Validation} of a {Multidimensional} {Dictionary} for {CSR} and {Application} {Using} {Prospectuses}},
	volume = {13},
	url = {https://doi.org/10.2308/jeta-51615},
	doi = {10.2308/jeta-51615},
	number = {2},
	journal = {Journal of Emerging Technologies in Accounting},
	author = {Pencle, Nadra and Mălăescu, Irina},
	year = {2016},
	pages = {109--127},
}

@article{hazell2022national,
  title={National wage setting},
  author={Hazell, Jonathon and Patterson, Christina and Sarsons, Heather and Taska, Bledi},
  journal={University of Chicago, Becker Friedman Institute for Economics Working Paper},
  number={2022-150},
  year={2022}
}

@book{arnold2022impact,
  title={The impact of pay transparency in job postings on the labor market},
  author={Arnold, David and Quach, Simon and Taska, Bledi},
  year={2022},
  publisher={SSRN}
}

@online{careeronestop_2021,
  title = {Career One Stop},
  author = {{U.S. Department of Labor, Employment and Training Administration}},
  year = {2022},
  howpublished = {{https://www.careeronestop.org/}},
  urldate = {2025-09-12}
}

@techreport{batra2023online,
  title={Online job posts contain very little wage information},
  author={Batra, Honey and Michaud, Amanda and Mongey, Simon},
  year={2023},
  institution={National Bureau of Economic Research}
}

@article{azar_labor_2022,
	title = {Labor {Market} {Concentration}},
	volume = {57},
	issn = {0022-166X, 1548-8004},
	url = {http://jhr.uwpress.org/content/57/S/S167},
	doi = {10.3368/jhr.monopsony.1218-9914R1},
	language = {en},
	number = {S},
	urldate = {2022-08-03},
	journal = {Journal of Human Resources},
	author = {Azar, José and Marinescu, Ioana and Steinbaum, Marshall},
	month = apr,
	year = {2022},
	note = {Publisher: University of Wisconsin Press},
	pages = {S167--S199},
}

@article{littman_2021,
	title = {Why aren't restaurant workers coming back? {Here}'s what the data shows.},
	shorttitle = {Why aren't restaurant workers coming back?},
	url = {https://www.restaurantdive.com/news/why-arent-restaurant-workers-coming-back-heres-what-the-data-shows/606198/},
	journal = {{Restaurant} {Dive}},
        language = {en-US},
	urldate = {2025-08-04},
	author = {Littman, Julie},
	month = sep,
    year = {2021},
}

@article{gibbons_monopsony_2019,
	title = {Monopsony {Power} and {Guest} {Worker} {Programs}},
	volume = {64},
	issn = {0003-603X},
	url = {https://doi.org/10.1177/0003603X19875040},
	doi = {10.1177/0003603X19875040},
	language = {en},
	number = {4},
	urldate = {2022-03-05},
	journal = {The Antitrust Bulletin},
	author = {Gibbons, Eric M. and Greenman, Allie and Norlander, Peter and Sørensen, Todd},
	month = dec,
	year = {2019},
	pages = {540--565},
}

@article{devaro_wage_2021,
	title = {Wage {Theft}, {Economic} {Conditions}, and {Market} {Power}: {The} {Case} of {H}-{1B} {Workers}},
	journal = {SSRN},
	author = {DeVaro, Jed and Norlander, Peter},
	year = {2021},
}

@misc{hirsch_2024,
  title={Reimagining the Workforce and Labor Market Information
System for the 21st Century. Workforce Information Advisory Council Recommendations},
  author={Hirsch, Lesley and Hui, Anna},
year={2024},
  month= nov,
  url ={https://www.dol.gov/sites/dolgov/files/ETA/wioa/pdfs/WIAC%20Recommendations.pdf}
}

@misc{bell_job_2020,
	address = {Rochester, NY},
	type = {{SSRN} {Scholarly} {Paper}},
	title = {Job {Amenities} and {Earnings} {Inequality}},
	url = {https://papers.ssrn.com/abstract=4173522},
	doi = {10.2139/ssrn.4173522},
	language = {en},
	urldate = {2025-08-21},
	publisher = {Social Science Research Network},
	author = {Bell, Alex},
	month = may,
	year = {2020},
}

@misc{lewis2024supplementing,
  title={Supplementing the O* NET Sources of Additional Information: A Preliminary Exploration of the Use of ChatGPT},
  author={Lewis, Phil and Morris, Jeremiah},
  year={2024},
  publisher={Raleigh, NC: National Center for O* NET Development. https://www. onetcenter~…}
}

@article{klein2025identification,
  title={Identification of Emerging Tasks in the O* NET System: A Revised Approach},
  author={Klein, Katherine and Pontikes, Mitchell and Dahlke, Jeffrey A and Putka, Dan J and Crawford, Brittany F and Reeder, Matthew C and Lewis, Phil},
  year={2025}
}

@article{lewis2025adding,
  title={Adding Drone-Specific Tasks to the O* NET Database: Initial Identification of Emerging Tasks using ChatGPT},
  author={Lewis, Phil and Gregory, Christina and Morris, Jeremiah},
  year={2025}
}

@misc{nyc_job_ads,
	title = {NYC Careers Portal},
	url = {http://www.nyc.gov/html/careers/html/search/search.shtml},
	urldate = {2025-03-24},
	author = {{City of New York}},
	month = mar,
        year = {2025},
    note = {http://www.nyc.gov/html/careers/html/search/search.shtml. Accessed March 24, 2025.}
}

@article{yourish_these_2025,
	chapter = {U.S.},
	title = {These {Words} {Are} {Disappearing} in the {New} {Trump} {Administration}},
	issn = {0362-4331},
	url = {https://www.nytimes.com/interactive/2025/03/07/us/trump-federal-agencies-websites-words-dei.html},
	language = {en-US},
	urldate = {2025-08-01},
	journal = {The New York Times},
	author = {Yourish, Karen and Daniel, Annie and Datar, Saurabh and White, Isaac and Gamio, Lazaro},
	month = mar,
	year = {2025},
}

@article{lou_ai_2021,
	title = {{AI} on {Drugs}: {Can} {Artificial} {Intelligence} {Accelerate} {Drug} {Development}? {Evidence} from a {Large}-{Scale} {Examination} of {Bio}-{Pharma} {Firms}},
	volume = {45},
	issn = {ISSN 0276-7783/ISSN 2162-9730},
	shorttitle = {{AI} on {Drugs}},
	url = {https://aisel.aisnet.org/misq/vol45/iss3/17},
	number = {3},
	journal = {Management Information Systems Quarterly},
	author = {Lou, Bowen and Wu, Lynn},
	month = sep,
	year = {2021},
	pages = {1451--1482},
}

@article{mckenny_what_2018,
author = {Aaron F. McKenny and Herman Aguinis and Jeremy C. Short and Aaron H. Anglin},
title ={What Doesn’t Get Measured Does Exist: Improving the Accuracy of Computer-Aided Text Analysis},
journal = {Journal of Management},
volume = {44},
number = {7},
pages = {2909-2933},
note = {10.1177/0149206316657594},
year = {2018},
doi = {10.1177/0149206316657594},
URL = { 
        https://doi.org/10.1177/0149206316657594
},
eprint = { 
        https://doi.org/10.1177/0149206316657594
}
}

@techreport{carnevale_understanding_2014,
	title = {Understanding online job ads data},
	institution = {A technical report. MS o. PP Center on Education and the Workforce. https …},
	author = {Carnevale, Anthony P. and Jayasundera, Tamara and Repnikov, Dmitri},
	year = {2014},
}

@article{choi_data_2024,
	title = {Data for labor market concentration using {Lightcast} (formerly {Burning} {Glass} {Technologies})},
	volume = {55},
	journal = {Data in Brief},
	author = {Choi, Hyeri and Marinescu, Ioana},
	year = {2024},
	pages = {110647},
}

@article{cifuentes_use_2010,
	title = {Use of {O}*{NET} as a job exposure matrix: {A} literature review},
	volume = {53},
	copyright = {Copyright © 2010 Wiley-Liss, Inc.},
	issn = {1097-0274},
	shorttitle = {Use of {O}*{NET} as a job exposure matrix},
	url = {https://onlinelibrary.wiley.com/doi/abs/10.1002/ajim.20846},
	doi = {10.1002/ajim.20846},
	language = {en},
	number = {9},
	urldate = {2025-06-09},
	journal = {American Journal of Industrial Medicine},
	author = {Cifuentes, Manuel and Boyer, Jon and Lombardi, David A. and Punnett, Laura},
	year = {2010},
	note = {10.1002/ajim.20846},
	pages = {898--914},
}

@article{horton_death_2025,
	title = {The {Death} of a {Technical} {Skill}},
	issn = {1047-7047},
	url = {https://pubsonline.informs.org/doi/full/10.1287/isre.2022.0709},
	doi = {10.1287/isre.2022.0709},
	urldate = {2025-07-15},
	journal = {Information Systems Research},
	author = {Horton, John J. and Tambe, Prasanna},
	month = mar,
	year = {2025},
}

@misc{decorte_extreme_2023,
	title = {Extreme {Multi}-{Label} {Skill} {Extraction} {Training} using {Large} {Language} {Models}},
	url = {http://arxiv.org/abs/2307.10778},
	doi = {10.48550/arXiv.2307.10778},
	urldate = {2025-07-10},
	publisher = {arXiv},
	author = {Decorte, Jens-Joris and Verlinden, Severine and Hautte, Jeroen Van and Deleu, Johannes and Develder, Chris and Demeester, Thomas},
	month = jul,
	year = {2023},
	note = {arXiv:2307.10778 [cs]},
}

@misc{BLS_OES,
  author = {{U.S. Bureau of Labor Statistics}},
  title = {{Occupational Employment Statistics Data}},
  howpublished = {{https://www.bls.gov/oes/}},
  note = {Data extracted on July 24, 2025},
  year = {2025},
  publisher = {{U.S. Department of Labor}}
}

@article{chen_is_2023,
	title = {Is hiring fast a good sign? {The} informativeness of job vacancy duration for future firm profitability},
	volume = {28},
	issn = {1573-7136},
	shorttitle = {Is hiring fast a good sign?},
	url = {https://doi.org/10.1007/s11142-023-09797-2},
	doi = {10.1007/s11142-023-09797-2},
	language = {en},
	number = {3},
	urldate = {2025-07-27},
	journal = {Review of Accounting Studies},
	author = {Chen, Ciao-Wei and Li, Laura Yue},
	month = sep,
	year = {2023},
	pages = {1316--1353},
}

@article{clemens2021dropouts,
  title={Dropouts need not apply? The minimum wage and skill upgrading},
  author={Clemens, Jeffrey and Kahn, Lisa B and Meer, Jonathan},
  journal={Journal of Labor Economics},
  volume={39},
  number={S1},
  pages={S107--S149},
  year={2021},
  publisher={The University of Chicago Press Chicago, IL}
}

@misc{hashizume_timing,
	title = {Timing {Fortune} 500 {Job} {Opportunities}},
	url = {https://nlxresearchhub.org/spotlights/timing-fortune-500-job-opportunities},
	author = {Hashizume, Marissa},
        language = {en-US},
	urldate = {2025-07-27},
	journal = {NLx Research Hub},
        year = {2024},
        month = {sep},
}

@article{bassier_vacancy_2025,
	title = {Vacancy {Duration} and {Wages}},
	issn = {0034-6535},
	url = {https://doi.org/10.1162/rest_a_01580},
	doi = {10.1162/rest_a_01580},
	urldate = {2025-07-27},
	journal = {The Review of Economics and Statistics},
	author = {Bassier, Ihsaan and Manning, Alan and Petrongolo, Barbara},
	month = mar,
	year = {2025},
	pages = {1--28},
}

@article{davis_establishment-level_2013,
	title = {The establishment-level behavior of vacancies and hiring},
	volume = {128},
	number = {2},
	journal = {The Quarterly Journal of Economics},
	author = {Davis, Steven J. and Faberman, R. Jason and Haltiwanger, John C.},
	year = {2013},
	pages = {581--622},
}

@article{mueller_vacancy_2024,
	title = {Vacancy durations and entry wages: {Evidence} from linked vacancy–employer–employee data},
	volume = {91},
	number = {3},
	journal = {Review of Economic Studies},
	author = {Mueller, Andreas I. and Osterwalder, Damian and Zweimüller, Josef and Kettemann, Andreas},
	year = {2024},
	pages = {1807--1841},
}

@misc{BLS_JOLTS,
  author = {{Bureau of Labor Statistics, U.S. Department of Labor}},
  title = {{Job Openings and Labor Turnover Survey (JOLTS)}},
  howpublished = {{https://www.bls.gov/jlt/}},
  year = {2025}, 
  note = {Accessed: 2025-07-24} 
}

@misc{QCEW_data,
  author = {{U.S. Bureau of Labor Statistics}},
  title = {{Quarterly Census of Employment and Wages (QCEW)}},
  howpublished = {{https://www.bls.gov/cew/downloadable-data-files.htm}},
  year = {2025}, 
  note = {Accessed on July 24, 2025}
}

@article{horton2015labor,
  title={Labor economists get their microscope: Big data and labor market analysis},
  author={Horton, John J and Tambe, Prasanna},
  journal={Big data},
  volume={3},
  number={3},
  pages={130--137},
  year={2015},
  publisher={Mary Ann Liebert, Inc. 140 Huguenot Street, 3rd Floor New Rochelle, NY 10801 USA}
}

@article{eloundou2024gpts,
  title={GPTs are GPTs: Labor market impact potential of LLMs},
  author={Eloundou, Tyna and Manning, Sam and Mishkin, Pamela and Rock, Daniel},
  journal={Science},
  volume={384},
  number={6702},
  pages={1306--1308},
  year={2024},
  publisher={American Association for the Advancement of Science}
}

@techreport{hansen2023remote,
  title={Remote work across jobs, companies, and space},
  author={Hansen, Stephen and Lambert, Peter John and Bloom, Nicholas and Davis, Steven J and Sadun, Raffaella and Taska, Bledi},
  year={2023},
  institution={National Bureau of Economic Research}
}

@article{alabdulkareem_unpacking_2018,
	title = {Unpacking the polarization of workplace skills},
	volume = {4},
	url = {https://www.science.org/doi/full/10.1126/sciadv.aao6030},
	doi = {10.1126/sciadv.aao6030},
	number = {7},
	urldate = {2025-06-09},
	journal = {Science Advances},
	author = {Alabdulkareem, Ahmad and Frank, Morgan R. and Sun, Lijun and AlShebli, Bedoor and Hidalgo, César and Rahwan, Iyad},
	month = jul,
	year = {2018},
}

@article{sauerwald_political_2024,
	title = {Political {Directors} and the {Recruitment} of {Foreign} {Workers}},
	copyright = {All rights reserved},
	issn = {0149-2063},
	url = {https://doi.org/10.1177/01492063241300311},
	doi = {10.1177/01492063241300311},
	language = {EN},
	urldate = {2025-03-31},
	journal = {Journal of Management},
	author = {Sauerwald, Steve and Norlander, Peter},
	month = dec,
	year = {2024},
	note = {Publisher: SAGE Publications Inc},
}

@article{schubert2024employer,
  title={Employer concentration and outside options},
  author={Schubert, Gregor and Stansbury, Anna and Taska, Bledi},
  journal={SSRN},
  year={2024}
}

@article{hershbein_recessions_2018,
	title = {Do {Recessions} {Accelerate} {Routine}-{Biased} {Technological} {Change}? {Evidence} from {Vacancy} {Postings}},
	volume = {108},
	issn = {0002-8282},
	shorttitle = {Do {Recessions} {Accelerate} {Routine}-{Biased} {Technological} {Change}?},
	url = {https://pubs.aeaweb.org/doi/10.1257/aer.20161570},
	doi = {10.1257/aer.20161570},
	language = {en},
	number = {7},
	urldate = {2025-07-16},
	journal = {American Economic Review},
	author = {Hershbein, Brad and Kahn, Lisa B.},
	month = jul,
	year = {2018},
	note = {10.1257/aer.20161570},
	pages = {1737--1772},
}

@article{heyden_conjoint_2015,
	title = {The {Conjoint} {Influence} of {Top} and {Middle} {Management} {Characteristics} on {Management} {Innovation}},
	volume = {44},
	url = {https://doi.org/10.1177/0149206315614373},
	doi = {10.1177/0149206315614373},
	number = {4},
	journal = {Journal of Management},
	author = {Heyden, Mariano L. M. and Sidhu, Jatinder S. and Volberda, Henk W.},
	year = {2015},
	pages = {1505--1529},
}

@article{banks_strategic_2018,
	title = {Strategic {Recruitment} {Across} {Borders}: {An} {Investigation} of {Multinational} {Enterprises}},
	volume = {45},
	url = {https://doi.org/10.1177/0149206318764295},
	doi = {10.1177/0149206318764295},
	number = {2},
	journal = {Journal of Management},
	author = {Banks, George C. and Woznyj, Haley M. and Wesslen, Ryan S. and Frear, Katherine A. and Berka, Gregory and Heggestad, Eric D. and Gordon, Heather L.},
	year = {2018},
	pages = {476--509},
}

@article{Adadi2018,
  doi = {10.1109/access.2018.2870052 },
  url = {https://doi.org/10.1109/access.2018.2870052},
  year = {2018},
  publisher = {Institute of Electrical and Electronics Engineers ({IEEE})},
  volume = {6},
  pages = {52138--52160},
  author = {Amina Adadi and Mohammed Berrada},
  title = {Peeking Inside the Black-Box: A Survey on Explainable Artificial Intelligence ({XAI})},
  journal = {{IEEE} Access}
}

@book{neuendorf2017content,
  title={The content analysis guidebook},
  author={Neuendorf, Kimberly A},
  year={2017},
  publisher={Sage}, 
  address= {Thousand Oaks, CA}
}

@book{abend_words_2023,
	title = {Words and {Distinctions} for the {Common} {Good}: {Practical} {Reason} in the {Logic} of {Social} {Science}},
	isbn = {978-0-691-24706-9},
	shorttitle = {Words and {Distinctions} for the {Common} {Good}},
	language = {en},
	publisher = {Princeton University Press},
	author = {Abend, Gabriel},
	month = jul,
	year = {2023},
}

@book{bowker_sorting_2000,
	address = {Cambridge, MA, USA},
	series = {Inside {Technology}},
	title = {Sorting {Things} {Out}: {Classification} and {Its} {Consequences}},
	isbn = {978-0-262-52295-3},
	shorttitle = {Sorting {Things} {Out}},
	language = {en},
	publisher = {MIT Press},
	author = {Bowker, Geoffrey C. and Star, Susan Leigh},
	editor = {Bijker, Wiebe and Jones-Imhotep, Edward and Slayton, Rebecca},
	month = aug,
	year = {2000},
}

@article{gephart1997hazardous,
  title={Hazardous measures: An interpretive textual analysis of quantitative sensemaking during crises},
  author={Gephart, Robert},
  journal={Journal of Organizational Behavior: The International Journal of Industrial, Occupational and Organizational Psychology and Behavior},
  volume={18},
  number={S1},
  pages={583--622},
  year={1997},
  publisher={Wiley Online Library}
}

@article{https://doi.org/10.48550/arxiv.1811.10154,     
  doi = {10.48550/ARXIV.1811.10154 },
  
  url = {https://arxiv.org/abs/1811.10154},
  
  author = {Rudin, Cynthia},
  
  title = {Stop Explaining Black Box Machine Learning Models for High Stakes Decisions and Use Interpretable Models Instead},
  
  publisher = {arXiv},
  
  year = {2018},
  
  copyright = {Creative Commons Attribution Share Alike 4.0 International}
}

@article{doi:10.1126/scirobotics.aay7120,      
author = {David Gunning  and Mark Stefik  and Jaesik Choi  and Timothy Miller  and Simone Stumpf  and Guang-Zhong Yang },
title = {XAI: Explainable artificial intelligence},
journal = {Science Robotics},
volume = {4},
number = {37},
year = {2019},
doi = {10.1126/scirobotics.aay7120 },
URL = {https://www.science.org/doi/abs/10.1126/scirobotics.aay7120}, 
eprint = {https://www.science.org/doi/pdf/10.1126/scirobotics.aay7120}}

@article{desjardine_one_2019,
	title = {One {Step} {Forward}, {Two} {Steps} {Back}: {How} {Negative} {External} {Evaluations} {Can} {Shorten} {Organizational} {Time} {Horizons}},
	volume = {30},
	url = {https://doi.org/10.1287/orsc.2018.1259},
	doi = {10.1287/orsc.2018.1259},
	number = {4},
	journal = {Organization Science},
	author = {DesJardine, Mark and Bansal, Pratima},
	year = {2019},
	pages = {761--780},
}

@article{moss_funding_2018,
	title = {Funding the story of hybrid ventures: {Crowdfunder} lending preferences and linguistic hybridity},
	volume = {33},
	url = {https://doi.org/10.1016/j.jbusvent.2017.12.004},
	doi = {10.1016/j.jbusvent.2017.12.004},
	number = {5},
	journal = {Journal of Business Venturing},
	author = {Moss, Todd W. and Renko, Maija and Block, Emily and Meyskens, Moriah},
	year = {2018},
	pages = {643--659},
}

@article{zachary_family_2011,
	title = {Family {Business} and {Market} {Orientation}},
	volume = {24},
	url = {https://doi.org/10.1177/0894486510396871},
	doi = {10.1177/0894486510396871},
	number = {3},
	journal = {Family Business Review},
	author = {Zachary, Miles A. and McKenny, Aaron and Short, Jeremy Collin and Payne, G. Tyge},
	year = {2011},
	pages = {233--251},
}

@article{short_construct_2009,
	title = {Construct {Validation} {Using} {Computer}-{Aided} {Text} {Analysis} ({CATA})},
	volume = {13},
	url = {https://doi.org/10.1177/1094428109335949},
	doi = {10.1177/1094428109335949},
	number = {2},
	journal = {Organizational Research Methods},
	author = {Short, Jeremy C. and Broberg, J. Christian and Cogliser, Claudia C. and Brigham, Keith H.},
	year = {2009},
	pages = {320--347},
}

@article{atalay_evolution_2020,
	title = {The {Evolution} of {Work} in the {United} {States}},
	volume = {12},
	issn = {1945-7782},
	url = {https://www.aeaweb.org/articles?id=10.1257/app.20190070},
	doi = {10.1257/app.20190070},
	language = {en},
	number = {2},
	urldate = {2025-03-25},
	journal = {American Economic Journal: Applied Economics},
	author = {Atalay, Enghin and Phongthiengtham, Phai and Sotelo, Sebastian and Tannenbaum, Daniel},
	month = apr,
	year = {2020},
	pages = {1--34},
}

@article{atalay_new_2018,
	title = {New technologies and the labor market},
	volume = {97},
	issn = {0304-3932},
	url = {https://www.sciencedirect.com/science/article/pii/S0304393218301636},
	doi = {10.1016/j.jmoneco.2018.05.008},
	urldate = {2025-03-25},
	journal = {Journal of Monetary Economics},
	author = {Atalay, Enghin and Phongthiengtham, Phai and Sotelo, Sebastian and Tannenbaum, Daniel},
	month = aug,
	year = {2018},
	pages = {48--67},
}

@article{david2013task,
  title={The `task approach' to labor markets: an overview},
  author={Autor, David, H},
  journal={Journal for Labour Market Research},
  volume={46},
  number={3},
  pages={185--199},
  year={2013},
  publisher={SpringerOpen}
}

@article{dingel_how_2020,
	title = {How many jobs can be done at home?},
	volume = {189},
	issn = {0047-2727},
	url = {https://www.sciencedirect.com/science/article/pii/S0047272720300992},
	doi = {10.1016/j.jpubeco.2020.104235},
	language = {en},
	urldate = {2021-06-23},
	journal = {Journal of Public Economics},
	author = {Dingel, Jonathan I. and Neiman, Brent},
	month = sep,
	year = {2020},
}

@article{marinescu_opening_2020,
	title = {Opening the {Black} {Box} of the {Matching} {Function}: {The} {Power} of {Words}},
	volume = {38},
	issn = {0734-306X},
	shorttitle = {Opening the {Black} {Box} of the {Matching} {Function}},
	url = {https://www.journals.uchicago.edu/doi/abs/10.1086/705903},
	doi = {10.1086/705903},
	number = {2},
	urldate = {2025-03-25},
	journal = {Journal of Labor Economics},
	author = {Marinescu, Ioana and Wolthoff, Ronald},
	month = apr,
	year = {2020},
	note = {Publisher: The University of Chicago Press},
	pages = {535--568},
}

@article{deming_qje,
    author = {Deming, David J.},
    title = {The Growing Importance of Social Skills in the Labor Market*},
    journal = {The Quarterly Journal of Economics},
    volume = {132},
    number = {4},
    pages = {1593-1640},
    year = {2017},
    month = {06},
    issn = {0033-5533},
    doi = {10.1093/qje/qjx022},
    url = {https://doi.org/10.1093/qje/qjx022},
    eprint = {https://academic.oup.com/qje/article-pdf/132/4/1593/30637898/qjx022.pdf},
}

@article{gathmann2024ai,
  title={AI, Task Changes in Jobs, and Worker Reallocation},
  author={Gathmann, Christina and Grimm, Felix and Winkler, Erwin},
  year={2024},
  journal={CESifo Working Paper Series 11585}
}

@article{borgman2021takes,
  title={Why it takes a village to manage and share data},
  author={Borgman, Christine L and Bourne, Philip E},
  journal={arXiv preprint arXiv:2109.01694},
  year={2021}
}

@article{gerhart2000measurement,
  title={Measurement error in research on human resources and firm performance: how much error is there and how does it influence effect size estimates?},
  author={Gerhart, Barry and Wright, Patrick M and McMahan, Gary C and Snell, Scott A},
  journal={Personnel psychology},
  volume={53},
  number={4},
  pages={803--834},
  year={2000},
  publisher={Wiley Online Library}
}

@article{stall_make_2019,
	title = {Make scientific data {FAIR}},
	volume = {570},
	copyright = {2021 Nature},
	url = {https://www.nature.com/articles/d41586-019-01720-7},
	doi = {10.1038/d41586-019-01720-7},
	language = {en},
	number = {7759},
	urldate = {2023-09-23},
	journal = {Nature},
	author = {Stall, Shelley and Yarmey, Lynn and Cutcher-Gershenfeld, Joel and Hanson, Brooks and Lehnert, Kerstin and Nosek, Brian and Parsons, Mark and Robinson, Erin and Wyborn, Lesley},
	month = jun,
	year = {2019},
	note = {Bandiera\_abtest: a
Cg\_type: Comment
Number: 7759
Publisher: Nature Publishing Group
Subject\_term: Databases, Research management, Policy, Publishing},
	pages = {27--29},
}

@inproceedings{zhang_skillspan_2022,
	address = {Seattle, United States},
	title = {{SkillSpan}: {Hard} and {Soft} {Skill} {Extraction} from {English} {Job} {Postings}},
	shorttitle = {{SkillSpan}},
	url = {https://aclanthology.org/2022.naacl-main.366/},
	doi = {10.18653/v1/2022.naacl-main.366},
	urldate = {2025-01-14},
	booktitle = {Proceedings of the 2022 {Conference} of the {North} {American} {Chapter} of the {Association} for {Computational} {Linguistics}: {Human} {Language} {Technologies}},
	publisher = {Association for Computational Linguistics},
	author = {Zhang, Mike and Jensen, Kristian and Sonniks, Sif and Plank, Barbara},
	editor = {Carpuat, Marine and de Marneffe, Marie-Catherine and Meza Ruiz, Ivan Vladimir},
	month = jul,
	year = {2022},
	pages = {4962--4984},
}

@book{holland_making_1997,
	address = {Odessa, FL.},
	title = {Making vocational choices: {A} theory of vocational personalities and work environments, 3rd ed},
	isbn = {978-0-911907-27-8},
	shorttitle = {Making vocational choices},
	publisher = {Psychological Assessment Resources},
	author = {Holland, John L.},
	year = {1997},
}

@article{rounds_updating_2022,
	title = {Updating vocational interests information for the {O}* {NET} content model},
	journal = {National Center for O*NET Development.},
        note = {{https://www. onetcenter. org/dl\_files/Voc\_Interests.pdf}},
	author = {Rounds, James and Putka, Dan J. and Lewis, P.},
	year = {2022},
}

@article{Lane2024Democratizing,
	author = {Lane, Julia and Potok, Nancy},
	journal = {Harvard Data Science Review},
	number = {Special Issue 4},
	year = {2024},
	month = {apr 2},
	note = {https://hdsr.mitpress.mit.edu/pub/m1o4oblm},
	publisher = {The MIT Press},
	title = {
{Democratizing} {Data}: Our {Vision}

},
	volume = { },
}

@book{lane2020democratizing,
  title={Democratizing our data: A manifesto},
  author={Lane, Julia},
  year={2020},
  publisher={MIT Press}
}

@online{bls2025unions,
    title = {Union Members},
    author = {{Bureau of Labor Statistics}},
    year = {2025},
    howpublished = {{https://www.bls.gov/news.release/pdf/union2.pdf}},
}

@book{committee_on_using_machine_learning_in_safety-critical_applications_setting_a_research_agenda_machine_2025,
	address = {Washington, D.C.},
	title = {Machine {Learning} for {Safety}-{Critical} {Applications}: {Opportunities}, {Challenges}, and a {Research} {Agenda}},
	isbn = {978-0-309-72666-5},
	shorttitle = {Machine {Learning} for {Safety}-{Critical} {Applications}},
	author = {National{\ }Academy{\ }of{\ }Sciences},
	url = {https://nap.nationalacademies.org/catalog/27970},
	urldate = {2025-09-30},
	publisher = {National Academies Press},
	collaborator = {{Committee on Using Machine Learning in Safety-Critical Applications: Setting a Research Agenda} and {Computer Science and Telecommunications Board} and {Division on Engineering and Physical Sciences} and {National Academies of Sciences, Engineering, and Medicine}},
	year = {2025},
	doi = {10.17226/27970},
}

@article{haselhuhn_investors_2022,
	title = {Investors respond negatively to executives’ discussion of creativity},
	volume = {171},
	issn = {0749-5978},
	url = {https://www.sciencedirect.com/science/article/pii/S0749597822000395},
	doi = {10.1016/j.obhdp.2022.104155},
	urldate = {2025-06-02},
	journal = {Organizational Behavior and Human Decision Processes},
	author = {Haselhuhn, Michael P. and Wong, Elaine M. and Ormiston, Margaret E.},
	month = jul,
	year = {2022},
	note = {10.1016/j.obhdp.2022.104155},
}

@article{kindermann_digital_2021,
	title = {Digital orientation: {Conceptualization} and operationalization of a new strategic orientation},
	volume = {39},
	issn = {0263-2373},
	shorttitle = {Digital orientation},
	url = {https://www.sciencedirect.com/science/article/pii/S0263237320301584},
	doi = {10.1016/j.emj.2020.10.009},
	number = {5},
	urldate = {2025-06-02},
	journal = {European Management Journal},
	author = {Kindermann, Bastian and Beutel, Sebastian and Garcia de Lomana, Gonzalo and Strese, Steffen and Bendig, David and Brettel, Malte},
	month = oct,
	year = {2021},
	pages = {645--657},
}

@article{levy_influence_2005,
	title = {The influence of top management team attention patterns on global strategic posture of firms},
	volume = {26},
	copyright = {Copyright © 2005 John Wiley \& Sons, Ltd.},
	issn = {1099-1379},
	url = {https://onlinelibrary.wiley.com/doi/abs/10.1002/job.340},
	doi = {10.1002/job.340},
	language = {en},
	number = {7},
	urldate = {2025-06-02},
	journal = {Journal of Organizational Behavior},
	author = {Levy, Orly},
	year = {2005},
	note = {10.1002/job.340},
	pages = {797--819},
}

@article{mckenny_strategic_2018,
	title = {Strategic entrepreneurial orientation: {Configurations}, performance, and the effects of industry and time},
	volume = {12},
	copyright = {Copyright © 2018 Strategic Management Society},
	issn = {1932-443X},
	shorttitle = {Strategic entrepreneurial orientation},
	url = {https://onlinelibrary.wiley.com/doi/abs/10.1002/sej.1291},
	doi = {10.1002/sej.1291},
	language = {en},
	number = {4},
	urldate = {2025-06-02},
	journal = {Strategic Entrepreneurship Journal},
	author = {McKenny, Aaron F. and Short, Jeremy C. and Ketchen Jr., David J. and Payne, G. Tyge and Moss, Todd W.},
	year = {2018},
	note = {10.1002/sej.1291},
	pages = {504--521},
}

@article{vaupel_role_2023,
	title = {The {Role} of {Share} {Repurchases} for {Firms}’ {Social} and {Environmental} {Sustainability}},
	volume = {183},
	issn = {1573-0697},
	url = {https://doi.org/10.1007/s10551-022-05076-3},
	doi = {10.1007/s10551-022-05076-3},
	language = {en},
	number = {2},
	urldate = {2025-06-02},
	journal = {Journal of Business Ethics},
	author = {Vaupel, Mario and Bendig, David and Fischer-Kreer, Denise and Brettel, Malte},
	month = mar,
	year = {2023},
	pages = {401--428},
}

@article{brochet_speaking_2015,
	title = {Speaking of the short-term: disclosure horizon and managerial myopia},
	volume = {20},
	issn = {1573-7136},
	shorttitle = {Speaking of the short-term},
	url = {https://doi.org/10.1007/s11142-015-9329-8},
	doi = {10.1007/s11142-015-9329-8},
	language = {en},
	number = {3},
	urldate = {2025-06-02},
	journal = {Review of Accounting Studies},
	author = {Brochet, Francois and Loumioti, Maria and Serafeim, George},
	month = sep,
	year = {2015},
	pages = {1122--1163},
}

@article{brigham_researching_2014,
	title = {Researching {Long}-{Term} {Orientation}: {A} {Validation} {Study} and {Recommendations} for {Future} {Research}},
	volume = {27},
	issn = {0894-4865},
	shorttitle = {Researching {Long}-{Term} {Orientation}},
	url = {https://doi.org/10.1177/0894486513508980},
	doi = {10.1177/0894486513508980},
	language = {EN},
	number = {1},
	urldate = {2025-06-02},
	journal = {Family Business Review},
	author = {Brigham, Keith H. and Lumpkin, G. T. and Payne, G. Tyge and Zachary, Miles A.},
	month = mar,
	year = {2014},
	note = {Publisher: SAGE Publications Inc},
	pages = {72--88},
}

@article{putka_evaluating_2023,
	title = {Evaluating a {Natural} {Language} {Processing} {Approach} to {Estimating} {KSA} and {Interest} {Job} {Analysis} {Ratings}},
	volume = {38},
	issn = {1573-353X},
	url = {https://doi.org/10.1007/s10869-022-09824-0},
	doi = {10.1007/s10869-022-09824-0},
	language = {en},
	number = {2},
	urldate = {2025-06-01},
	journal = {Journal of Business and Psychology},
	author = {Putka, Dan J. and Oswald, Frederick L. and Landers, Richard N. and Beatty, Adam S. and McCloy, Rodney A. and Yu, Martin C.},
	month = apr,
	year = {2023},
	pages = {385--410},
}

@article{cunningham2021value,
  title={A Value-Driven Approach to Building Data Infrastructures: The Example of the MidWest Collaborative},
  author={Cunningham, Jessica and Hui, Anna and Lane, Julia and Putnam, George},
  journal={Harvard Data Science Review},
  volume={4},
  year={2021}
}

@article{rounds_using_2023,
	title = {Using {Machine} {Learning} to {Develop} {Occupational} {Interest} {Profiles} and {High}-{Point} {Codes} for the {O}* {NET} {System}},
	journal = {National Center for O*NET Development.},
	author = {Rounds, James},
    note = {{https://www.onetcenter.org/reports/ML\_OIPs.html}},
	year = {2023},
}

@online{onet,
	title = {O*NET OnLine},
	language = {en},
	author = {National{\ }Center{\ }for O{\ }*NET{\ }Development},
        year = {2025},
    howpublished = {{https://www.onetcenter.org/}}
}

@incollection{acemoglu2011skills,
  title={Skills, tasks and technologies: Implications for employment and earnings},
  author={Acemoglu, Daron and Autor, David},
  booktitle={Handbook of labor economics},
  volume={4},
  pages={1043--1171},
  year={2011},
  publisher={Elsevier}
}

@misc{meisenbacher_transforming_2023,
	title = {Transforming {Unstructured} {Text} into {Data} with {Context} {Rule} {Assisted} {Machine} {Learning} ({CRAML})},
	url = {http://arxiv.org/abs/2301.08549},
	language = {en},
	urldate = {2024-06-07},
	publisher = {arXiv},
	author = {Meisenbacher, Stephen and Norlander, Peter},
	month = jan,
	year = {2023},
	note = {arXiv:2301.08549 [cs]},
}

@article{russ_evaluation_2023,
	title = {Evaluation of the updated {SOCcer} v2 algorithm for coding free-text job descriptions in three epidemiologic studies},
	volume = {67},
	issn = {2398-7308},
	url = {https://doi.org/10.1093/annweh/wxad020},
	doi = {10.1093/annweh/wxad020},
	number = {6},
	urldate = {2024-07-17},
	journal = {Annals of Work Exposures and Health},
	author = {Russ, Daniel E and Josse, Pabitra and Remen, Thomas and Hofmann, Jonathan N and Purdue, Mark P and Siemiatycki, Jack and Silverman, Debra T and Zhang, Yawei and Lavoué, Jerome and Friesen, Melissa C},
	month = jul,
	year = {2023},
	pages = {772--783},
}

@misc{decorte_jobbert_2021,
	title = {{JobBERT}: {Understanding} {Job} {Titles} through {Skills}},
	shorttitle = {{JobBERT}},
	url = {http://arxiv.org/abs/2109.09605},
	doi = {10.48550/arXiv.2109.09605},
	urldate = {2025-09-20},
	publisher = {arXiv},
	author = {Decorte, Jens-Joris and Hautte, Jeroen Van and Demeester, Thomas and Develder, Chris},
	month = sep,
	year = {2021},
	note = {arXiv:2109.09605 [cs]},
}

@misc{decorte_multilingual_2025,
	title = {Multilingual {JobBERT} for {Cross}-{Lingual} {Job} {Title} {Matching}},
	url = {http://arxiv.org/abs/2507.21609},
	doi = {10.48550/arXiv.2507.21609},
	urldate = {2025-09-20},
	publisher = {arXiv},
	author = {Decorte, Jens-Joris and Lange, Matthias De and Hautte, Jeroen Van},
	month = jul,
	year = {2025},
	note = {arXiv:2507.21609 [cs]},
}

@article{decorte_efficient_2025,
	title = {Efficient {Text} {Encoders} for {Labor} {Market} {Analysis}},
	volume = {13},
	issn = {2169-3536},
	url = {http://arxiv.org/abs/2505.24640},
	doi = {10.1109/ACCESS.2025.3589147},
	urldate = {2025-09-20},
	journal = {IEEE Access},
	author = {Decorte, Jens-Joris and Hautte, Jeroen Van and Develder, Chris and Demeester, Thomas},
	year = {2025},
	note = {arXiv:2505.24640 [cs]},
	pages = {133596--133608},
}

@misc{decorte_skillmatch_2024,
	title = {{SkillMatch}: {Evaluating} {Self}-supervised {Learning} of {Skill} {Relatedness}},
	shorttitle = {{SkillMatch}},
	url = {http://arxiv.org/abs/2410.05006},
	doi = {10.48550/arXiv.2410.05006},
	urldate = {2025-09-20},
	publisher = {arXiv},
	author = {Decorte, Jens-Joris and Hautte, Jeroen Van and Demeester, Thomas and Develder, Chris},
	month = oct,
	year = {2024},
	note = {arXiv:2410.05006 [cs]},
}

@misc{decorte_career_2023,
	title = {Career {Path} {Prediction} using {Resume} {Representation} {Learning} and {Skill}-based {Matching}},
	url = {http://arxiv.org/abs/2310.15636},
	doi = {10.48550/arXiv.2310.15636},
	urldate = {2025-09-20},
	publisher = {arXiv},
	author = {Decorte, Jens-Joris and Hautte, Jeroen Van and Deleu, Johannes and Develder, Chris and Demeester, Thomas},
	month = oct,
	year = {2023},
	note = {arXiv:2310.15636 [cs]},
}

@article{maffie_mythology_2023,
	title = {The mythology of ‘{Big} {Data}’ as a source of corporate power},
	volume = {61},
	copyright = {© 2023 John Wiley \& Sons Ltd.},
	issn = {1467-8543},
	url = {https://onlinelibrary.wiley.com/doi/abs/10.1111/bjir.12728},
	doi = {10.1111/bjir.12728},
	language = {en},
	number = {3},
	urldate = {2025-09-23},
	journal = {British Journal of Industrial Relations},
	author = {Maffie, Michael David},
	year = {2023},
	note = {10.1111/bjir.12728},
	pages = {674--696},
}

@incollection{dey_rise_2025,
	title = {The {Rise} of the {Contract} {Workforce} in {US} {Manufacturing}},
	url = {https://www.nber.org/books-and-chapters/changing-nature-work/rise-contract-workforce-us-manufacturing},
	urldate = {2025-09-28},
	booktitle = {The {Changing} {Nature} of {Work}},
	publisher = {University of Chicago Press},
	author = {Dey, Matthew and Houseman, Susan},
	month = may,
	year = {2025},
}

@book{zweig_job_2026,
	address = {S.l.},
	title = {Job {Architecture}: {Building} a {Language} for {Workforce} {Intelligence}},
	isbn = {978-1-394-36906-5},
	shorttitle = {Job {Architecture}},
	language = {English},
	publisher = {Wiley},
	author = {Zweig, Ben},
	year = {2026},
}

@misc{meisenbacher_classifiers_2022,
	title = {Machine learning classifiers of job advertisement text described in `Creating Data from Unstructured Text with Context Rule Assisted Machine Learning (CRAML).'},
	urldate = {2025-09-27},
	publisher = {Zenodo},
	author = {Meisenbacher, Stephen and Norlander, Peter},
	month = dec,
	year = {2022},
	note = {10.5281/zenodo.7454652},
}

@book{weil_fissured_2014,
	title = {The {Fissured} {Workplace}},
	isbn = {978-0-674-72612-3},
	language = {en},
	publisher = {Harvard University Press},
	author = {Weil, David},
	month = feb,
	year = {2014},
}

@misc{chatterji_how_2025,
	type = {Working {Paper}},
	series = {Working {Paper} {Series}},
	title = {How {People} {Use} {ChatGPT}},
	url = {https://www.nber.org/papers/w34255},
	doi = {10.3386/w34255},
	urldate = {2025-09-20},
	publisher = {National Bureau of Economic Research},
	author = {Chatterji, Aaron and Cunningham, Thomas and Deming, David J. and Hitzig, Zoe and Ong, Christopher and Shan, Carl Yan and Wadman, Kevin},
	month = sep,
	year = {2025},
	doi = {10.3386/w34255},
}

@misc{gasco_overview_2025,
	title = {Overview of the {TalentCLEF} 2025: {Skill} and {Job} {Title} {Intelligence} for {Human} {Capital} {Management}},
	shorttitle = {Overview of the {TalentCLEF} 2025},
	url = {http://arxiv.org/abs/2507.13275},
	doi = {10.48550/arXiv.2507.13275},
	urldate = {2025-09-20},
	publisher = {arXiv},
	author = {Gasco, Luis and Fabregat, Hermenegildo and García-Sardiña, Laura and Estrella, Paula and Deniz, Daniel and Rodrigo, Alvaro and Zbib, Rabih},
	month = jul,
	year = {2025},
	note = {arXiv:2507.13275 [cs]},
}

@misc{anand_is_2022,
	title = {Is it {Required}? {Ranking} the {Skills} {Required} for a {Job}-{Title}},
	shorttitle = {Is it {Required}?},
	url = {http://arxiv.org/abs/2212.08553},
	doi = {10.48550/arXiv.2212.08553},
	urldate = {2025-09-20},
	publisher = {arXiv},
	author = {Anand, Sarthak and Decorte, Jens-Joris and Lowie, Niels},
	month = nov,
	year = {2022},
	note = {arXiv:2212.08553 [cs]},
}

@inproceedings{russ_computer-based_2014,
	address = {New York, NY, USA},
	title = {Computer-{Based} {Coding} of {Occupation} {Codes} for {Epidemiological} {Analyses}},
	isbn = {978-1-4799-4435-4},
	url = {http://ieeexplore.ieee.org/document/6881904/},
	doi = {10.1109/CBMS.2014.79},
	urldate = {2024-07-17},
	booktitle = {2014 {IEEE} 27th {International} {Symposium} on {Computer}-{Based} {Medical} {Systems}},
	publisher = {IEEE},
	author = {Russ, Daniel E. and Ho, Kwan-Yuet and Johnson, Calvin A. and Friesen, Melissa C.},
	month = may,
	year = {2014},
	pages = {347--350},
}

@misc{howison_replication_2022,
	title = {Replication files for: `{Occupational} models from 42 million unstructured job postings'},
	shorttitle = {Replication files for},
	url = {https://zenodo.org/records/7319953},
	doi = {10.5281/zenodo.7319953},
	urldate = {2024-07-17},
	publisher = {Zenodo},
	author = {Howison, Mark},
	month = nov,
	year = {2022},
}

@misc{howison_recommending_2023,
	address = {Rochester, NY},
	type = {{SSRN} {Scholarly} {Paper}},
	title = {Recommending {Career} {Transitions} to {Job} {Seekers} {Using} {Earnings} {Estimates}, {Skills} {Similarity}, and {Occupational} {Demand}},
	url = {https://papers.ssrn.com/abstract=4371445},
	doi = {10.2139/ssrn.4371445},
	language = {en},
	urldate = {2024-07-17},
	author = {Howison, Mark and Long, Joe and Hastings, Justine},
	month = feb,
	year = {2023},
}

@article{howison_extracting_2024,
  title={Extracting structured labor market information from job postings with generative ai},
  author={Howison, Mark and Ensor, William O and Maharjan, Suraj and Parikh, Rahil and Sengamedu, Srinivasan H and Daniels, Paul and Gaither, Amber and Yeats, Carrie and Reddy, Chandan K and Hastings, Justine S},
  journal={Digital Government: Research and Practice},
  volume={6},
  number={1},
  pages={1--12},
  year={2025},
  publisher={ACM New York, NY}
}

@misc{nguyen_rethinking_2024,
	title = {Rethinking {Skill} {Extraction} in the {Job} {Market} {Domain} using {Large} {Language} {Models}},
	url = {http://arxiv.org/abs/2402.03832},
	doi = {10.48550/arXiv.2402.03832},
	urldate = {2025-01-14},
	publisher = {arXiv},
	author = {Nguyen, Khanh Cao and Zhang, Mike and Montariol, Syrielle and Bosselut, Antoine},
	month = feb,
	year = {2024},
	note = {arXiv:2402.03832 [cs]},
}

@article{lazer_computational_2020,
	title = {Computational social science: {Obstacles} and opportunities},
	volume = {369},
	shorttitle = {Computational social science},
	url = {https://www.science.org/doi/full/10.1126/science.aaz8170},
	doi = {10.1126/science.aaz8170},
	number = {6507},
	urldate = {2022-04-11},
	journal = {Science},
	author = {Lazer, David M. J. and Pentland, Alex and Watts, Duncan J. and Aral, Sinan and Athey, Susan and Contractor, Noshir and Freelon, Deen and Gonzalez-Bailon, Sandra and King, Gary and Margetts, Helen and Nelson, Alondra and Salganik, Matthew J. and Strohmaier, Markus and Vespignani, Alessandro and Wagner, Claudia},
	month = aug,
	year = {2020},
	pages = {1060--1062},
}

@article{choudhury_machine_2020,
	title = {Machine learning and human capital complementarities: {Experimental} evidence on bias mitigation},
	volume = {41},
	copyright = {© 2020 John Wiley \& Sons, Ltd.},
	issn = {1097-0266},
	shorttitle = {Machine learning and human capital complementarities},
	url = {https://onlinelibrary.wiley.com/doi/abs/10.1002/smj.3152},
	doi = {10.1002/smj.3152},
	language = {en},
	number = {8},
	urldate = {2025-01-24},
	journal = {Strategic Management Journal},
	author = {Choudhury, Prithwiraj and Starr, Evan and Agarwal, Rajshree},
	year = {2020},
	note = {10.1002/smj.3152},
	pages = {1381--1411},
}

@article{handa2025economic,
  title={Which Economic Tasks are Performed with AI? Evidence from Millions of Claude Conversations},
  author={Handa, Kunal and Tamkin, Alex and McCain, Miles and Huang, Saffron and Durmus, Esin and Heck, Sarah and Mueller, Jared and Hong, Jerry and Ritchie, Stuart and Belonax, Tim and others},
  journal={arXiv preprint arXiv:2503.04761},
  year={2025}
}

@article{balducchi_labor_2004,
	title = {Labor {Exchange} {Policy} in the {United} {States}},
	url = {https://research.upjohn.org/up_press/143},
	doi = {10.17848/9781417550005},
	journal = {Upjohn Press},
	author = {Balducchi, David and Eberts, Randall and O'Leary, Christopher},
	month = jan,
	year = {2004},
}

@article{eberts_frontline_2003,
	title = {A {Frontline} {Decision} {Support} {System} for {Georgia} {Career} {Centers}},
	url = {https://research.upjohn.org/bookchapters/43},
	journal = {Book Chapters},
	author = {Eberts, Randall and O'Leary, Christopher},
	month = jan,
	year = {2003},
}

@article{blinder_how_2009,
	title = {How many {US} jobs might be offshorable?},
	volume = {10},
	number = {2},
	journal = {World Economics},
	author = {Blinder, Alan S.},
	year = {2009},
	pages = {41--78},
}

@website{aea_ethics,
	title = {AEA `Data Legality Policy and Explanations' and `AEA Data and Code Availability Policy'},
	url = {https://www.aeaweb.org/journals/data/data-legality-policy},
	urldate = {2025-01-14},
	publisher = {American Economics Association},
	author = {American{\ }Economics{\ }Association},
    note = {https://www.aeaweb.org/journals/data/data-legality-policy. Accessed 2025-01-14.},
	month = jan,
	year = {2023},
}

@book{committee_on_automation_and_the_us_workforce_an_update_artificial_2024,
	address = {Washington, D.C.},
	title = {Artificial {Intelligence} and the {Future} of {Work}},
	isbn = {978-0-309-71714-4},
	url = {https://nap.nationalacademies.org/catalog/27644},
	urldate = {2025-01-24},
	author = {National{\ }Academy{\ }of{\ }Sciences},
	publisher = {National Academies Press},
	collaborator = {{Committee on Automation and the U.S. Workforce: An Update} and {Computer Science and Telecommunications Board} and {Division on Engineering and Physical Sciences} and {Board on Human-Systems Integration} and {Division of Behavioral and Social Sciences and Education} and {National Academies of Sciences, Engineering, and Medicine}},
	year = {2024},
	doi = {10.17226/27644}
}

@article{lancaster2019technology,
  title={Technology report review of burning glass job-ad data},
  author={Lancaster, Vicki and Mahoney-Nair, Devika and Ratcliff, Nathaniel J},
  journal={University of Virginia, Biocomplexity Institute and Initiative Social and Decision Analytics Division},
  year={2019}
}

@article{frauenheim_what_2007,
	title = {What {Killed} {America}'s {Job} {Bank}?},
	volume = {86},
	copyright = {Copyright Crain Communications, Incorporated Jul 23, 2007},
	issn = {1547-5565},
	language = {eng},
	number = {13},
	journal = {Workforce Management},
	author = {Frauenheim, Ed},
	year = {2007},
	pages = {33--},
}

@article{wasi_record_2015,
	title = {Record linkage using {Stata}: {Preprocessing}, linking, and reviewing utilities},
	volume = {15},
	number = {3},
	journal = {The Stata Journal},
	author = {Wasi, Nada and Flaaen, Aaron},
	year = {2015},
	note = {ISBN: 1536-867X
Publisher: SAGE Publications Sage CA: Los Angeles, CA},
	pages = {672--697},
}

@misc{price_systems_2024,
	title = {Systems and {Methods} for {Processing} {Data} {Using} {Interference} and {Analytics} {Engines}},
	url = {https://patents.google.com/patent/US20240289641A1/en},
	urldate = {2025-01-27},
	publisher = {Google Patents},
	author = {Price, Ronald N. and Boyda, Jason and Bobay, Kathleen},
	month = aug,
	year = {2024},
        note = {U.S. Patent Application US20240289641A1.}
}

@inproceedings{
he2021deberta,
title={{DeBERTa}: Decoding-enhanced {BERT} with Disentangled Attention},
author={Pengcheng He and Xiaodong Liu and Jianfeng Gao and Weizhu Chen},
booktitle={International Conference on Learning Representations},
year={2021},
url={https://openreview.net/forum?id=XPZIaotutsD}
}

@inproceedings{10.5555/3666122.3668142,
author = {Zheng, Lianmin and Chiang, Wei-Lin and Sheng, Ying and Zhuang, Siyuan and Wu, Zhanghao and Zhuang, Yonghao and Lin, Zi and Li, Zhuohan and Li, Dacheng and Xing, Eric P. and Zhang, Hao and Gonzalez, Joseph E. and Stoica, Ion},
title = {Judging LLM-as-a-judge with MT-bench and Chatbot Arena},
year = {2023},
publisher = {Curran Associates Inc.},
address = {Red Hook, NY, USA},
booktitle = {Proceedings of the 37th International Conference on Neural Information Processing Systems},
articleno = {2020},
numpages = {29},
location = {New Orleans, LA, USA},
series = {NIPS '23}
}

\clearpage

\appendix
\counterwithin{figure}{section} % Makes figure counter depend on section counter
\counterwithin{table}{section}  % Makes table counter depend on section counter
\renewcommand{\thefigure}{\thesection.\arabic{figure}} % Redefine figure numbering (e.g., A.1)
\renewcommand{\thetable}{\thesection.\arabic{table}}   % Redefine table numbering (e.g., A.1)
% --- End of important commands ---

\clearpage 
\section{Appendix: Mapping O*NET Features}
\label{sec:app_onet_data}

For the dataset we sought to create, O*NET provides the scaffolding, and in many cases the detail necessary for creating standard structured data. We initiated our work by inspecting the O*NET database scheme. We directly incorporated many of the O*NET tables and structures into the design of our work.  The Content Reference Model is the ``conceptual foundation of O*NET'' that joins all major features of the O*NET database into a single unified taxonomy for  understanding the most important aspects of work. We pursued elements of the O*NET Content Reference Model with the opportunities provided by rich detail from frequently updated job ads data. 

\subsection{O*NET 1: Worker Characteristics}
\label{sec:wrkr_chars}

O*NET includes measures of worker characteristics. Job ads often state a desire for worker characteristics. We did not seek to extract Abilities (1.A.), measures of which can be constructed with a cross-walk provided by O*NET that links Abilities to Work Activities (O*NET 4.A), (described in Section \ref{sec:taskmatch}). Interests (1.B.) include general occupational interests (1.B.1) based on Holland's (\citeyear{holland_making_1997})  RIASEC framework, which we capture using a dictionary of RIASEC keywords provided by O*NET \citep{rounds_updating_2022}. We do not construct a separate model to complete elements of Work Values (1.B.2.) or Basic Occupational Interests (1.B.3) or Work Styles (1.C.). Some of these might be ascertained theoretically or by examining other sections of O*NET. For example, Dependability (1.C.5.a) as a worker characteristic is likely to be implicated in a job's description of a required shift or schedule (4.C.3.d.4), which is described below in Section \ref{sec:onet_occ_reqs}.  

\subsection{O*NET 2: Worker Requirements}
\label{sec:wrkr_reqs}

Basic and Cross-Functional Skills (2.A. and 2.B.) can be constructed with a cross-walk provided by O*NET that links Skills to Work Activities (O*NET 4.A, described in Section \ref{sec:taskmatch}). However, to increase the level of detail and directly capture empirical data on skills, we also develop a skill matching model (SkillMatch) described in Section \ref{sec:skillmatch} that is based upon the more elaborately detailed European Skills, Competences, and Occupations (ESCO) database v 1.20. 

We do not construct a separate model for Knowledge requirements (2.C.). We build a dictionary (see Table \ref{tab:dict_educ}) to capture the Required Level of Education (2.D.1). We do not capture field of education requirements (2.D.3.). We capture jobs that reference Spanish language skills, but not other languages. 

\subsection{O*NET 3: Experience Requirements}
\label{sec:exp_reqs}

We do not capture the number of years of related work experience required (3.A.1). We capture whether training is described in a job ad (3.A.3.). We do not systematically capture which basic or cross-functional skills are specifically mentioned as requirements for entry into the occupation (3.B. or 3.C.); these could perhaps be calculated from data we provide on skills in entry-level jobs within an occupation.  We do calculate a Flesch-Kincaid readability score for each job posting, which informs reading comprehension entry requirements (3.B.1.a). We capture whether an occupational license is indicated in the job posting (3.D.), but not which one (3.D.2.), or whether it is preferred or required. We capture several entry requirements that are imposed by the organization (3.D.5) or government (3.D.5.a), such as drug and criminal background checks (see Table \ref{tab:dict_dbc}) and requirements related to work authorization and visa sponsorship. We capture presence of a labor union and/or a professional association (3.D.5.C).

\subsection{O*NET 4: Occupational Requirements}
\label{sec:onet_occ_reqs}

O*NET hierarchically structures generalized work activities (4.A.) into intermediate activities (4.E.) that span occupations, and detailed (4.D.) work activities that reside within an occupation, which are further detailed in over 20,000 occupation-specific task statements (5.A). We identify task statements within each job advertisement separately from identifying occupational codes from titles. Technical detail is provided in Section \ref{sec:taskmatch}. Therefore, we do not impose a requirement built into O*NET that a task lies within only a single occupation. 

Organizational Context (4.B.) includes ``characteristics of the organization that influence how people do their work.'' Organizational context includes structural characteristics (4.B.1) such as human resources systems and practices (4.B.1.b.) and recruitment and selection processes (4.B.1.b.1).  We develop novel dictionaries for  Diversity, Equity, Inclusion, and Belonging (DEIB) by examining an existing list of terms \citep{yourish_these_2025}, elements of the rewards system (4.B.1.b.3), including benefits from a dictionary we build. We develop a custom NER model (``WageExtract'') to extract wage information from job ad text (see section \ref{sec:wageextract}).

Social Processes (4.B.2) include culture, values and principles of the organization.  A number of the dictionaries that relate generally to organizational context (4.B.) are standard dictionaries from the business and management literature. Standard dictionaries we also run include recruiting signals  \citep{banks_strategic_2018}; time horizon  \citep{brigham_researching_2014,brochet_speaking_2015,desjardine_one_2019};  innovation \citep{heyden_conjoint_2015}; digital orientation \citep{kindermann_digital_2021}; creativity \citep{haselhuhn_investors_2022}; attention \citep{levy_influence_2005}; ambidexterity \citep{mckenny_what_2018}; entrepreneurial orientation \citep{mckenny_strategic_2018,short_construct_2009}; Value Orientation \citep{moss_funding_2018}; Corporate Social Responsibility (CSR) \citep{pencle_whats_2016}; Sustainability Orientation \citep{vaupel_role_2023}; Market Orientation  \citep{zachary_family_2011}. Given the volume of data extraction related to these dictionaries and the nuance required, we are in the process of reviewing and interpreting aggregate data. 

Work Context (4.C.) includes features we capture, including whether work must be done physically close to others, and required work schedules for the position (see Appendix Table \ref{tab:dict_shift}). 

\subsection{O*NET 5: Occupation-Specific Requirements}
\label{sec:onet_occ_spec_reqs}

Task (5.A.) extraction is described further in the methods section \ref{sec:taskmatch} on TaskMatch. Titles (5.C.) and Alternate Titles (5.E.) are described above as related to occupation, with additional technical detail on ``TitleMatch'' in Section \ref{sec:titlematch}. Many features from titles are extracted and described in Appendix \ref{sec:dict_job_titles}. A dictionary of technologies (5.F.1.) and tools (5.G.1) is included in O*NET. 

The O*NET database includes United Nations Standard Products and Services Codes (UNSPSC) for a list of bespoke tools ranging from ‘abdominal binders’ to ‘Zylonite files’, as well as  technologies such as ‘Microsoft Office.’ We remove terms that generate many false positives. For example, we removed 96 items that generated excessive false positives (e.g., “scale”, “range”) from the O*NET tools and technologies dictionaries. We add Artificial Intelligence keywords from \citet{lou_ai_2021} and a novel list of AI related terms.

\subsection{O*NET 6: Workforce Characteristics}
\label{sec:work_chars}

Labor Market Information (6.A.) and Occupational Statistics (6.A.1.) incorporate ``information related to economic conditions and labor force characteristics of occupations.''  While we do not attempt an Occupational Outlook (6.B.) or Projections (6.B.1.), data extracted here could be well-suited for that purpose. 

%We utilize job advertisements to calculate occupational and industry labor market concentration within commuting zones using the Herfindahl Hirschman Index (HHI), as done in earlier research \citep{azar_labor_2022} with CareerBuilder.com data.

\clearpage

\clearpage
\section{Detailed Methodology and Validation Steps}
\label{sec:detailed_methods}

Here, we provide additional details of the model training procedures, and additional detail on procedures undertaken for several of the key models, as well as supplemental information on dictionaries and trial-and-error learning to complement the summary in Section \ref{sec:methods}. 

\subsubsection{Dictionaries}

The foundation for all models we develop originate from initial taxonomies or ``dictionaries.'' These data are lists of terms, chunks, or sentences that have standardized labels. We adopt standard dictionaries from O*NET and build novel dictionaries following its structure. Before augmentation and iterative processing, we typically run dictionaries and focus manual review efforts on the most frequently appearing observations. Before processing the corpus, we validate each model iteration by strategically auditing results binned by frequency of appearance of features in a large random sample of job ad text and by auditing small random samples of the results to ensure accuracy. We are satisfied when over 90\% are accurate in a small sample. For example, we found 98/100 identifications of a tool or technology were true positives in a random audit. In strategic auditing of tools, we found 88 high-frequency tools including `levels', `ranges', and 'scales' that were frequently false positives. We remove false positives from the dictionaries. 

A limitation of reliance on benchmark / standard dictionaries is that emerging tools and technologies, e.g., ‘ChatGPT’, will not have a commodity code or entry in the UNSPSC list of technologies and tools included in O*NET. Other parts of O*NET are not elaborated. Where possible, we build custom dictionaries for such cases, and make direct edits to the results until satisfied.

\subsubsection{From Dictionaries to Augmented Training Data}

Dictionaries often lack context and absent NLP methods, can only yield results that are based on exact text string matching. The augmentation method described in Section \ref{sec:skillmatch} provides illustration and detail of the process we followed to produce our results. Here, we briefly describe other approaches that failed to work, and provide some additional detail on future directions.

We attempted a dictionary of tasks based on analysis of parts of speech in the O*NET task statements. Identifying unique noun-verb pairs within the text of O*NET task statements, and authoring rules requiring pairs appear within a narrow context window, proved unsatisfying in both recall and precision. 

% We first deployed the augmentation procedure to the task verbs and nouns from the O*NET to boost recall by adding highly similar verbs and nouns suggested by embeddings. However, many positive identifications from this approach proved to be false. Sentences labeled in this approach often did not truly contain information relevant to the duties to be performed by the worker, i.e., were not sentences that contained relevant task information. 

We then attempted to split the job advertisements into sections for targeted extraction, e.g., 'skill requirements' or 'task' sections. However, job advertisements are free-form and no set of rules were consistently able to break apart job ads into sections. Had that been successful, dictionaries operating within specific sections might have been a fruitful approach to context-specific results. 

We then developed the augmentation process and labeled hundreds of thousands of sentences as 'task'/'not-task' and 'skill'/'not-skill' sentences, and experimented with embeddings models. We developed audit routines focusing on manual analysis of a small number of results binned by similarity score. The success of this in scaling and screening out false positives led us to further adopt and refine our approach. 

%Future work can combine LLM-as-Judge methods with training data and results from existing versions of SkillMatch and TaskMatch to pursue a sentence-level multi-classification approach, labeling each sentence in a job ad first, and only then pursuing classification with a specific model. 

\subsubsection{Model Training Procedures}
All procedures involving the fine-tuning of pretrained Language Models were performed using the HuggingFace Trainer library\footnote{\url{https://huggingface.co/docs/transformers/en/main_classes/trainer}}, using default parameters -- including a learning rate of 5e-5 and the Adam optimizer. Training was performed for one epoch on the 90\% train split, which was always obtained using a random seed of 42. All training was run on a single Nvidia RTX A6000 GPU, and the utilized batch size was tailored to the maximum sequence length and pretrained model size.

Models requiring the entire job ad text as context (i.e., FirmExtract) were limited to 1024 tokens, whereas models trained on sentences (TaskMatch, SkillMatch, binary WageExtract model) were trained with a maximum length of 64 (roughly 3x the average English sentence length), or 128 in the case of the two larger WageExtract models. TitleMatch models were similarly restricted to 32 tokens, since only the title text is given as input to these models.

\subsubsection{LLM-as-a-Judge Validation Prompts}
In Table \ref{tab:prompt1}, we provide the prompt used to validation the binary classification stage of SkillMatch. Few-shot examples were curated from the human-coded dataset used to train the classification model. Similarly, in Table \ref{tab:prompt2}, we provide the prompt used for the second part of the LLM-as-a-Judge validation, namely on the matching results for Skill Match.

\begin{table*}[htbp]
\centering
\footnotesize
\begin{tabular}{p{0.95\linewidth}}
\hline
\textbf{Prompt}
\\ \hline
You will be given a sentence.
Your task is to decide whether the given sentence contains a skill statement or not.
A skill is the ability to perform a specific task and apply knowledge, particularly in the work context.
Answer simply with SKILL or NOT\_SKILL, denoting that the sentence contains a skill statement or not, respectively.

Provide your feedback as follows:

Output:::\\
Classification: (SKILL or NOT\_SKILL)

Here are some examples:

sentence: listens to what other people are saying and asks questions as appropriate.

Output:::\\
Classification: SKILL

sentence: *position summary: * customer service - country code top-level domain (cctld) specialist’s work within the client service organization supporting our client base of primarily fortune 100 companies.

Output:::\\
Classification: NOT\_SKILL

sentence: responsible for delivering and serving food and beverage to guests in a friendly, prompt and efficient manner.

Output:::\\
Classification: SKILL

sentence: this position will travel onsite to the client's charlotte, nc location as needed for client meetings.

Output:::\\
Classification: NOT\_SKILL

Now here is the sentence.

sentence: [INPUT SENTENCE]

Output:::\\
Classification:\\
\hline
\end{tabular}
\caption{Prompt for LLM-as-a-Judge Validation of Binary Skill Classification.}
\label{tab:prompt1}
\end{table*}

\begin{table*}[htbp]
\centering
\footnotesize
\caption{Prompt for LLM-as-a-Judge Validation of Skill Matching.}
\begin{tabular}{p{0.95\linewidth}}
\hline
\textbf{Prompt}
\\ \hline

You will be given a SENTENCE and a CANDIDATE.
Your task is to decide whether the given CANDIDATE is an appropriate match for the given SENTENCE.
To be an appropriate match, the CANDIDATE should accurately represent what skills are described in the SENTENCE.
A skill is the ability to perform a specific task and apply knowledge, particularly in the work context.
Answer simply with MATCH or NOT\_MATCH, denoting that the CANDIDATE matches the skill statement in the SENTENCE or not, respectively.

Provide your feedback as follows:

Output:::
\\Classification: (MATCH or NOT\_MATCH)

Here are some examples:

SENTENCE: operate, calibrate, and maintain lab equipment.
\\CANDIDATE:  operating scientific and laboratory equipment 

Output:::
\\Classification: MATCH

SENTENCE: workers will be exposed to all types of weather conditions.
\\CANDIDATE:  operating lifting or moving equipment 

Output:::
\\Classification: NOT\_MATCH

SENTENCE: problem solving .
\\CANDIDATE:  solving problems

Output:::
\\Classification: MATCH

SENTENCE: uses time effectively to manage workload/tasks.
\\CANDIDATE:  documenting technical designs, procedures, problems or activities 

Output:::
\\Classification: NOT\_MATCH

Now here are the actual texts.

SENTENCE: TEST
\\CANDIDATE: TEST

Output:::
\\Classification: \\
\hline
\end{tabular}
\caption{Prompt for LLM-as-a-Judge Validation of Skill Matching.}
\label{tab:prompt2}
\end{table*}

\subsection{FirmExtract}

This training dataset was prepared by randomly sampling 200,000 job ads from the NLx corpus in which the self-report firm name was present. An algorithm was created to match all instances of this firm name in the text (lower case, shortened title, etc.), thereby creating quality tagged NER data. The firmNER model was trained for one epoch on a 90\% split of our training dataset, with a sequence classification objective (all tokens are labeled as either part of a firm name or not), and this model achieved an F1 score of 94.5 on the validation set.

 To match firms to industries, we first standardize Data Axle firm names using the same cleaners discussed above, and then calculate the Levenshtein Distance similarity ratio between standardized firm names. We accept a match when the match score is above 0.8.  We leverage the availability of zip code and state information in both job ads and Data Axle to match firms to establishments at the zip code level if possible. Based on three rounds of manual auditing, we use 0.8 as a threshold for fuzzy matching. If no match is above 0.8 at the zip code level, we match firm names at the state level. If no match is above 0.8 at the state level, we match against the national list of all firm names, and expedite the search by looking only at firms with the first character in common with the job ad firm name. Although we retain the best match and match score, in the dataset we construct, we treat all results below 0.8 as missing industry.

 %If no match is found at the zip code level, we seek the highest-scoring match within state, and if no match above 0.8 is found at the state level, we retrieve the highest-scoring match nationally above 0.8. We developed this process after auditing three rounds of firm-matching results, focusing on accurate identification of the largest firms.

\subsection{TitleMatch}
\label{sec:titlematch_detail}

We initially ran O*NETs sample of titles as a dictionary against a large random sample of job ad titles and found that 80\% of the online job titles can be exact matched to at least one O*NET occupation code. A random audit revealed that 96.8\% of the positive identifications based on an exact match of an occupation were true positives.

After the initial round of exact matching between reported job ad titles from O*NET and job ad titles, we augmented the training data of reported titles with high-match scoring results from a random sample. We manually labeled high-frequency titles that could not be exact matched. We manually deleted high-frequency false positives from the sample of reported titles in O*NET (``helper'', ``laborer''), and corrected high-frequency false positives in the training data when multiple occupation codes were exact matched based on a dictionary approach. For example, audits revealed that ``delivery'' in isolation is often associated with the courier occupation, but the co-appearance of ``delivery'' and ``nurse'' in a title is corrected to always associate with the nursing occupation. We compared results between Version 1 (sample of reported titles embeddings) and Version 2 (sample of reported titles augmented by real job titles suggested by TitleMatchV1). We find an exact match for 89.5\% of job ad titles.

%, and manually audit a strategic sample of 100 job titles from across the distribution of match scores.

Table \ref{tab:app_data_analyst} illustrates the major remaining problem with this approach: using the illustrative title ``data analyst,'' with 9 candidate labels and without additional information, exact-matching of job titles cannot guarantee accurate results. In this case, TitleMatch codes all ``data analysts'' to the occupation Financial Quantitative Analysts. 

\begin{table}[ht]
\centering
\small
\begin{tabularx}{\textwidth}{l l}
\toprule
\textbf{O*NET-SOC Code} &
{\textbf{Occupation }} \\
\midrule
13-2099.01	& Financial Quantitative Analysts \\	
15-1243.00	& Database Architects \\ 
15-2041.00	& Statisticians	\\
15-2051.00	& Data Scientists	\\
15-2051.01	& Business Intelligence Analysts \\
15-2099.01	& Bioinformatics Technicians \\
19-1029.01	& Bioinformatics Scientists \\	
19-3022.00	& Survey Researchers \\
19-4061.00	& Social Science Research Assistants \\
\midrule
\end{tabularx}
\caption{Occupations for the ``Data Analyst'' job title in the sample of reported titles and alternate titles.}
\label{tab:app_data_analyst}
\end{table}

While we follow others in matching titles to occupation, and TitleMatch performs well in independent testing,  there is a need for further research. Future work that combines data from titles with extracted data on industry, tools, tasks and other information in online job ads can improve occupational matching accuracy with the text of online job ads, and could in the future potentially be used to capture emerging occupational definitions. 

\paragraph{Additional Check of Convergent Validity.} We compare TitleMatch's output to results from a newspaper corpus of help wanted ads from the 1950s through 2000. \citet{atalay_new_2018,atalay_evolution_2020} predict standard occupation codes from job titles and provide labeled title and occupation data in the paper's online \href{https://occupationdata.github.io/}{data repository} . The underlying job titles in newspapers have challenges not faced in online job ad data: drawn from OCR scans of newspapers, these are subject to transcription errors, and titles could be misidentified due to layout and parsing issues as well. Newspaper job ad titles are also much shorter than online job ad titles, and the job ad titles are older in newspapers, reflecting an earlier time period \citep{marinescu_opening_2020}. Nevertheless, in a 1\% sample of job titles, for 1,124 titles where the Atalay et al. model returns a confidence level of 1, TitleMatch retrieves the same major occupational group in 91\% of observations and the same six-digit occupational code for 74\% of the observed job titles. This suggests a high degree of out of sample correspondence. 

\subsection{Hierarchy from Titles.} 

To extract the hierarchy values, we built a knowledge map of terms that indicate hierarchy. We exhaustively searched for but do not display in the excerpts all the variant keywords we include in the production-use map. Null results default to zero. A second `stepper' knowledge map returns a value ranging from -7 to +4. Words such as ``assistant'' and ``vice'' decrement the base level, and words such as ``senior'' or ``chief'' increment the base level. The sum of the base and the `stepper' maps is equal to the hierarchy level value, and we construct a training dataset of nearly 245k text examples from NLx using this base knowledge map and stepper logic.

The training dataset maps title texts to hierarchy values, and we use this data to train BERT-based regression model, i.e., to predict the hierarchy value given a text input. Due to the semantic intricacy of the task, we choose to fine-tune a \textsc{DeBERTa-v3-base} model \citep{he2021deberta} for regression, which we perform on a 90\% train split for one epoch. 

\begin{table}[htbp!]
\centering
\medskip 
\setlength{\tabcolsep}{2pt}
\begin{tabular}{@{}p{0.48\textwidth} p{0.47\textwidth}@{}}
\begin{minipage}[t]{0.48\textwidth}
\centering
\footnotesize
\begin{tabular}{|l|r|l|}
\toprule
Keyword & Value & Label \\ \hline
Internship & -10 & Intern level \\
Trainee & -10 & Intern level \\
Entry-Level & 0 & Base level \\
Manager & 10 & First-Level Supervisor \\
Supervisor & 10 & First-Level Supervisor \\
Team Leader & 10 & First-Level Supervisor \\
Territory Manager & 20 & Second-Level Supervisor \\
Division Leader & 30 & Third-Level Supervisor \\
General Manager & 30 & Third-Level Supervisor \\
Director & 40 & First-level Executive \\
CHRO & 50 & Senior Executive  \\
CEO & 60 & Top Management \\
\bottomrule
\end{tabular}
\medskip
\\
1A. Excerpt of Base Hierarchy Map.
%\label{tab:firstlevel}
\end{minipage} &
\begin{minipage}[t]{0.47\textwidth}
\centering
\footnotesize
\begin{tabular}{|l|r|l|}
\toprule
Helper & -7 & Helper \\
Junior & -6 & Junior \\
Asst & -5 & Assistant \\
Associate & -3 & Associate \\
Vice & -2 & Vice \\
Deputy & -1 & Deputy \\
Lead & 1 & Lead \\
Leader & 1 & Leader \\
Sr & 2 & Senior \\
Exec & 3 & Executive \\
Chief & 4 & Executive \\
\bottomrule
\end{tabular}
\medskip
\\
1B. Excerpt of Hierarchy Stepper Map
\label{tab:HierarchyBaseMap}
\end{minipage} \
\caption{Mapping Hierarchy to Job Titles}
\label{tab:combined_hierarchy}
\end{tabular}
\end{table}

\subsection{Features from Job Titles}
\label{sec:dict_job_titles}
We developed a list of features in titles by examining a random sample of job ad titles and noticing words that were not indicators of hierarchy or occupation. For example, we notice urgent hiring, sign-on bonuses, travel, remote work, seasonal, part-time, and full-time work. In Appendix Table \ref{sec:dict_job_titles}, we list unique features we extract and  associated codes. 

Figure \ref{fig:featurestitles} illustrates cyclical hiring trends for positions with titles that indicate seasonal and college graduate hiring. To extract such features from job titles, we fine-tuned an additional \textsc{DeBERTa-v3-base} model in a multi-label setting, and the resulting model achieves an 81.4\% accuracy in correctly identifying the labels present in a title.

\begin{figure}[htbp!]
\centering
\includegraphics[width=\textwidth]{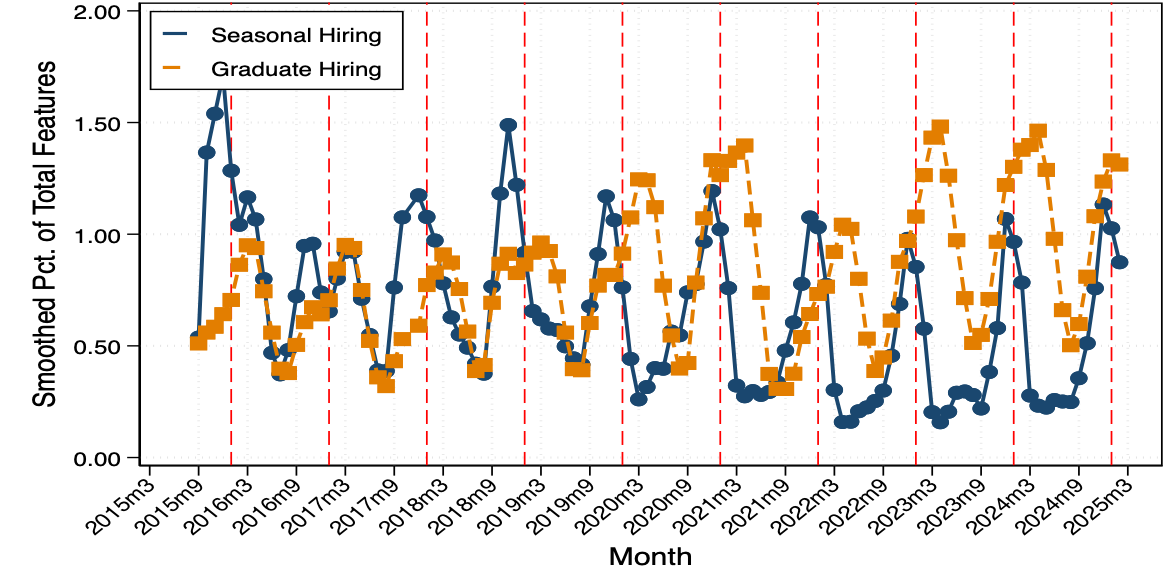}
\small
\justifying
\emph{Note: } These figures use a 3-month moving average (t, t-1, t-2) by date compiled. Seasonal and Graduate Hiring (January vertical line).
\caption{TitleMatch Features: Seasonal and Graduate Hiring}
\label{fig:featurestitles}
\end{figure}

\begin{table}[htbp]
\centering 
\scriptsize
\begin{filecontents*}{csvs/titles_features.csv}
Illustrative Term,Value,Label
Allowance,All,Allowance
Benefits,Bnf,Benefits
Bilingual,Bi,Bilingual
Bonus,Bns,Bonus
Commission,Cm,Commission-Based
Contract,Ctr,Contract
Contract-to-Hire,C2H,Contract-to-Hire
Entry-Level,EL,Entry-Level
Experienced,Ex,Experienced
Flexible,Flx,Flexible Hours
F/T,FT,Full-Time
Grad,Grd,Graduate Hiring
Holiday,Holiday,Holiday
Home-Based,Hm,Home-Based
Hourly,hrly,Hourly
Hybrid,Hy,Hybrid
Immediate,Imm,Immediate Start
Incentive,Inc,Incentive
Intern,Int,Internship
Job Fair,JF,Job Fair
Evening ,ON,Nights
On Call,OC,On Call
Onsite,OS,Onsite
On-Site,OS,On-Site
Overnight,ON,Overnight
Paid,Pd,Paid
Paid Per,PPS,Paid Per Service
P/T,PT,Part-time
Peak Time,Peak,Peak Time
Per Diem,PD,Per Diem
Relocation,Rel,Relocation
Remote,Rem,Remote
Rotational,Rot,Rotational
Salary,Slry,Salary
Seasonal,Ssn,Seasonal
Security Clearance,Cl,Security Clearance
Shift,SW,Shift Work
Sign on,SO,Sign-on Bonus
Spanish,ES,Spanish language
Subcontract,Sub,Subcontract
Telecommute,Tc,Telecommute
Temporary,Temp,Temporary
Training,TP,Training Provided
Travel,Trv,Travel Required
Undergraduate,Ugrd,Undergraduate
Virtual Hiring,VJF,Virtual Job Fair
Volunteer,Vol,Volunteer
Weekend,WE,Weekend
\end{filecontents*}
\csvautobooktabular[respect all]{csvs/titles_features.csv}
\caption{Custom Dictionary of Features Extracted from Job Titles}
\label{tab:dict_title}
\end{table}

\clearpage
\section{Aggregating Data and Comparisons to Benchmark Data}
\label{sec:robust}

This appendix contains additional information comparing the total count of job ads, industry, occupation, and wage distributions dataset with official government statistical benchmarks (JOLTS, QCEW, and OES) for multiple time periods.

\subsection{Additional Detail on Dates and Comparison to JOLTS}
\label{sec:robust_dates}

Research on job vacancies provide a variety of findings and techniques to address online job posting durations. An analysis of JOLTS data from 2001-2009 estimates a mean vacancy length of 14-25 days, and that each job opening yields between 1 - 1.8 hires \citep{davis_establishment-level_2013}.  With LinkUp data, \cite{chen_is_2023} drop job ads that are posted for more than 180 days and find an average posting duration of 36.5 days and median of 23 days. \cite{mueller_vacancy_2024} find a mean posting duration in Austrian job postings of 30.5 days.  Using UK job postings data, \cite{bassier_vacancy_2025} find a mean posting duration in UK job postings of 17-18 days. 

Analyzing NLx data, \cite{hashizume_timing} describes a long tail of ``evergreen jobs'': 25\% of job postings are up for more than 90 days, and 10\% for more than 180 days. Employers often use evergreen postings to fill multiple vacancies for stable roles with steady demand over long durations -- these assist employers seeking to meet just-in-time hiring needs. At the extreme end, 20 job postings have been posted for the entire 16 years the NLx has data, dating back to 2007, leading some to call these ``phantom'' or ``ghost'' jobs. As \cite{hashizume_timing} describes, both states and employers address vacancy duration differently: while most states do not impose a time limit, several states cull their records after 30/60/90 day periods, and many employers set initial 30, 60, or 90 days as initial windows for postings, and then may extend these. \cite{hashizume_timing} finds that employer differences seem to drive the variation in job posting duration. 

% \begin{figure}[!ht]
%     \centering
%     \caption{Distribution of NLx Job Posting Duration in 2024}
%     \includegraphics[width=\textwidth]{pdfs/days_distribution_2024.pdf}
%     \label{fig:distr_nlx_23}
% \end{figure}

The best available and regularly updated analysis of job openings from NLx and the relationship to the Job Openings and Labor Turnover Survey (JOLTS) \citep{BLS_JOLTS} data is performed and described by the NLx Research Hub in its Job Openings Estimator (JOE) application on its website: \url{https://nlxresearchhub.org/nlx-joe}.

\subsubsection{Date Compiled and Date Acquired}
Here, we provide additional on the date compiled, date acquired, and adjusted date acquired we use to build lists of monthly active jobs for aggregation of extracted data.  As displayed in Figure \ref{fig:distr_date_compiled}, monthly job files begin September 2015. One month -- November 2017 -- is missing. Several months contain an unusually large numbers of job ads.

\begin{figure}[!ht]
    \centering
    \includegraphics[width=\textwidth]{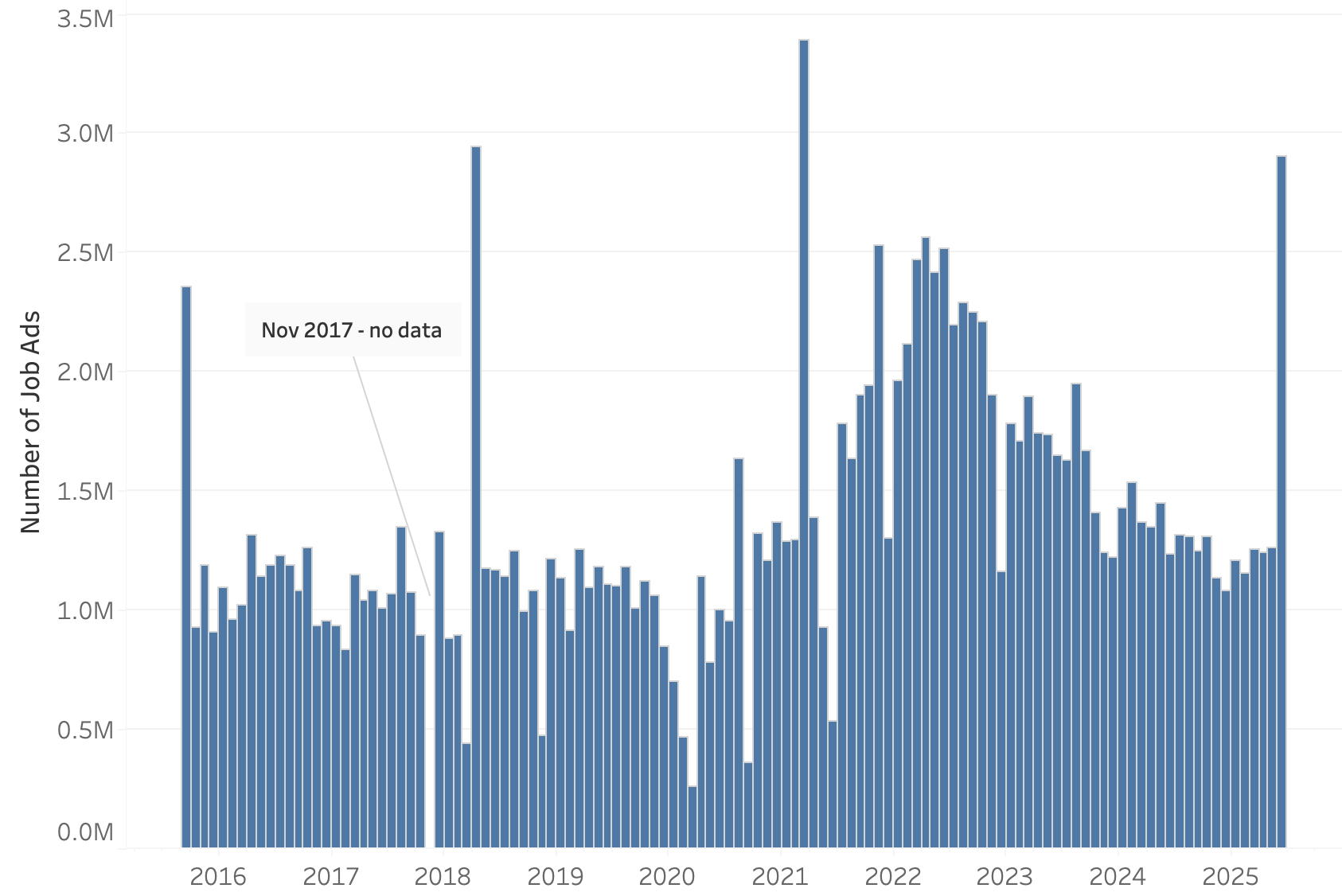}
    \caption{Number of Monthly Job Ads by Date Compiled}
    \label{fig:distr_date_compiled}
\end{figure}

The distribution of \emph{date\_acquired} is shown in Figure \ref{fig:distr_date_acquired}.  We observe that there are three large outliers corresponding to January 2015, January 2016, and January 2017.  These abnormally large values probably reflect quirks in the data collection process during those time periods and do not correspond to the actual starting dates of these job ads.  We analyze the monthly 'job history table' available since June 2021 to understand patterns of job duration in all NLx data. In 2024, we find that more than half of job postings compiled in a typical month will also appear in the prior month, and more than 25\% will appear 3 or more months prior to the month in which they are compiled. Looking at all postings since 2021, 86\% of postings compiled in a given month were acquired 2 months prior or less. In order to rectify this problem with the date acquired, we adjust the start month for all affected jobs to be 2 months prior to the end month (\emph{date\_compiled}). The adjusted distribution of start months is shown in Figure \ref{fig:distr_date_acquired_adj}.

\begin{figure}[!ht]
    \centering
    \includegraphics[width=\textwidth]{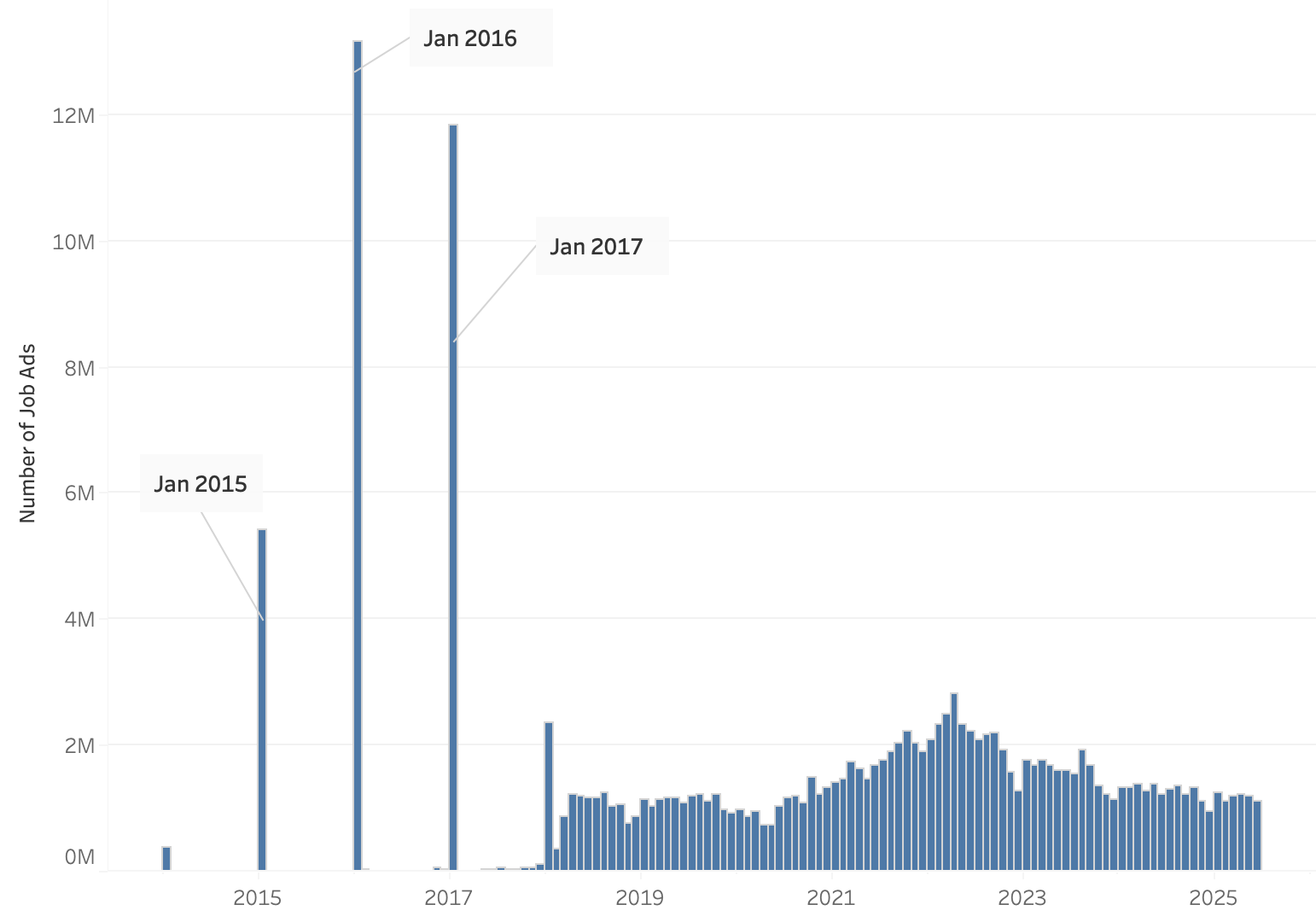}
    \caption{Number of Monthly Job Ads by Date Acquired}
    \label{fig:distr_date_acquired}
\end{figure}

\begin{figure}[!ht]
    \centering
    \includegraphics[width=\textwidth]{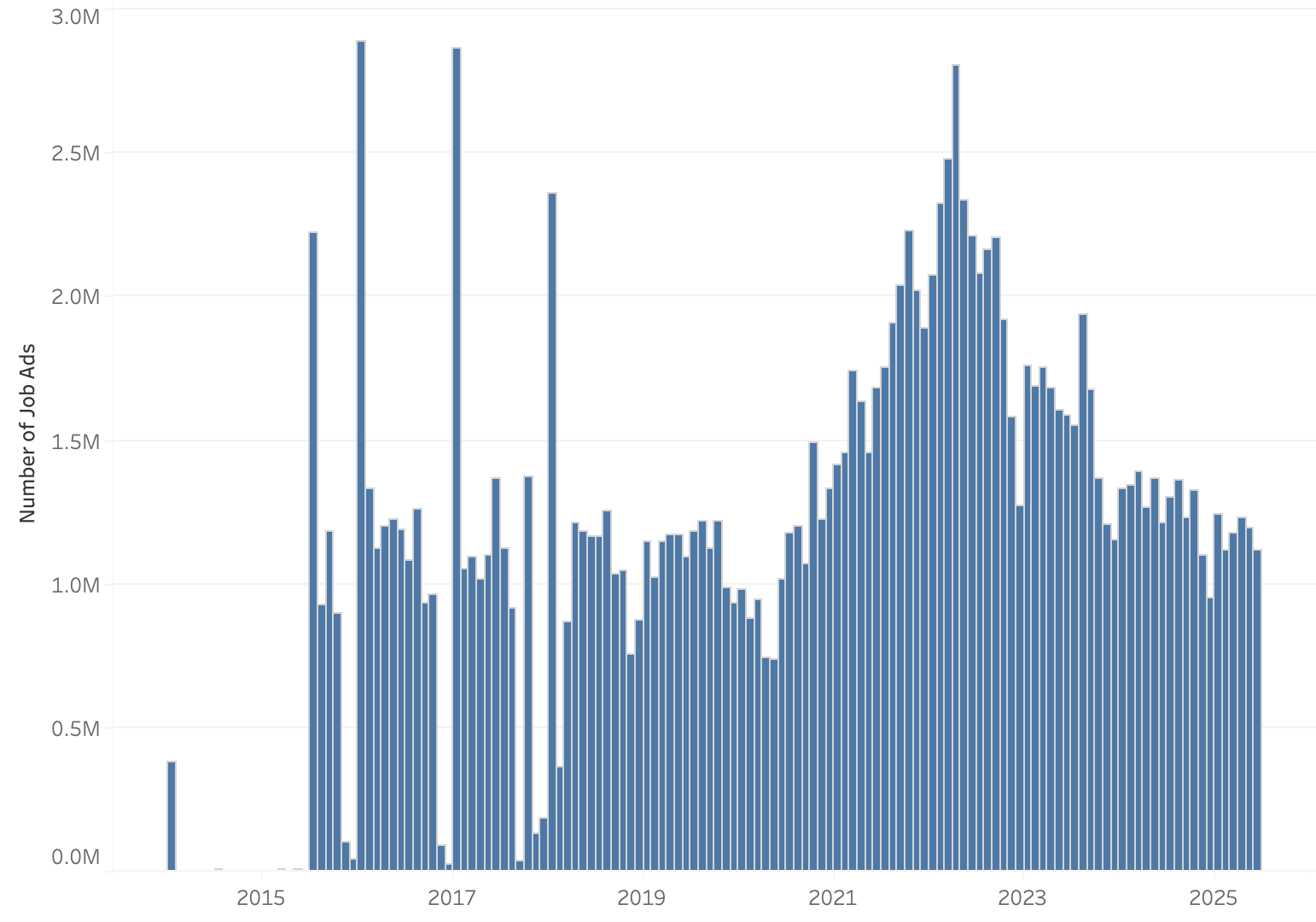}
    \caption{Number of Monthly Job Ads by Date Acquired (adjusted)}
    \label{fig:distr_date_acquired_adj}
\end{figure}
% We first analyze the monthly 'job history table' available since June 2021 to understand patterns of job duration in NLx data. We find the maximum duration for each job posting appearing in each monthly job history file. This answers the question: how long have job postings active in a particular month been posted? Figure \ref{fig:distr_nlx_23} provides boxplots of durations of job postings in the monthly job history files in calendar year 2023. The median ranges between 29 and 47 days, the mean between 125 and 158 days, and the 75th percentile of durations is between 80 and 120 days. It appears that more than half of job postings compiled in a typical month will also appear in the prior month, and more than 25\% will appear 3 or more months prior to the month in which they are compiled.

\clearpage
% ===================================================================
\subsection{FirmExtract: Annual Industry Distribution Comparison (vs. QCEW)}
% ===================================================================

To assess the representation of job postings by industry, we compare the distribution of job ads across major 2-digit industry groups in the NLx data with FirmExtract to the distribution for all employed workers in the U.S. with data from the Quarterly Census of Employment and Wages (QCEW) \citep{QCEW_data}, as in the \cite{hazell2022national} analysis using Lightcast data. We take the average share of employment by 2-digit industry (QCEW) and the average share of job postings (NLx with FirmMatch Industry) for each year (by date compiled) and plot the comparison. Figure \ref{fig:ind_bar_2024} provides the bar chart for 2024, Figure \ref{fig:ind_corr_2024} provides the correlation, and Figure \ref{fig:ind_corr_time} illustrates changes in the correlation over time.

While there are noticeable differences between job ad data from NLx and similar figures built with Lightcast data, there are similarities in that NLx postings with sector from FirmExtract significantly under-represent the manufacturing, construction, and accommodation and food services sectors, and over-represent information and professional, scientific, and technical services.  

\begin{figure}[h!]
  \centering
  \includegraphics[width=\textwidth]{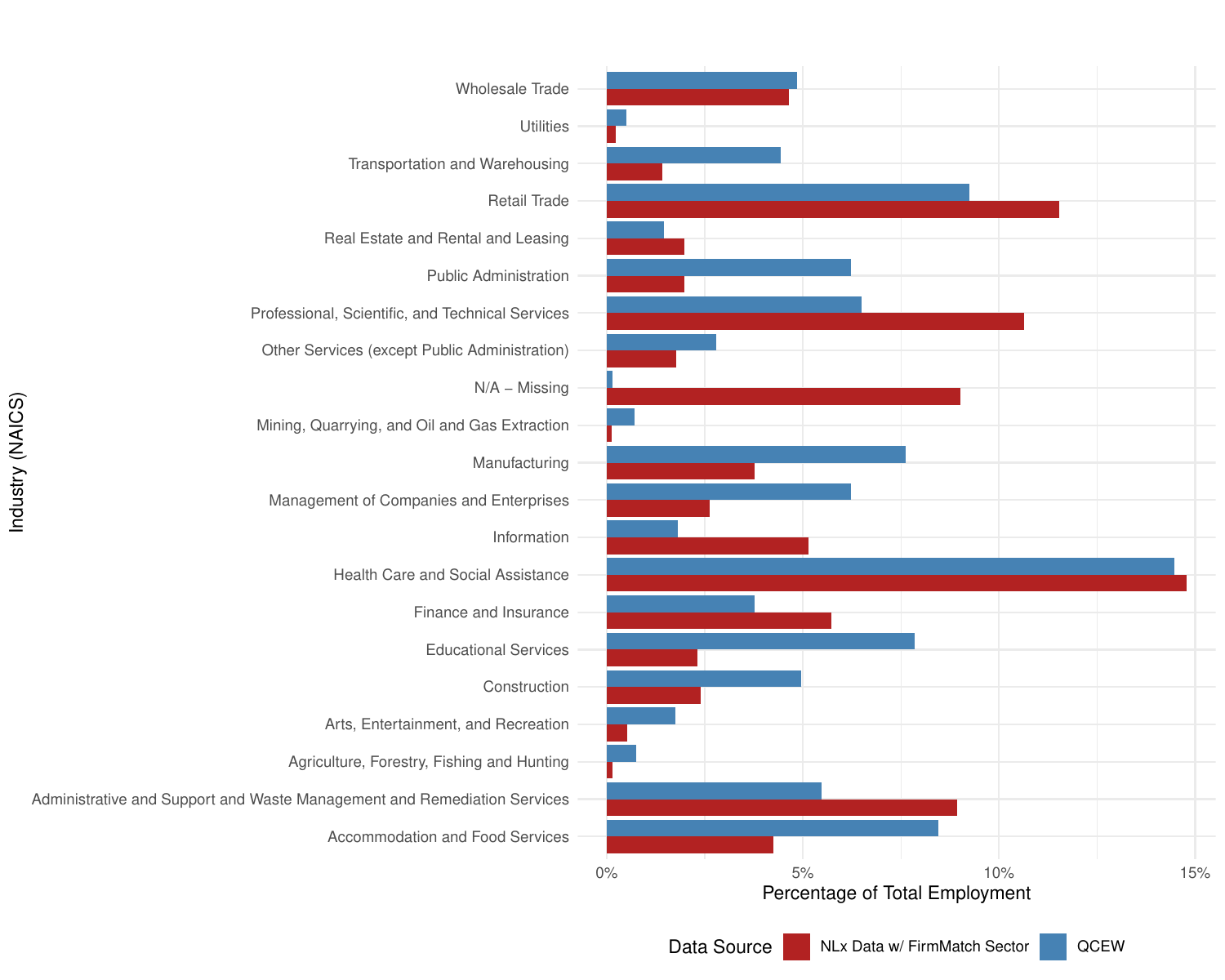}
      \caption{Industry Distribution Comparison, 2024.}
      \small
\justifying
\emph{Note: }This figure is based on aggregation of 2024 data by date compiled.
  \label{fig:ind_bar_2024}
\end{figure}

\begin{figure}[h!]
  \centering
  \includegraphics[width=1\textwidth]{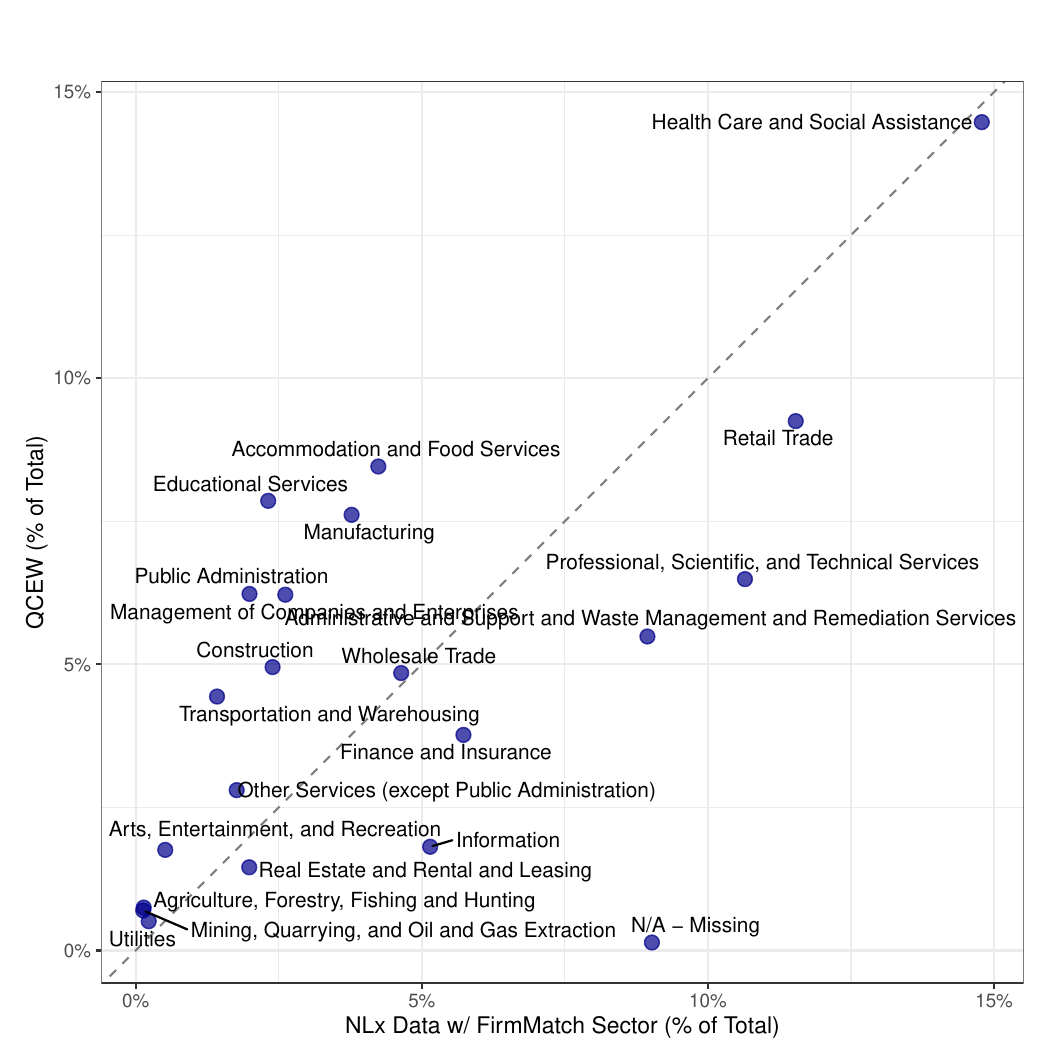}
        \small
\justifying
\emph{Note: }This figure is based on aggregation of 2024 data by date compiled. The Pearson correlation is 0.613.
    \caption{Industry Correlation, 2024.}
  \label{fig:ind_corr_2024}
\end{figure}

\begin{figure}[h!]
  \centering
  \includegraphics[width=1\textwidth]{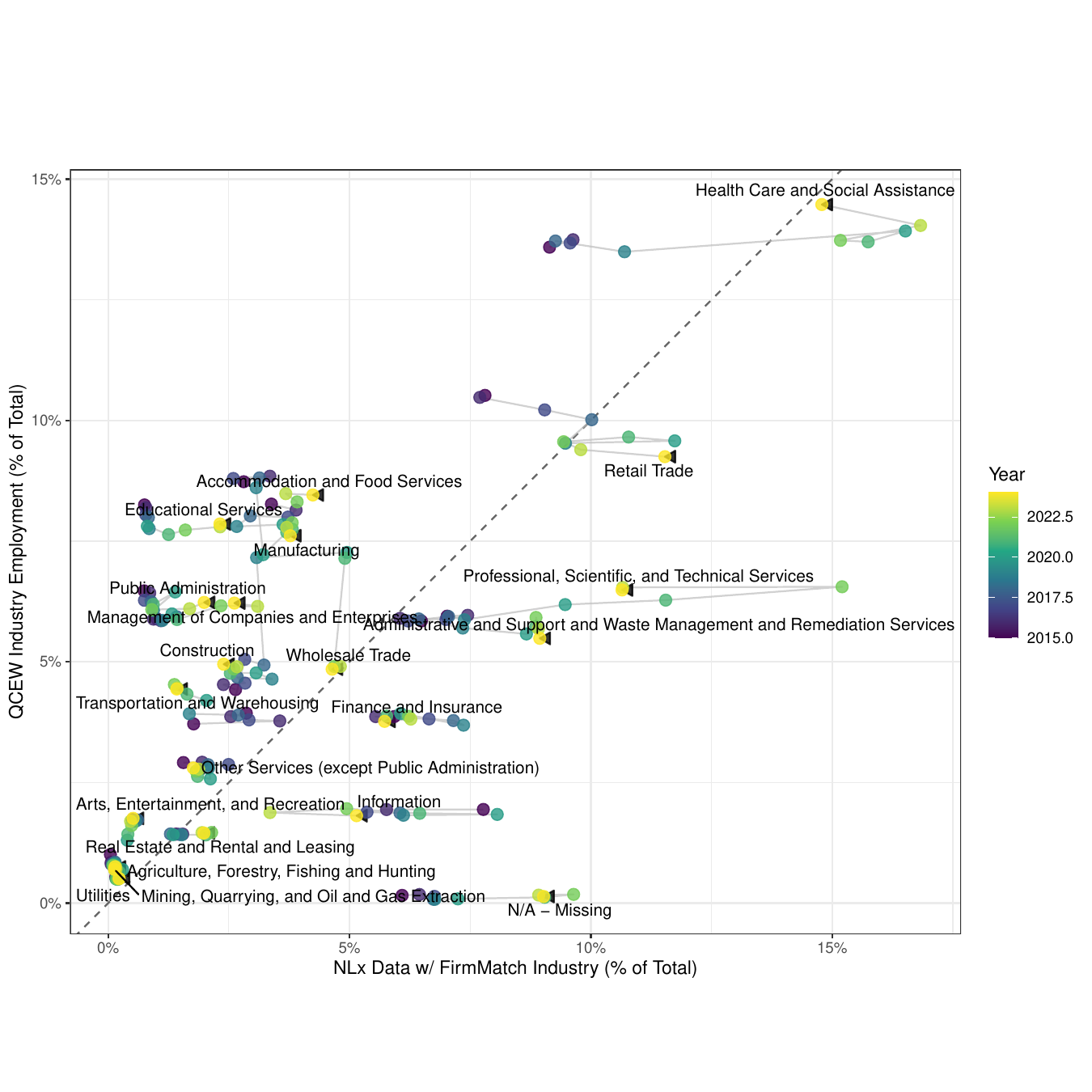}
            \small
\justifying
\emph{Note: }Paths show movement over years, and arrows show the final year change. This figure is based on aggregation of annual data by date compiled.
      \caption{Trajectory of Industry Correlation Over Time.}

  \label{fig:ind_corr_time}
\end{figure}

% \begin{figure}[h!]
%   \centering
%   \includegraphics[width=1\textwidth]{pdfs/industry_bar_comparison_2020.pdf}
%     \caption{Industry Distribution Comparison, 2020.}
%   \label{fig:ind_bar_2020}
% \end{figure}

% \begin{figure}[h!]
%   \centering
%   \includegraphics[width=1\textwidth]{pdfs/industry_correlation_2020.pdf}
%     \caption{Industry Correlation, 2020.}
%   \label{fig:ind_corr_2020}
% \end{figure}

% \begin{figure}[h!]
%   \centering
%   \includegraphics[width=1\textwidth]{pdfs/industry_bar_comparison_2015.pdf}
%       \caption{Industry Distribution Comparison, 2015.}
%   \label{fig:ind_bar_2015}
% \end{figure}

% \begin{figure}[h!]
%   \centering
%   \includegraphics[width=1\textwidth]{pdfs/industry_correlation_2015.pdf}
%     \caption{Industry Correlation, 2015.}
%   \label{fig:ind_corr_2015}
% \end{figure}

\clearpage 

% ===================================================================
\subsection{TitleMatch: Annual Occupation Distribution Comparison (vs. OES)}
% ===================================================================

To assess the representativeness of the NLx corpus coded by TitleMatch, we contrast the distribution of job ads across major 2-digit occupational groups in the NLx data with TitleMatch to the 2-digit occupational distribution for all employed workers in the U.S. using data from the BLS Occupational Employment Statistics (OES) survey \citep{BLS_OES}. OES data is collected in May and November of each year, and each annual OES report includes results from the prior 3 years of data collection. For each year, we first compare the percentage of jobs within each 2-digit occupation from OES to the comparable three-year data from NLx (by date compiled) with occupation coded by TitleMatch. 

%(see \url{https://www.bls.gov/oes/oes_ques.htm#qf1})

For the most recent year, Figure \ref{fig:occ_bar_2024} illustrates what studies of Lightcast / Burning Glass job ad data also show \citep{hershbein_recessions_2018}: job ad data do not reflect the distribution of jobs in the overall economy, and certain occupational groups are significantly over- or under- represented. Similar to Burning Glass data, NLx data coded by occupation with TitleMatch has a significantly higher proportion of  Management, Computer and Mathematical Occupations, Healthcare Practitioners, and Business and Financial Operations occupations than the labor market as a whole. 

\begin{figure}[h!]
  \centering
  \includegraphics[width=\textwidth]{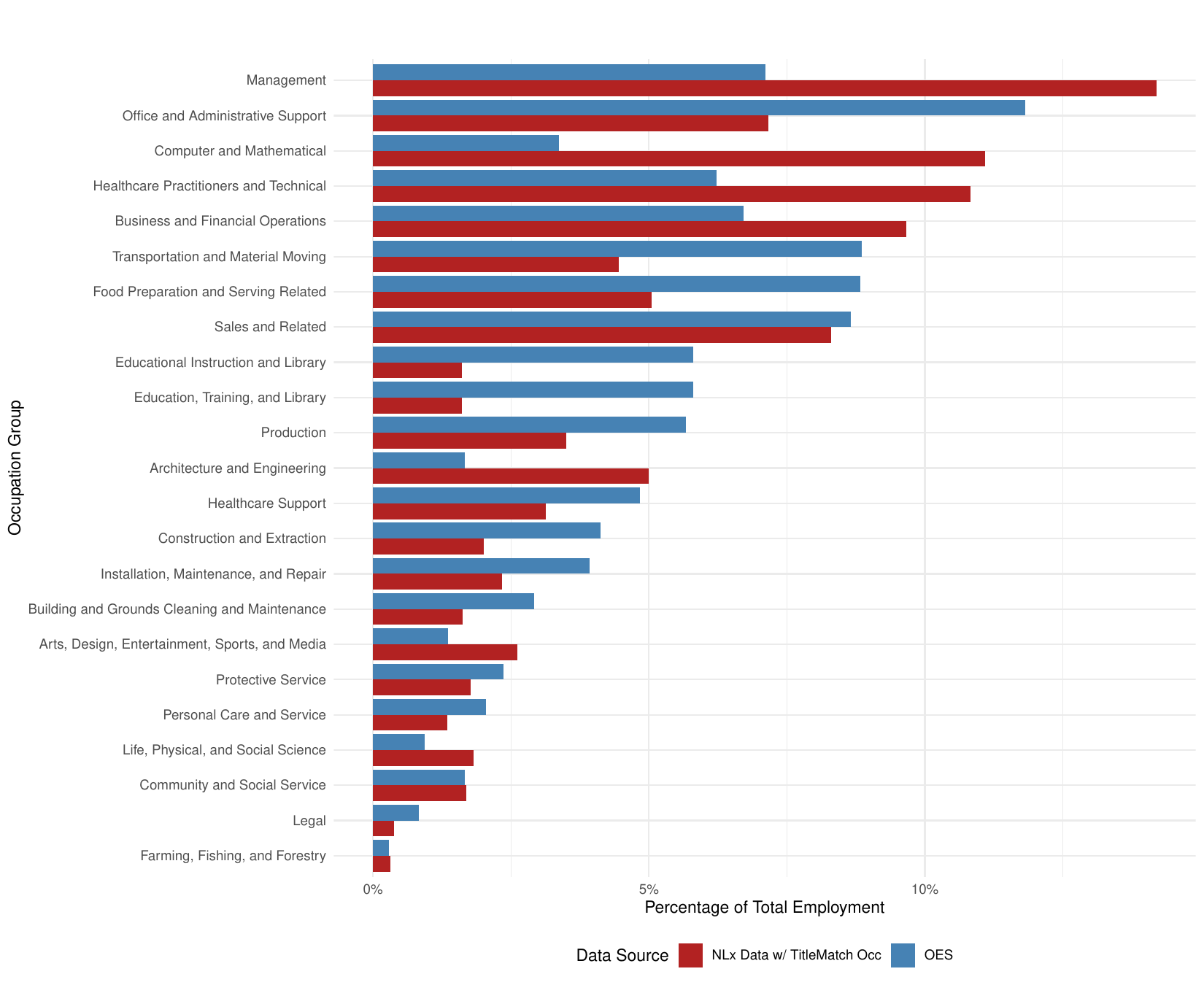}
        \small
\justifying
\emph{Note: }This figure is based on aggregation of 2024 data by date compiled.
      \caption{Occupation Distribution Comparison (3-Year Sample), 2024.}
  \label{fig:occ_bar_2024}
\end{figure}

Figure \ref{fig:occ_corr_2024} plots the correlation for 2024 data, and Figure \ref{fig:occ_corr_2024} plots changes in the correlation over the period observed. Earlier years are shaded darker, and arrows display the direction in the most recent year. It can be seen that management over-representation has become more significant over time, while under-representation of office and administrative support occupations has decreased over time. In several occupations, there is meaningful year-to-year fluctuation in the representativeness of the NLx data. As in \cite{hershbein_recessions_2018}, research designs with job ad data must carefully select appropriate weights, controls, and strategies given significant changes in underlying composition of the source. Additional figures in Appendix Section \ref{sec:robust} provide correlation plots and bar chart comparisons for earlier years.

\begin{figure}[h!]
  \centering
  \includegraphics[width=\textwidth]{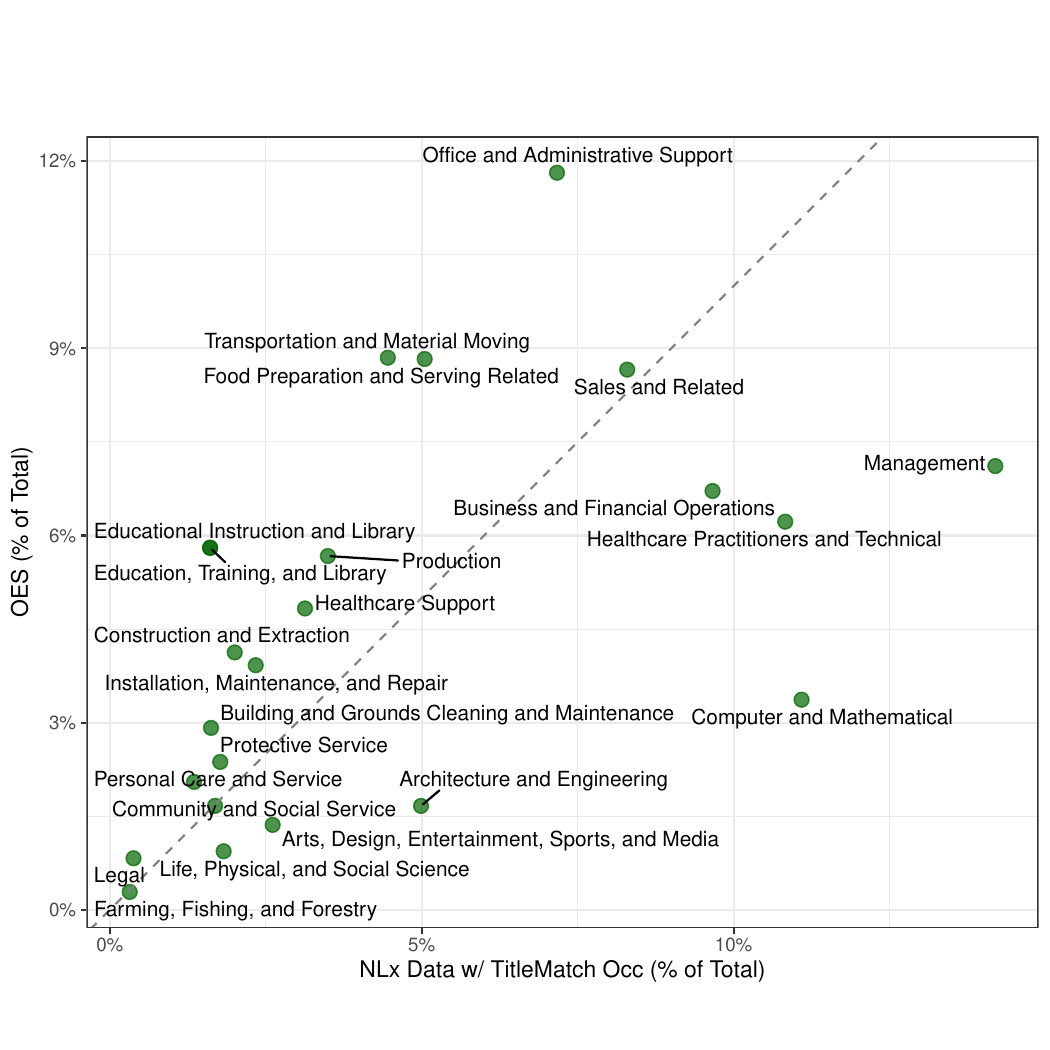}
              \small
\justifying
\emph{Note: }This figure is based on aggregation of 2024 data by date compiled. The Pearson correlation is 0.539.
      \caption{Occupation Correlation (3-Year Sample), 2024.}
  \label{fig:occ_corr_2024}
\end{figure}

\begin{figure}[h!]
  \centering
  \includegraphics[width=\textwidth]{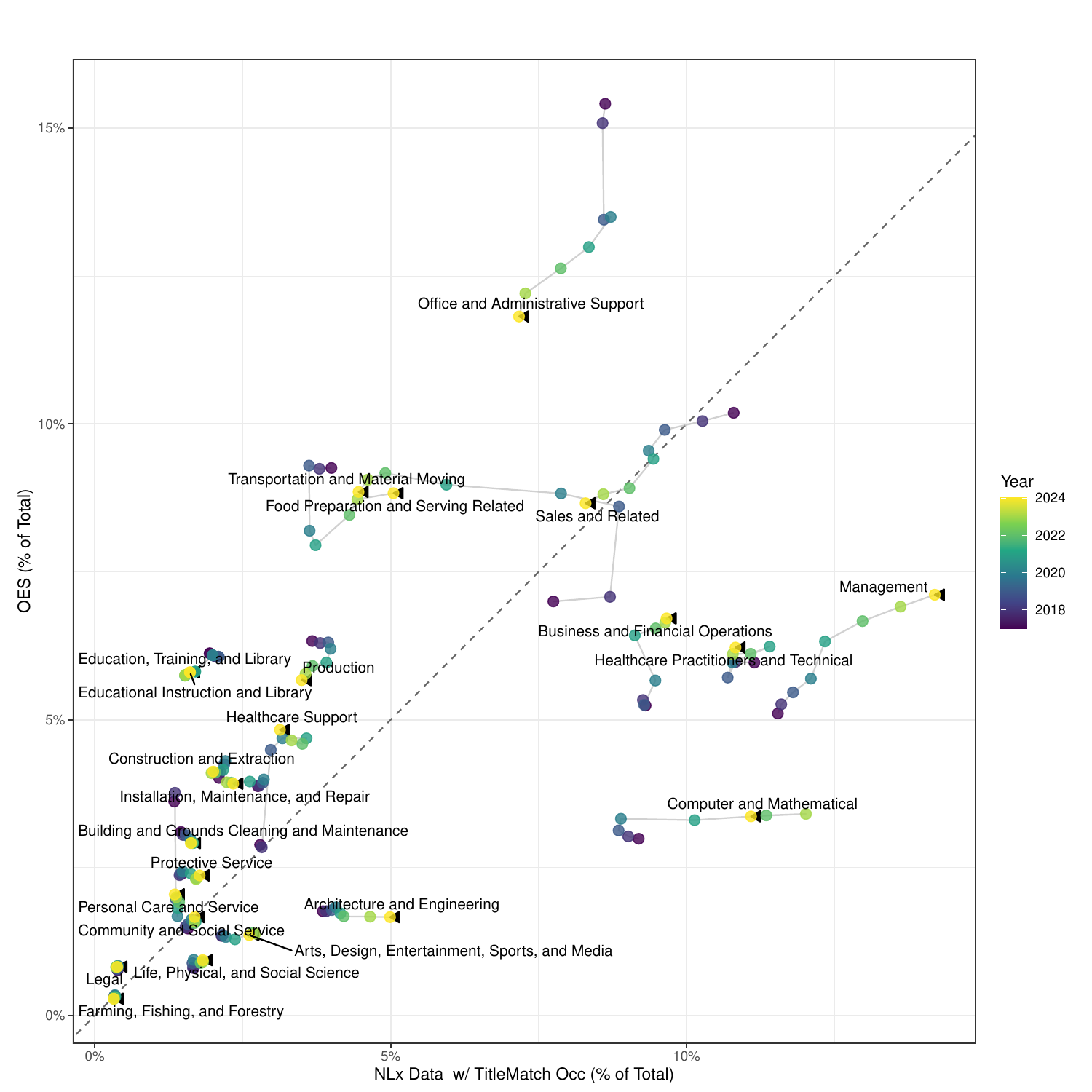}
              \small
\justifying
\emph{Note: }Paths show movement over years, and arrows show the final year change. This figure is based on aggregation of annual data by date compiled.
      \caption{Trajectory of Occupation Correlation Over Time.}
  \label{fig:occ_corr_time}
\end{figure}

% \begin{figure}[h!]
%   \centering
%   \caption{Occupation Distribution Comparison (3-Year Sample), 2021.}
%   \label{fig:occ_bar_2021}
%   \includegraphics[width=\textwidth]{pdfs/occupation_bar_comparison_2021.pdf}

% \end{figure}

% \begin{figure}[h!]
%   \centering
%     \caption{Occupation Correlation (3-Year Sample), 2021.}
%   \label{fig:occ_corr_2021}
%   \includegraphics[width=0.7\textwidth]{pdfs/occupation_correlation_2021.pdf}
% \end{figure}

% \begin{figure}[h!]
%   \centering
%   \caption{Occupation Distribution Comparison (3-Year Sample), 2018.}
%   \label{fig:occ_bar_2018}
%   \includegraphics[width=\textwidth]{pdfs/occupation_bar_comparison_2018.pdf}

% \end{figure}

% \begin{figure}[h!]
%   \centering
%   \caption{Occupation Correlation (3-Year Sample), 2018.}
%   \label{fig:occ_corr_2018}
%   \includegraphics[width=0.7\textwidth]{pdfs/occupation_correlation_2018.pdf}
% \end{figure}

\clearpage

% ===================================================================
\subsection{WageExtract: 2015-2019 Wage Distribution Comparison (vs. OES)}
% ===================================================================

Figures \ref{fig:wage_24_1} and \ref{fig:wage_24_2} display boxplots comparing OES data with wage results from WageExtract and occupation coding from TitleMatch. \cite{batra2023online} caution strongly against using wage information from job postings as a proxy for administrative wage data. While we repeat their finding that job ads consistently have wages that are ``lower in high wage occupations and higher in low wage occupations relative to the OES'', the availability of wage data in job postings has increased substantially in recent years, and recent years appear to be more aligned with benchmark data than earlier years. Further research is needed.

\begin{figure}[h!]
  \centering
  \includegraphics[width=\textwidth]{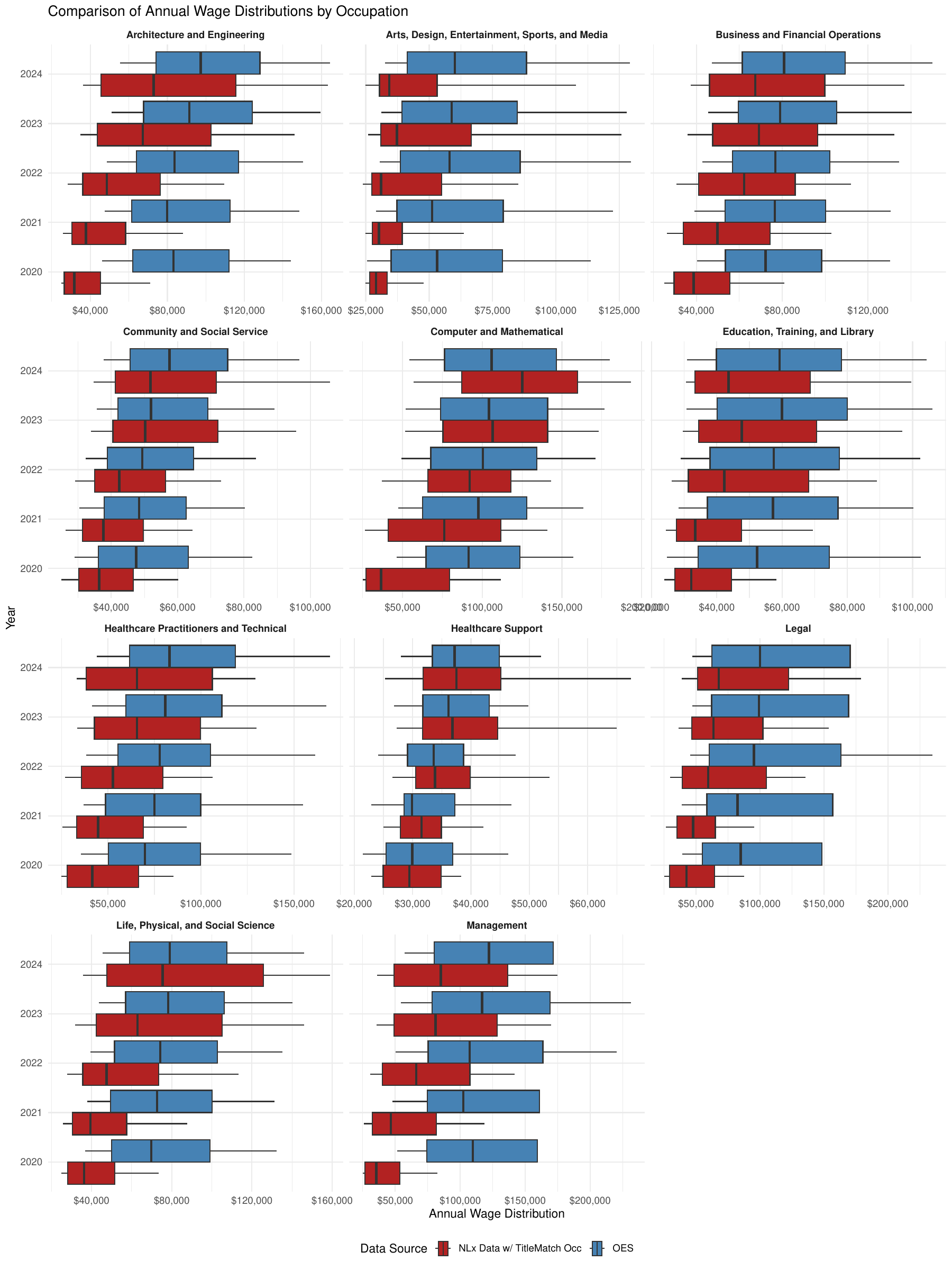}
      \caption{Wage Distributions, 2020-2024, for SOC codes below 32.}
  \label{fig:wage_24_1}
\end{figure}

\begin{figure}[h!]
  \centering
  \includegraphics[width=\textwidth]{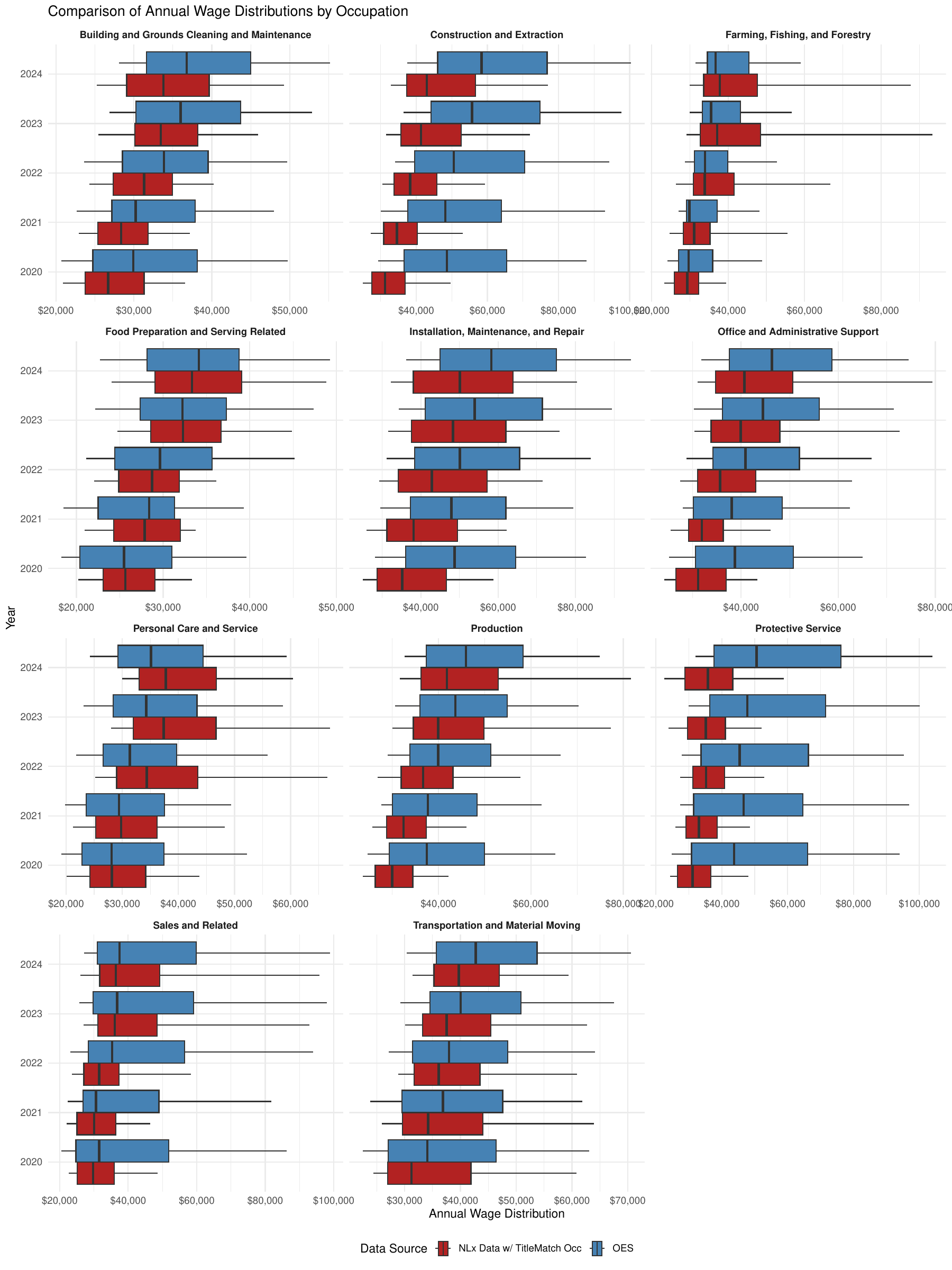}
      \caption{Wage Distributions, 2020-2024, for SOC codes above 32.}
  \label{fig:wage_24_2}
\end{figure}

\begin{figure}[h!]
  \centering
  \includegraphics[width=\textwidth]{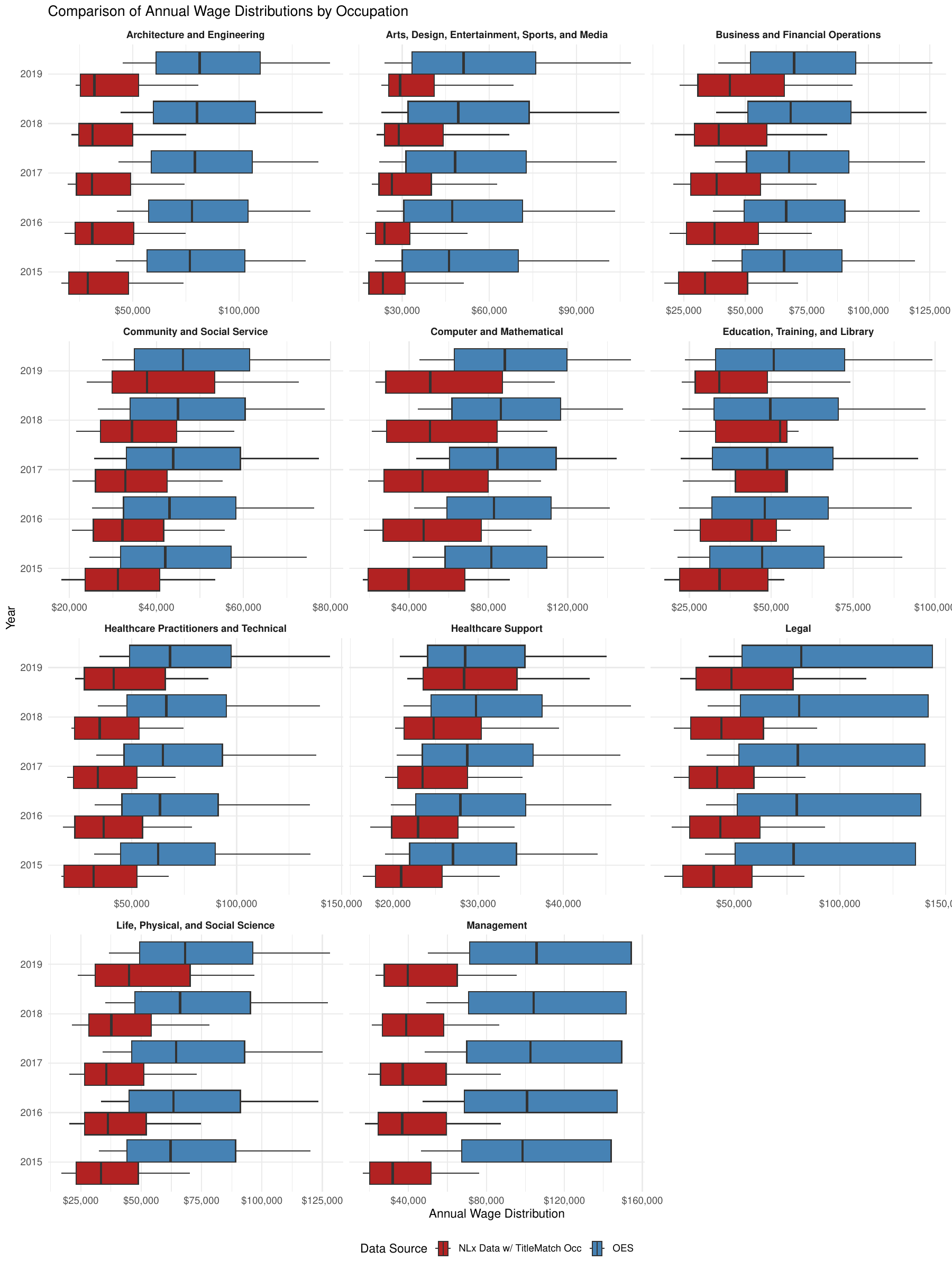}
      \caption{Wage Distributions, 2015-2019, for SOC codes below 32.}
  \label{fig:wage_19_1}
\end{figure}

\begin{figure}[h!]
  \centering
  \includegraphics[width=\textwidth]{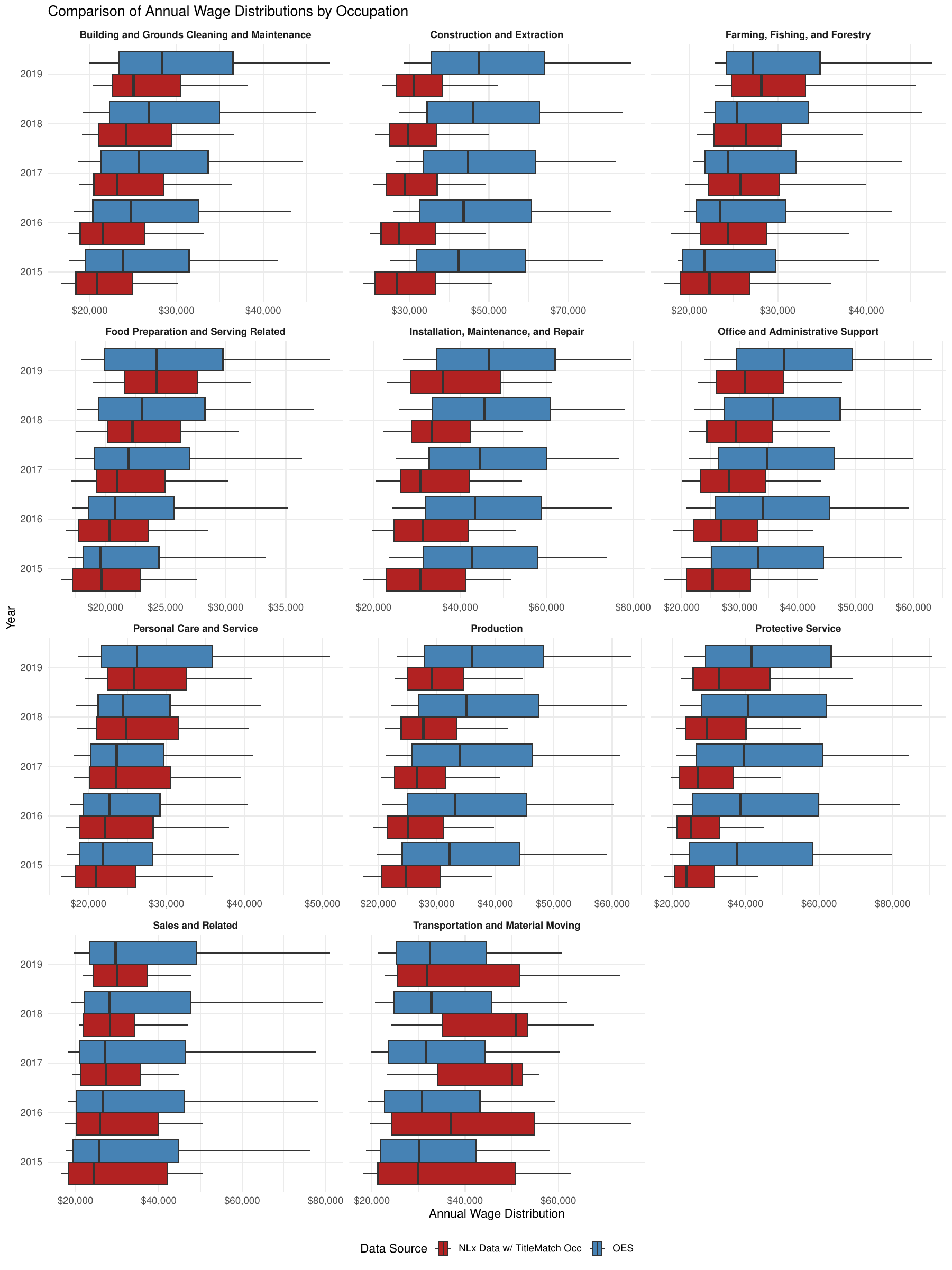}
    \caption{Wage Distributions, 2015-2019, for SOC codes above 32.}
  \label{fig:wage_19_2}
\end{figure}

%\section{Database}
\clearpage
\section{Custom Dictionaries}
\label{sec:all_dictionaries}

This section provides one illustrative term for each label inside a custom dictionary that we build. In addition to the illustrative terms, the full dictionary is available on our GitHub. 

\subsection{Custom Dictionary of Benefits}

From a random sample of job ads, we extracted sentences that contained an initial list of seed keywords plausibly related to employee benefits. From these sentences, we noted all ``interesting'' features, producing a novel list of benefits. We exact match items in the dictionary to the job ad corpus.

\label{sec:dict_benefits}
\begin{table}[htbp]
\centering 
\scriptsize
\caption{Dictionary of Benefits Features Extracted from Job Ads}
\label{tab:dict_benefits} 
\resizebox{\textwidth}{!}{
\begin{filecontents*}{csvs/benefits_9_29.csv}
Illustrative Term,Value,Additional Category
% discount,Discount,Additional_Compensation
Annual equity,Equity,Additional_Compensation
Employee Stock Ownership Plan,ESOP,Additional_Compensation
Employee Stock Purchase Plan (ESPP),ESPP,Additional_Compensation
Company match,Matching contribution,Additional_Compensation
Differential is paid,Pay differential,Additional_Compensation
Detention Pay,Special Pay,Additional_Compensation
Arcade games,arcade,Amenity
Company Vehicle,Auto,Amenity
BBQ,BBQ,Amenity
Catered breakfast,Breakfasts,Amenity
Fun Fridays,Fun Fridays,Amenity
Stocked micro kitchens,Kitchen,Amenity
Ping pong,Ping pong,Amenity
Basketball hoops,basketball,Amentiy
Benefits Information,Benefits,Benefits
Employee Assistance Program (EAP)* ,EAP,Benefits
Adoption Assistance ,Family,Benefits
Benefit for public transportation,Transit,Benefits
Full-time and Part-time benefits ,Benefits,Benfits
Identity Theft Protection,Theft,Benfits
Onboarding Bonus,SignonBonus,Bonuses
Opportunity for growth,Growth,Career
Equity plan,Equity,Compensation
Workplace culture,Culture,Culture
Career coaches ,Coaching,Education
Student loan repayment,College loan repayment,Education
Continuing Education Programs ,Continuing Education,Education
Comprehensive training ,Training,Education
Discounted tuition benefits ,Tuition Discount,Education
Tuition Assistance ,Tuition Reimbursement,Education
Computer and Cell Phone ,Computer/Cell,Expenses
Relocation,Relocation,Expenses
Catered lunches,Lunches,Food
"Health, dental, and vision",HealthFull,Health
Prescription drug,Rx,Health
Based on performance,Pay for performance,Incentives
Accident and Disability Insurance ,Accident,Insurance
Group Auto Insurance,Auto,Insurance
Paid Dental,Dental,Insurance
Dental and vision insurance,Dental and Vision,Insurance
Short/long term disability,Disability,Insurance
Disability and Life Insurance ,DisabilityLife,Insurance
BENEFITS Health Insurance,Health,Insurance
Medical and Dental Insurance ,HealthandDental,Insurance
Group Home Insurance,Home,Insurance
Income protection,Income,Insurance
Life Insurance,Life,Insurance
Pet Insurance,Pet,Insurance
Vision and Life Insurance,Vision and Life,Insurance
Military leave,Military Leave,Job Protected Leave
Paid parental leave ,Parental leave,Job Protected Leave
Legal Assistance,Assistance,Legal
Paid leave,PTO,Paid Leave
Vacation,Vacation,Paid Leave
Competitive base pay ,MarketPay,Pay
Wage will increase,Progression in job,payprogression
Wage will increase through training,training_progression,payprogression
Soccer,Soccer,Recreational
401 (k) ,401K,Retirement
401(k) plan with acompany match,401K Match,Retirement
403 (b),403b,Retirement
Pension,Pension,Retirement
Retirement savings plan,Retirement savings plan,Retirement
Retirement ,Retirement ,Retirement 
529 college,College savings plan,Saving
Deferred compensation,Deferredcomp,Saving
Savings,Savings,Saving
Series A stage startup,Startup,Startup
Dependent Care,FSA,Tax Shelter
Tax benefit,Tax benefit,Tax Shelter
Tax-free benefit for public transportation or parking expenses,Transit,Tax Shelter
FMLA,Disability,Unpaid Leave
Paid Holidays,Holidays,Vacation
Sick leave,Sick leave,Vacation
Week Home,Time Off Benefit,Vacation
\end{filecontents*}
\csvautobooktabular[respect all]{csvs/benefits_9_29.csv}
}
\end{table}

\clearpage
\subsection{Custom Dictionary of Education Levels}
From a random sample of job ads, we extracted sentences that contained an initial list of seed keywords plausibly related to levels of education. From these sentences, we noted bigrams and larger ``chunks'' of text that indicated a desired level of education and exact match them to the job ad corpus.

\label{sec:dict_education}
\begin{table}[htbp]
\centering 
\scriptsize
\caption{Dictionary of Education Features Extracted from Job Ads}
\label{tab:dict_educ} 
\begin{filecontents*}{csvs/education_9_29.csv}
Illustrative Term,Label,Additional Category
qualifications high school,HS,High School or GED
high school degree ged ,HSGED,High School or GED
ged equivalent,GED,High School or GED
associates degree ,Associates,Associates
accredited college or university ,Bachelors,College or University
bachelors masters,Masters,Postgraduate
bachelors masters or phd,Doctorate,Doctorate
\end{filecontents*}
\csvautobooktabular[respect all]{csvs/education_9_29.csv}
\end{table}

\subsection{Custom Dictionary of Drug, Background, and Criminal Checks}
\label{sec:dict_dbcs}

From a random sample of job ads, we extracted sentences that contained an initial list of seed keywords plausibly related to drug, background, and criminal checks, as well as recruitment policies related to hiring the formerly incarcerated. From the chunks of text we label, we exact match the dictionary to the job ad corpus.

\begin{table}[htbp]
\centering 
\small
\caption{Custom Dictionary of Background, Drug, and Criminal Check Features}
\label{tab:dict_dbc} 
\resizebox{\textwidth}{!}{
\begin{filecontents*}{csvs/background_9_29.csv}
Illustrative Term,Value,Additional Category
background,background,background
successfully complete a background,background check,background
criminal conviction,criminal background,background
maintaining a satisfactory criminal and credit record,criminal background,background
asked questions regarding any felony,criminal ask,background
no drug or alcohol related conviction,drug background,background
drug,drug,drug
drug free workplace,drugfree,drug
pre employment drug,drugtest,drug
dwi convictions,no dwi,drug
not be obligated to disclose sealed or expunged records of conviction,no background criminal,no background
offer pre employment,preemployment,preemployment
screening,screening,screening
test,test,test
fair chance employer,Fair Chance Employer,will_hire
fair chance hiring,Fair Chance Employer,will_hire
Criminal history will not automatically disqualify,Fair Chance Law,will_hire
Fair Chance Ordinance,Fair Chance Law,will_hire
Second Chance Act,Second Chance Act,will_hire
recognizes the value in second chances,Second Chance Employer,will_hire
second chance,Second Chance Employer,will_hire
Applicants with a felony conviction or pending/unresolved court cases may not qualify for licenses in all required states.,nohire_crime,will_not_hire
\end{filecontents*}
\csvautobooktabular[respect all]{csvs/background_9_29.csv}
}
\end{table}

\clearpage
\subsection{Custom Dictionary of Shift Features}
\label{sec:dict_shifts}

From a random sample of job ads, we extracted sentences that contained an initial list of seed keywords plausibly related to shifts and scheduling. We exact match items in the dictionary to the job ad corpus.

\begin{table}[htbp]
\centering 
\small
\caption{Dictionary of Shift Features Extracted from Job Ads}
\label{tab:dict_shift} 
\begin{filecontents*}{csvs/shifts_9_29.csv}
Illustrative Term,Value,Additional Category
Activity Schedule,Activity schedule, 
Daily schedule,Daily Schedule, 
Flexibility as to the hours,Flexible Hours,Flexible for Employer
Schedule allows,Flexible Schedule,Flexible for Employee
Flexibility as to the hours and schedule of work,Flexible Schedule,Flexible for Employer
Flexible scheduling,Flexible Schedule, 
Holiday schedule,Holiday Shifts, 
Hybrid Schedule,Hybrid Work Schedule, 
Night work,Night Shift , 
Closed most holidays ,No Holidays,Scheduling 
Overtime,Overtime, 
40 hour,Predictable Hours,Full Time
Weekly duties and schedule,Predictable Schedule,Weekly
Regular Schedule,Predictable Schedule, 
Project schedule,Project Schedule, 
Rotating schedule,Rotational Schedule, 
Retention Schedule for a minimum of three years,Salary Schedule, 
Schedule,Schedule, 
Schedule of appointments,Schedule others,Responsibility
Schedule appointments,Schedule Others, 
Shift,Shift, 
Shift pay,Shift Pay,Additional_Compensation
Schedule Per Diem,Shift Pay, 
Schedule Type,Specific Shfit, 
Shift Day,Specific shift,Day
Schedule including early morning,Specific Shift,First Shift
12 Hour Shift,Specific Shift,Long
End of the day shift,Specific Shift,Second Shift
During a shift,Specific Shift, 
3rd Shift,Specific Shift ,Night Shift
2nd Shift,Specific Shift ,Second Shift
Entry Level Shift,Specific Shift ,Seniority based shifts
Overnight shift ,Specific Shift ,Third Shift
End of shift,Specific Shift , 
Special shift,Specific Shifts,Flexible for Employer
Supervision of a shift,Supervise Shfits, 
Shift Lead,Supervise Shifts, 
Temp to hire position,Temp to Hire, 
Weekend Shift ,Weekend Shift, 
Friday schedule,Weekly Schedule, 
\end{filecontents*}
\csvautobooktabular[respect all]{csvs/shifts_9_29.csv}
\end{table}
\clearpage

\clearpage
%\section{Appendix References}
%\bibliography{onet.bib}

\end{document}